\documentclass[fleqn,usenatbib]{mnras}

\usepackage[T1]{fontenc}

\DeclareRobustCommand{\VAN}[3]{#2}
\let\VANthebibliography\thebibliography
\def\thebibliography{\DeclareRobustCommand{\VAN}[3]{##3}\VANthebibliography}

\usepackage{graphicx}	
\usepackage{amsmath}	
\usepackage{amssymb}	
\usepackage{color}
\usepackage[dvipsnames]{xcolor}
\usepackage{colortbl}
\usepackage{bm}
\usepackage{mathalpha}
\usepackage{newtxtext,newtxmath}
\usepackage[left]{lineno}
\newcommand{\grayline}{\arrayrulecolor{gray}\hline\arrayrulecolor{black}}

\title[Red giant -- jet collisions I]{Red giant -- jet collisions in galactic nuclei I:\\ 3D hydrodynamical model of a few stellar orbits}

\author[Kurf\"urst et al.]{
Petr Kurfürst,$^{1}$\thanks{E-mail: petrk@physics.muni.cz (PK)}
Michal Zaja\v{c}ek,$^{1}$\thanks{E-mail: zajacek@physics.muni.cz (MZ)}
Norbert Werner$^{1}$
and Ji\v{r}í Krti\v{c}ka$^{1}$
\\
$^{1}$Department of Theoretical Physics and Astrophysics, Faculty of Science, Masaryk University, Kotl\'a\v{r}sk\'a 2, 611 37 Brno, Czech Republic
}

\date{Accepted XXX. Received YYY; in original form ZZZ}

\pubyear{2015}

\begin{document}
\label{firstpage}
\pagerange{\pageref{firstpage}--\pageref{lastpage}}
\maketitle

\begin{abstract}
Several models have been proposed to explain missing red giants (RGs) near the Galactic centre. Recently, a scenario has been suggested that predicts, among other processes, a long-term ablation of the surface layers of RGs during their repetitive passages through the Galactic jet (Zaja\v{c}ek et al., 2020). In this study, we perform detailed three-dimensional numerical modelling of this phenomenon. We calculate the ablation rate of the surface layers of a RG orbiting the supermassive black hole (SMBH) as it passes through the nuclear jet. In particular, we model the jet-star interaction for approximately 10 passages for the closer orbital distance of $10^{-3}\,\text{pc}$ and 2 passages for $10^{-2}\,\text{pc}$. We find that the mass loss due to ablation by the jet behaves with time as $\Delta M_{\star}\propto \sqrt{t}$ and the total ablated mass during a single active galactic nucleus (AGN) phase ($10^5$ years) is $\sim 10^{-4}\,M_{\odot}$. We arrive at similar rates of the stellar ablation for the relatively smaller jet luminosity $10^{42}\,\text{erg}\,\text{s}^{-1}$ as in the previous analytical calculations. For larger jet luminosities of $10^{44}$ and $10^{48}\,\text{erg}\,\text{s}^{-1}$, the ablation rates inferred from $\sim 10$ interactions as well as extrapolated power-law fits are significantly lower than analytical values. Overall, the mass ablation rate per interaction and the extrapolated cumulative mass loss during the jet activity are comparable to the stellar-wind mass loss. For the smallest orbital distance of $10^{-3}\,\text{pc}$, we also track the thermal behaviour of the stellar surface layer, whose temperature appears to grow rapidly during the first 10 passages from $\sim 3600\,{\rm K}$ (spectral type M) to $\sim 8500\,{\rm K}$ (spectral type A). RG-jet interactions can thus lead to observable changes in the nuclear stellar population during the jet existence.
\end{abstract}




\section{Introduction}

A unified picture of active galactic nuclei \citep[AGN;][]{1985ApJ...297..621A,1995PASP..107..803U,2015ARA&A..53..365N,2021bhns.confE...1K,2023Ap&SS.368....8C,2024SSRv..220...29Z,2025arXiv250119365Z} consists typically of one or more supermassive black holes (SMBHs), an accretion disc, broad- and narrow-line regions, dusty molecular torus, and a jet for radio-loud sources. In this picture, stars are often not considered and their influence on the accretion onto SMBHs is neglected \citep[see, however, ][and references therein]{2002A&A...387..804V,2021ApJ...917...43S,2024MNRAS.532.2143K,2024arXiv241012090Z}. On the other hand, most galaxies with the total stellar masses larger than $\sim 10^8\,M_{\odot}$, including the Milky Way, host very dense nuclear star clusters (NSCs) in their centres \citep{2014CQGra..31x4007S,2020A&ARv..28....4N}. In our Galactic centre, fast-moving stars in the innermost part of the NSC have been identified that orbit the central SMBH (Sgr\,A*) on the scale of milliparsecs ($\sim 10^{-3}\,{\rm pc}\simeq 200\,{\rm au}$) or potentially even closer during their periapses \citep{2002Natur.419..694S,2012Sci...338...84M,2020ApJ...889...61P,2020ApJ...899...50P,2022ApJ...933...49P}. This implies that stellar orbits in Galactic nuclei could intersect the accretion flow as well as the jet. Several theoretical and computational studies have been dedicated to star-accretion disc interactions \citep{2014ApJ...781L..18A,2016ApJ...823..155K,2020MNRAS.492..250A}, which generally indicate that for inclined stellar orbits, the ablation of stellar atmospheres can only be significant if the disc is fragmenting, i.e. it would contain denser self-gravitating clumps. When stars are embedded in the accretion disc and corotating, they can actually gain mass by accretion of fresh material from the surrounding disc \citep{2020MNRAS.498.3452D,2021ApJ...910...94C}. However, the collisions of stars with jets and the associated impact on the nuclear stellar population have not been studied in detail until the seminal analytical study by \citet{2020ApJ...903..140Z}.

So far the interaction of stars with the jet has mainly been studied with the focus of studying the impact of stars on the jet flow and its emission, mainly in terms of the generated non-thermal radiation and cosmic rays in strong shocks, mass-loading, and the associated slowing down of jets, and their chemical enrichment via stellar winds \citep[see e.g.][]{1994MNRAS.269..394K,2012ApJ...749..119B,2012A&A...539A..69B,2013MNRAS.436.3626A,2015ApJ...807..168B,2016A&A...591A..15D}. In this study, we instead look at how jets can affect stars whose orbits intersect the jet cone. We focus on red giants (RGs) since the jet generally reaches the photosphere of evolved stars because of their slow stellar winds with velocities in the range of $\sim 10-100\,{\rm km\,s^{-1}}$ that cannot balance the jet kinetic pressure above the stellar photosphere. This leads to a more profound ablation of stellar material in comparison with early-type OB stars with powerful stellar winds \citep{2020ApJ...903..140Z}.  

Another reason for studying RG-jet collisions is that most massive galaxies host nuclear star clusters \citep[][]{2020A&ARv..28....4N} that seem to consist predominantly of late-type, evolved stars, such as RGs and supergiants \citep{2020A&A...641A.102S}. In the nearby Universe, these galaxies are mostly quiescent, i.e. they accrete via the radiatively inefficient accretion flows, such as those represented by the class of advection-dominated accretion flows (ADAFs). ADAFs are geometrically thick, optically thin, and strongly magnetised, hence they can support the launching of collimated magnetised outflows -- jets \citep{2014ARA&A..52..529Y} -- that are associated with the radio-mechanical feedback on the scales from the Bondi radius to galaxy groups and clusters \citep{2019SSRv..215....5W,2022NatAs...6.1008Z}. It is thus natural that in the nuclei of galaxies, RGs repeatedly interact with jets on subparsec scales, yet the impact of the jet on the stellar properties in NSCs has not been studied and quantified in detail previously. 

This is also related to the observed dearth of RGs in the inner $\sim 0.5\,{\rm pc}$ of the Galactic centre \citep{1990ApJ...359..112S,1996ApJ...472..153G}. Recently, the lack of late-type stars has been confirmed mainly for brighter RGs and supergiants as manifested by their core-like surface-density distribution \citep{2020A&A...641A.102S}, while fainter late-type stars do exhibit a cusp-like distribution. Specifically, \citet{2018A&A...609A..26G} report about 100 missing giants within $\sim 0.3$ pc and \citet{2019ApJ...872L..15H} infer the number of missing bright giant stars to be $\sim 4-5$ within $\sim 0.04$ pc. This points towards a mechanism acting preferentially on bigger stars that orbit closer to the SMBH. Suggested processes include (partial) tidal disruption of RGs by the SMBH, star-accretion disc collisions, star-jet interactions, star-star collisions, and the depletion of the inner region due to the inspiral of a second massive black hole \citep[see][and references therein]{2020ApJ...903..140Z}. It is plausible that different processes act simultaneously and their efficiency varies as a function of distance from the SMBH \citep{2020bhns.work..357Z}. Close to the SMBH, tidal disruption of red-giant envelopes takes place, while star-accretion disc interactions become more relevant on larger length-scales where the disc is gravitationally unstable and forms denser, self-gravitating clumps where the Toomre stability criterion drops below unity \citep{2004ApJ...604L..45M}. The ablation of RGs by the jet seems to be efficient enough on intermediate scales of $\lesssim 0.04\,{\rm pc}$ \citep[i.e. within the S cluster;][]{2020ApJ...903..140Z} and the star-star and star-compact remnant collisions can complement the other red-giant depletion mechanisms at all scales \citep{2009MNRAS.393.1016D}, depending on the stellar distribution and star-formation history. It is not clear yet whether Sgr~A* launches a faint jet in its current, extremely low-luminous state, though it is quite likely based on the current models of galaxy evolution that the jet activity and the outflows could have been significantly more powerful in the past. This is also hinted by bipolar $\gamma$-ray Fermi bubbles \citep{2010ApJ...724.1044S}, X-ray eROSITA bubbles \citep{2020Natur.588..227P}, as well as other multiwavelength data \citep{2019ApJ...886...45B,2019Natur.573..235H}, which point towards a potential enhanced jet activity a few million years ago, when the jet kinetic luminosity could have reached $\sim 10^{44}\,{\rm erg\,s^{-1}}$ \citep{2012ApJ...756..181G,2022NatAs...6..584Y}.

The ablation of stars by the jet is a special and so far an unexplored case among stellar ablation mechanisms \citep{2018A&A...615A..78G}. Envelope stripping significantly impacts stellar evolution, e.g. producing Wolf-Rayet stars from massive stars and hot sub-dwarfs from less massive stars \citep{2016PASP..128h2001H}. In the context of the Galactic centre, collisions of evolved stars with the jet active in the past is expected to increase their effective surface temperature and thus affect the stellar spectral type. This could help address the problem of missing evolved cooler stars and at the same time the overabundance of young hot stars of spectral types O and B in the central $\sim 0.5$ parsecs. In addition, the interaction of late-type stars with a fast outflow or a jet in a galactic nucleus, which would result in the partial ablation of the envelope, could lead to the apparent decrease in the metallicity that is detected in the Galactic centre \citep{2025arXiv250311856F}.

In this manuscript, we study the collisions between a RG orbiting the SMBH with a nuclear jet from a numerical perspective. We perform 3D hydrodynamical simulations with the aim to quantify the amount of red-giant ablation based on a few ($\sim 5$) stellar orbits around the SMBH. The results are compared to a previous analytical analysis \citep{2020ApJ...903..140Z} and we discuss the implications of RG-jet interaction in the context of a short-term as well as a long-term AGN activity.

The paper is structured as follows. In Section~\ref{modsetup} we describe the model set-up including the jet, ambient medium, and the star. In Section~\ref{numsetup}, we describe the computational method, including the numerical scheme, codes employed, and the initial and the boundary conditions. Then, in Section~\ref{sec_results}, we include the main results of the numerical simulations of a few stellar orbits. Specifically, in Subsection~\ref{subsec_mass_ablation}, we quantify the ablation rates of a RG for different jet luminosities and distances. In addition, we also present calculations of the temperature evolution of the red-giant surface layers in Subsection~\ref{subsec_eff_temp}. The main results are discussed in Section~\ref{sec_discussion}. In Subsection~\ref{subsec_comparison_mass_loss}, we compare the ablation rates inferred from the simulations with the analytical model. Then in Subsection~\ref{subsec_mag_field}, we comment on the effects of the magnetic field and in Subsection~\ref{subsec_star_outflow}, we discuss the connection to observational data, mainly concerning the Galactic center. Furthermore, in Subsection~\ref{subsec_narrower_jet}, we discuss the effect of a jet opening angle on the mass-ablation rate and in Subsection~\ref{subsec_statistics} we assess the statistics of the red giant-jet interactions in the Milky Way-like nuclear star clusters. We provide a concise summary of the main results in Section~\ref{sec_conclusions}.  

\section{Model set-up}
\label{modsetup}
\begin{figure}
    \centering
    \includegraphics[width=\columnwidth]{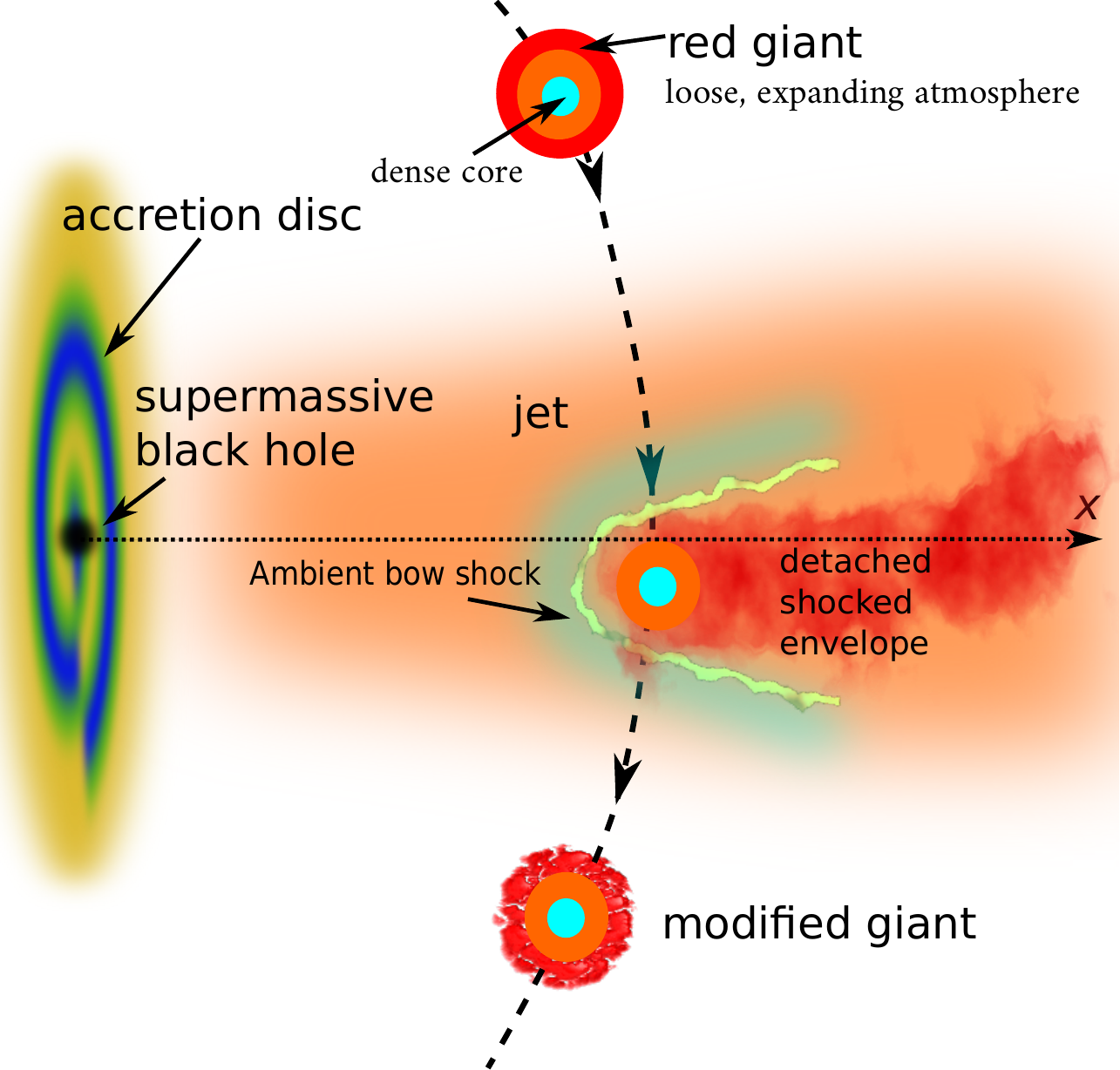}
    \caption{Illustration of the interaction between an evolved star (RG) and the jet in the vicinity of the SMBH. As the star consisting of a dense core (blue) and an envelope (orange and red) collides with the fast jet, two shocks are formed: the ambient shock (yellow) and shock propagating through the envelope that are separated by a contact discontinuity. See also \citet{2020ApJ...903..140Z} for the basic set-up.}
    \label{fig_illustration}
\end{figure}

The numerical model studies the encounters of an evolved, late-type star -- a RG or a supergiant  -- with the jet associated with an AGN at different distances from the SMBH, see Fig.~\ref{fig_illustration} for an illustration. We assume that for most of the jet lifetime in the range $t_{\rm jet}\sim 10-10^5$ years, the RG passes repeatedly through the jet sheath within the same orbital plane. This may not be the case in a realistic galactic nucleus due to coherent torques by other stars, i.e. the resonant relaxation \citep{1996NewA....1..149R}, whose vector form can change the orbital angular momentum orientation on a relatively short timescale of a few million years \citep{2006ApJ...645.1152H}. However, the timescale of the vector resonant relaxation depends on the number of enclosed stars inside a certain radius $r$, $N_{\star}(r)$, which is generally uncertain, see \citet{2020ApJ...903..140Z} for the analysis of this effect in the Galactic centre. We also assume the star is non-rotating for computational simplicity.

Galaxies similar to the Milky Way undergo typically several AGN phases with the duty cycle of the order of $t_{\rm jet}$. Considering the total SMBH growth time of $t_{\rm growth}\sim 10^7-10^9\,{\rm yr}$ \citep{2015MNRAS.451.2517S}, the number of AGN events is $n_{\rm AGN}\sim t_{\rm growth}/t_{\rm jet}\sim 100-10000$, assuming that SMBHs grow mainly during relatively short AGN phases of $10^5$ years, which is also inferred based on the size of proximity zones and Ly$\alpha$ nebulae surrounding quasars at $z=6$ \citep{2025arXiv250500080D}. For the Galactic centre, the AGN phase could be associated with the infalls of giant molecular clouds every $10^6$ years, which trigger star-formation episodes whose imprints may currently be traced using the detected, kinematically coherent stellar disks and streamers \citep{paumard2006,2008ApJ...683L..37W,2014ApJ...787L..14W,2020ApJ...896..100A,2022ApJ...932L...6V}. In contrast, in early-type, radio elliptical galaxies, the jets are active most of the time and only their intensity appears to change on the timescale of $\sim 10^7$ years \citep{2019SSRv..215....5W,2022ApJS..258...30G}. Given the uncertainties on the duration of $t_{\rm jet}$, we consider $t_{\rm jet}=10^5$ years for simplicity for a longer, AGN phase. For some estimates, we also take into account a potentially short-lived jet phase following e.g. a tidal disruption event \citep[TDE;][]{2011Natur.476..421B}, for which we set $t_{\rm jet}$ to 10 years. 

For computational limitations, we focus here on a small number ($\sim 10$) of the collisions between an evolved star and a jet. The results concerning the cumulative mass loss due to the ablation by the jet are then extrapolated for both the shorter jet lifetime of 10 years (TDE phase) and the longer jet activity of $10^5$ years (AGN phase). This interaction has three key components that need to be considered in the model: (i) duration of the stellar interaction with the jet and the timescale outside the jet, (ii) jet model, and (iii) a model of the RG. These components and the values used for numerical simulations are discussed below.

\subsection{Repetitive star interactions with the jet}

\begin{figure*}
    \centering
    \includegraphics[width=\columnwidth]{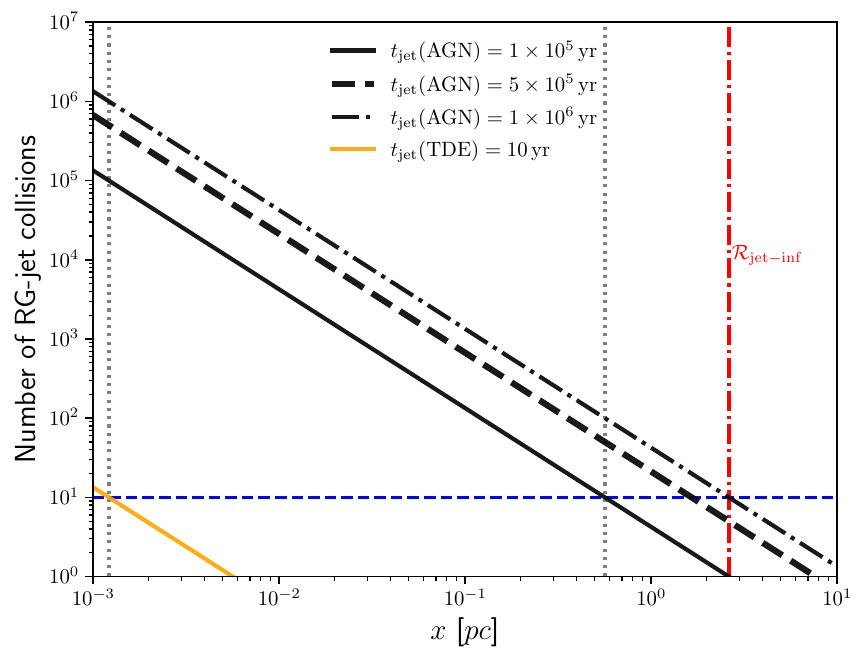}
    \includegraphics[width=\columnwidth]{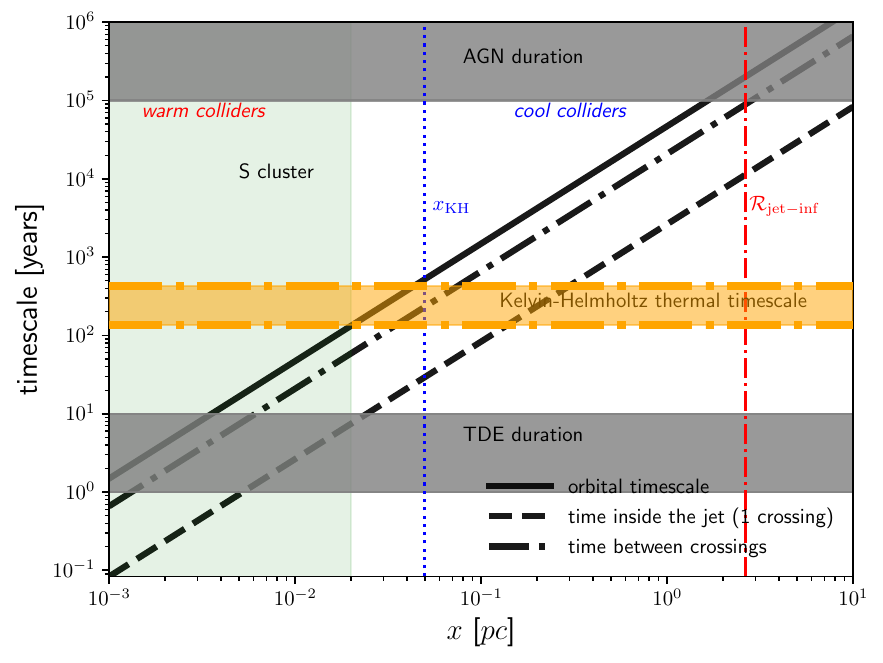}
    \caption{Number of jet-star interactions and the basic timescales. {\bf Left panel:} A number of RG-jet passages as a function of distance (in parsecs) from the SMBH. The lines stand for different jet lifetimes according to the legend. In particular, we plot the expected number of collisions for an AGN phase ($\sim 10^5-10^6$ years; black lines) and a short post-TDE phase ($\sim 10$ years; orange line). The blue dashed horizontal line stands for 10 passages, i.e. the order of magnitude we consider in this study. For an AGN phase ($10^5$ years), a star undergoes 10 collisions at $\sim 0.57$ parsecs (dotted vertical line), while for a shorter TDE phase (10 years), a star collides with the jet about 10 times at $\sim 1.2 \times 10^{-3}$ parsecs (dotted vertical line). The dot-dashed red line denotes the expected sphere of influence of an AGN jet defined by the distance from the SMBH where the star undergoes less than one collision with the jet, see also Eq.~\eqref{eq_jet_influence_radius}.  {\bf Right panel:} Orbital period, the time inside the jet for one passage, and the time between the jet crossings (all expressed in years) as a function of the distance (in parsecs) from the SMBH. All the calculations were performed for the SMBH mass of $M_{\bullet}=4\times 10^6\,M_{\odot}$ corresponding to Sgr\,A*. Gray-shaded regions depict the timescales of an active jet during the AGN phase (top) and the TDE phase (bottom). The Kelvin-Helmholtz thermal timescale is represented by an orange-shaded rectangle. It is estimated for a RG of $M_{\star}=1\,M_{\odot}$, $R_{\star}=100\,R_{\odot}$, and $T_{\star}=3000-4000\,{\rm K}$. Based on whether $t_{\rm KH}>t_{\rm out}$ or $t_{\rm KH}<t_{\rm out}$, we can distinguish the regions where warm colliders or cool colliders are expected. These are separated by $x_{\rm KH}$, see Eq.~\eqref{eq_KH_distance} (blue dotted line). The green-shaded area depicts the radial scale of the S cluster with the radius of $\sim 0.02$ pc.}
    \label{fig_timescales}
\end{figure*}

The basic dynamical timescale is the orbital timescale of the star at the distance $x$ from the SMBH. If not stated otherwise, we set the SMBH mass to the mass of Sgr\,A*, $M_{\bullet}\simeq 4\times 10^6\,M_{\odot}$ \citep{2010RvMP...82.3121G,2017FoPh...47..553E,2019Sci...365..664D,2022ApJ...933...49P}. Assuming that the star is gravitationally bound to the SMBH, we obtain the orbital timescale from the third Kepler law,

\begin{align}
   P_{\star} &= 2\pi \frac{x^{3/2}}{\sqrt{GM_{\bullet}}}\,\notag\\
   &=46.8\,\left(\frac{x}{0.01\,{\rm pc}} \right)^{3/2} \left(\frac{M_{\bullet}}{4\times 10^6\,M_{\odot}}\right)^{-1/2}\,{\rm yr}\,,
   \label{eq_orbital_time}
\end{align}
where we denote the jet axis coordinate as $x$ due to the horizontal left-right representation of the jet down-stream direction in most of the following figures of the individual models (see also Fig.~\ref{fig_illustration} for the illustration).

The orbital stellar velocity $v_{\star}$ can be estimated simply as the Keplerian velocity for a circular orbit,
\begin{align}
    v_{\star} &= \sqrt{\frac{GM_{\bullet}}{x}}\,\notag\\
    &=1311.7\,\left(\frac{x}{0.01\,{\rm pc}} \right)^{-1/2} \left(\frac{M_{\bullet}}{4\times 10^6\,M_{\odot}} \right)^{1/2}\,{\rm km\,s^{-1}}\,,
    \label{eq_star_velocity}
\end{align}
where we neglected the extended mass due to a distributed stellar and compact remnant mass. However, since the mass of the NSC $M_{\rm NSC}(r<r_{\rm m})\simeq 2M_{\bullet}$  at the gravitational sphere of influence of the SMBH with the radius $r_{\rm m}$, which is $r_{\rm m}\sim 2.5\,{\rm pc}$ for Sgr\,A* \citep{2013degn.book.....M}, the extended mass does not contribute significantly at the scales of the fraction of a parsec we consider here. 

When the star enters the jet, the contact discontinuity develops between the shock driven into the jet and the shock driven into the stellar wind. Adopting the balance between the wind kinetic pressure and the jet ram pressure, the stagnation radius is approximately located at \citep{2020ApJ...903..140Z},
\begin{equation}
    R_{\rm stag}=\sqrt{\frac{\dot{m}_{\rm w}v_{\rm w}c}{4L_{\rm j}}}x \tan{\theta}
    \label{eq_Rstag}
\end{equation}
where $\dot{m}_{\rm w}$ is the stellar mass-loss rate, $v_{\rm w}$ is the terminal stellar-wind velocity, $L_{\rm j}$ is the jet kinetic luminosity, and $\theta$ is the jet half-opening angle. For $R_{\rm stag}\gtrsim R_{\star}\sim 10\,R_{\odot}$, the stellar-wind velocity must be larger than,
\begin{align}
    v_{\rm w} &\gtrsim \frac{4L_{\rm j}}{\dot{m}_{\rm w}c}\left(\frac{R_{\star}}{x\tan{\theta}} \right)^2\,\notag\\
    &\sim 346\left(\frac{L_{\rm j}}{10^{42}\,{\rm erg\,s^{-1}}} \right)\left(\frac{\dot{m}_{\rm w}}{10^{-7}\,{\rm M_{\odot}\,yr^{-1}}} \right)^{-1}\,\notag\\
    & \times \left(\frac{R_{\star}}{10\,R_{\odot}} \right)^2 \left(\frac{x}{10^{-3}\,{\rm pc}} \right)^{-2}\,{\rm km\,s^{-1}}\,
    \label{eq_stellar_wind}
\end{align}
where we adopted $\theta=10^{\circ}$ that will be used in the following calculations as well. Hence, for hot main-sequence stars with stellar-wind speeds of a few 100\,${\rm km\,s^{-1}}$, the stagnation radius develops generally above the photosphere. On the other hand, for a RG with a slow wind of a few $10\,{\rm km\,s^{-1}}$, the stagnation radius reaches the photoshere with the radius of $R_{\star}\sim 100\,R_{\odot}$. Hence, in this work, we consider evolved stars with slow winds, for which the ablation by the jet is anticipated to be the most profound. 

The bow shock develops as the RG enters the jet due to the supersonic motion of the jet material with respect to the RG atmosphere. The bow shock orientation is given by the relative velocity of the star with respect to the ambient medium $\mathbf{v}_{\rm rel}=\mathbf{v}_{\star}-\mathbf{v}_{\rm amb}$. Outside the jet, $\mathbf{v}_{\rm rel}\simeq \mathbf{v}_{\star}$ if the medium outside the jet is stationary or slowly moving, i.e. $v_{\rm amb}\ll v_{\star}$. Inside the jet with the mean velocity of $v_{\rm jet}=\eta c$, where $\eta\leq 1$, the relative velocity $\mathbf{v}_{\rm rel} \simeq -\mathbf{v}_{\rm j}$ since typically $v_{\star}\ll v_{\rm jet}$. More precisely, the relative velocity deviates from the jet velocity direction by angle $\beta=\arctan{[v_{\star}/(\eta c)]}$, which is $0.8^{\circ}$ for $\eta=0.3$. This also determines the direction of the symmetry axis of the formed bow shock. The magnitude of the relative velocity is $v_{\rm rel}\simeq \sqrt{v_{\star}^2+\eta^2c^2}\gtrsim \eta c$, largely independent of the distance $x$ in the sphere of gravitational influence of the SMBH unless $v_{\rm jet}$ becomes comparable to $v_{\star}(x)$.

The number of jet-star interactions on the length-scales of the Galactic centre S cluster can reach a few $10^3$ passages, assuming that the stellar orbit intersects the jet cone, as follows from,
\begin{align}
    n_{\rm cross}&=2\frac{t_{\rm jet}}{P_{\star}}=\frac{t_{\rm jet}\sqrt{GM_{\bullet}}}{\pi x^{3/2}}\,\notag\\
    &\simeq 4200 \left(\frac{t_{\rm jet}}{0.1\,{\rm Myr}} \right)\left(\frac{M_{\bullet}}{4\times 10^6\,M_{\odot}}\right)^{1/2}\left(\frac{x}{0.01\,{\rm pc}} \right)^{-3/2}\,.
    \label{eq_number_crossings}
\end{align}
According to Eq.~\eqref{eq_number_crossings}, the number of interactions depends mainly on the distance as well as on the AGN-phase duration. For $x\leq 0.5\,{\rm pc}$, $n_{\rm cross}\geq 10$, see also Fig.~\ref{fig_timescales} (left panel), which implies the potential relevance of repetitive encounters in the central parts of the Galactic centre. The sphere of jet influence on the stars can be defined by the radius inside which $n_{\rm cross} \geq 1$, see Eq.~\eqref{eq_number_crossings},
\begin{align}
   \mathcal{R}_{\rm jet-inf} &\leq \frac{t_{\rm jet}^{2/3}(GM_{\bullet})^{1/3}}{\pi^{2/3}}\,\notag\\
   &\simeq 2.6\,\left(\frac{t_{\rm jet}}{0.1\,{\rm Myr}}\right)^{2/3}\left(\frac{M_{\bullet}}{4\times 10^6\,M_{\odot}} \right)^{1/3}\,{\rm pc}\,,
   \label{eq_jet_influence_radius}
\end{align}
that approximately coincides with the sphere of influence of the Sgr~A* SMBH \citep{2013degn.book.....M} and is about twice smaller than a half-light radius of the Galactic centre NSC, $r_{\rm hl}=4.2 \pm 0.4\,{\rm pc}$ \citep{2014CQGra..31x4007S}. \footnote{As a caveat, the Keplerian velocity profile breaks down at the outskirts of the NSC due to an extended mass. In the first approximation, we may still use the Keplerian approximation for basic estimates, but the central mass could effectively approach the mass of the NSC, $M_{\rm NSC}=(2.5 \pm 0.4)\times 10^7\,M_{\odot}$ \citep{2014CQGra..31x4007S}, which would make the radius given by Eq.~\eqref{eq_jet_influence_radius} larger.}

The duration of each passage of the RG through the jet can be estimated as,
\begin{align}
    t_{\rm in}&=\frac{2R_{\rm j}}{v_{\star}}=\frac{2x^{3/2}\tan{\theta}}{\sqrt{GM_{\bullet}}}\,\notag\\
    &\simeq 2.6\,\left(\frac{x}{0.01\,{\rm pc}} \right)^{3/2}\left(\frac{\theta}{0.17} \right)\left(\frac{M_{\bullet}}{4\times 10^6\,M_{\odot}}\right)^{-1/2}{\rm yr}\,,
    \label{eq_inside_timescale}
\end{align}
 The ratio $t_{\rm in}/P_{\star}=\tan{(\theta})/\pi\simeq 0.056(\tan{\theta}/0.176)$ just depends on the half-opening angle. The time between jet-star interactions then is,
\begin{align}
    t_{\rm out}&=\frac{P_{\star}}{2}-t_{\rm in}=\frac{x^{3/2}}{\sqrt{GM_{\bullet}}}(\pi-2\tan{\theta})\,\notag\\
    &\simeq 20.8 \left(\frac{x}{0.01\,{\rm pc}} \right)^{3/2}\left(\frac{M_{\bullet}}{4\times 10^6\,M_{\odot}} \right)^{-1/2}\,{\rm yr}\,,
    \label{eq_outside_timescale}
\end{align}
where $t_{\rm out}$ was evaluated for $\theta=10^{\circ}$.The ratio $t_{\rm out}/P_{\star}=1/2-\tan(\theta)/\pi\simeq 0.444$. Naturally, the crossings through the jet occur twice during the orbital period once the stellar inclination is $\iota\gtrsim 90^{\circ}-\theta$ with respect to the plane perpendicular to the jet axis.  

We plot the orbital timescale $P_{\star}$ and timescales $t_{\rm in}$ and $t_{\rm out}$ as functions of the distance from the SMBH in Fig.~\ref{fig_timescales}, taking into account the range $(10^{-3},10)$ pc. Across this range, all the timescales are mostly shorter than the jet lifetime (shaded grey area), which implies multiple, repetitive encounters, especially inside the S cluster. For the distance range $(10^{-3},2.6)$ pc and the assumed jet lifetime $t_{\rm jet}=10^5\,{\rm yr}$, the number of jet-star interactions is in the range from $\sim 1.35\times 10^5$ down to one.

The character of the jet-star interaction is determined by the ability of a perturbed RG to cool down before the following collision with the jet. The giant can be heated because of (i) the envelope ablation, i.e. warmer, deeper layers are exposed that expand adiabatically, (ii) the propagating shock through the envelope. The duration of the perturbed puffed-up state of a RG due to the collisional heat is given by the thermal Kelvin-Helmholtz timescale \citep{2020ApJ...903..140Z},

\begin{equation}
    t_{\rm KH}\approx \frac{G m_{\star}^2}{R_{\star} L_{\star}}\,,
    \label{eq_KH_timescale}
\end{equation}
where the stellar luminosity can be calculated as $L_{\star}=\sigma T_{\star}^4 4 \pi R_{\star}^2$. Assuming the typical RG parameters, $T_{\star}\sim 3500\,{\rm K}$ and $R_{\star}\sim 100\,R_{\odot}$, we obtain $L_{\star}\sim 1300\,L_{\odot}$. Depending on whether $t_{\rm KH}>t_{\rm out}$ or $t_{\rm KH}<t_{\rm out}$, one can distinguish \textit{warm} or \textit{cool colliders}, respectively, as proposed by \citet{2020ApJ...903..140Z}. This puts a limit on the distance from the SMBH,
\begin{equation}
    x_{\rm KH}\approx \frac{GM_{\star}^{4/3}M_{\bullet}^{1/3}}{R_{\star}^{2/3}L_{\star}^{2/3}(\pi-2\tan{\theta})^{2/3}}\,,
    \label{eq_KH_distance}
\end{equation}
which for the parameters assumed above and $m_{\star}=1M_{\odot}$ yields, $x_{\rm KH}\sim 0.05\,{\rm pc}$, which within a factor of two coincides with the radius of the cluster of fast-moving B-type S stars \citep[S cluster;][]{1996Natur.383..415E,1997MNRAS.284..576E,1998ApJ...509..678G,2009ApJ...692.1075G}. Hence inside the S cluster, we expect late-type stars to be \textit{warm colliders} that get more extended than the interacting RGs further away that have had time to cool down before the subsequent collision. In the 3D hydrodynamical model presented here, we consider stellar distances of $10^{-3}$ and $10^{-2}\,\text{pc}$, hence the RGs are expected to behave as warm colliders. 

\subsection{Jet and ambient medium model}
\label{jetambmod}
When the star orbits outside the jet, we assume that it moves through Bondi-like hot flow whose density and temperature profiles are typically described as power-law functions \citep{2003ApJ...591..891B,2013Sci...341..981W,2015A&A...581A..64R},
\begin{align}
    n_{\rm a}&\approx n_{\rm B}\left(\frac{d_{\rm orb}}{r_{\rm B}} \right)^{-1}\,,\label{eq_densityB}\\
    T_{\rm a}&\approx T_{\rm B}\left(\frac{d_{\rm orb}}{r_{\rm B}} \right)^{-1}\,,\label{eq_tempB}
\end{align}
where $d_{\rm orb}$ is the distance of the star from the SMBH. The normalisation coefficients are set to the values inferred by \citet{2003ApJ...591..891B} for the Galactic centre, $n_{\rm B}=26\,{\rm cm^{-3}}$ and $T_{\rm B}=1.5\times 10^7\,{\rm K}$, which correspond to the number density and temperature approximately at the Bondi radius. The Bondi radius as such is approximately located at $r_{\rm B}=2GM_{\bullet}/c_{\rm s}^2=2GM_{\bullet}\mu m_{\rm H}/(k_{\rm B}T_{\rm B})\sim 0.14\,(T_{\rm B}/1.5\times 10^7\,{\rm K})^{-1}\,{\rm pc}$.

In the following, we assume that the jet plasma is matter-dominated, consisting of electrons and protons. The jet exerts the pressure on the passing star mainly in the form of the bulk motion of the jet plasma at the velocity of $v_{\rm jet}$, which results in the ram pressure of $P_{\rm j}=\Gamma \rho_{\rm j} v_{\rm jet}^2$, where $\Gamma$ is the Lorentz factor and $\rho_{\rm j}$ is the mass density inside the jet. The number density inside the hadronic jet can then be estimated as,
\begin{align}
    n_{\rm j}&=\frac{L_{\rm j}}{\mu  m_{\rm H}\left(\Gamma-1\right)c^2 v_{\rm jet} \pi x^2 \tan^2{\theta}}\,\notag \\
    &\simeq 53 \left(\frac{L_{\rm j}}{10^{42}\,{\rm erg\,s^{-1}}}\right)\left(\frac{x}{0.01\,{\rm pc}}\right)^{-2}\,{\rm cm^{-3}}\,,
    \label{eq_jet_number_density}
\end{align}
where we evaluated the number density for $\Gamma\sim 10$, which implies $v_{\rm jet}\simeq 0.995\,c\sim c$, and the kinetic jet luminosity $L_{\rm j}$ was scaled to $10^{42}\,{\rm erg\,s^{-1}}$ since we expect the jet luminosity in the range of $10^{41}-10^{44}\,{\rm erg\,s^{-1}}$ for Sgr~A*, where the upper limit is set by its Eddington luminosity, 
\begin{equation}
    L_{\rm Edd}=5.03 \times 10^{44} \left(\frac{M_{\bullet}}{4\times 10^6\,M_{\odot}} \right)\,{\rm erg\,s^{-1}}\,,
    \label{eq_Edd_lum}
\end{equation}
where we adopted Thomson cross-section for the scattering of photons on electrons. See also the discussion in \citet{2020ApJ...903..140Z} and references therein.

In the classical limit, i.e. for $v_{\rm jet}\ll c$, we obtain the jet number density from the definition of the jet power,
\begin{align}
    \frac{1}{2}\rho_{\rm j} v_{\rm jet}^3&=\frac{L_{\rm j}}{\pi R_{\rm j}^2},\notag\\
    n_{\rm j}&=\frac{2L_{\rm j}}{\mu m_{\rm H}v_{\rm jet}^3 \pi x^2 \tan^2{\theta}}\,,
    \label{eq_jet_number_density_nerel}
\end{align}
which follows from Eq.~\eqref{eq_jet_number_density} by setting $\Gamma-1\approx v_{\rm jet}^2/(2c^2)$.

The jet temperature is assumed to be $T_{\rm j}=10^{10}\,{\rm K}$, i.e. close to the inverse Compton limit, see \citet{2012A&A...539A..69B}. Outside the jet, the Mach number $\mathcal{M}=v_{\rm rel}/c_{\rm s}\sim v_{\star}/c_{\rm s}(T_{\rm a})\sim \sqrt{GM_{\bullet}\mu m_{\rm H}/(k_{\rm B}T_{\rm B}r_{\rm B})}\sim 0.704$ is approximately constant and indicates a subsonic or only mildly supersonic motion. Within the jet, $\mathcal{M}=v_{\rm rel}/c_{\rm s}(T_{\rm j})\sim c\sqrt{\mu m_{\rm H}/(k_{\rm B}T_{\rm j})}\sim 23.1$, and hence the relative motion is clearly supersonic, which leads to the formation of the bow-shock that is driven into the ambient medium.

In the numerical setup, apart from the Bondi spherical flow with the temperature and density profiles given by Eq.~\eqref{eq_densityB} and Eq.~\eqref{eq_tempB}, which mimics hot ADAF flow that is geometrically thick, we do not consider red-giant interactions with a denser, standard accretion disc. We neglect such interactions in this work for (i) jets and associated radio-mechanical feedback are generally found in galaxies with lower Eddington-ratio accretion flows that are hot and geometrically thick \citep{2022ApJS..258...30G,2022MNRAS.517.3682P}; in Appendix~\ref{app_ADAF}, we estimate the expected amount of stripping for such a hot flow during one RG-ADAF collision for the Eddington ratio of $\dot{m}=0.01$, where $\dot{m}\equiv \dot{M}/\dot{M}_{\rm Edd}$ is the relative accretion rate and represents the ratio of the actual accretion rate to the Eddington rate. The result is that the stripped mass could be comparable or rather lower than the ablation during one RG-jet collision for the jet luminosity of $L_{\rm j}=10^{42}\,{\rm erg\,s^{-1}}$. Hence, the effect can be enhanced by a maximum factor of two or rather less due to a smaller ADAF ram pressure in comparison with the jet. This effect is partially mitigated by the implemented Bondi-like density and temperature profiles in our model. The second reason (ii) is related to the innermost orbital radius of the RG considered in this work is $d_{\rm orb}=10^{-3}\,{\text pc}$, which translates to $d_{\rm orb}=5256\,(M_{\bullet}/10^{6.6}\,M_{\odot})^{-1}\,r_{\rm g}$ gravitational radii, and at this distance any standard accretion disc is expected to be diluted given the density profile $\rho_{\rm D}\propto x^{-15/8}$ \citep{2002apa..book.....F} or totally absent due to gravitational instability \citep{1994A&A...290...19H}. Even if there still is a continuous disc, the ram pressure due to collisions is not expected to be high enough to strip RGs of their envelope \citep[][]{2020MNRAS.492..250A}.  Hence, in this work, we do not find a need to include a standard accretion disc, which naturally saves a substantial computational time. However, in the future work, a combined interaction between the RG and the disc/jet system is relevant since it can lead to an even larger amount of the ablated mass than inferred here for the RG -- jet/Bondi flow interaction.

\subsection{Model of a colliding star}
\label{rgmod}
We focus on the evolved stars or RGs with slow stellar winds with terminal velocities of the order of $100\,{\rm km\,s^{-1}}$ or less \citep[e.g.][]{2024ApJ...967..120W}. Hence, the stagnation radius of the bow shock then effectively reaches stellar photosphere at $R_{\star}\sim 100\,R_{\odot}$ for $x\lesssim 0.02\,{\rm pc}$ considering $L_{\rm j}=10^{42}\,{\rm erg\,s^{-1}}$ and $\dot{m}_{\rm w}=10^{-7}\,{\rm M_{\odot}\,yr^{-1}}$ (at $\sim 0.027\,\text{pc}$ for $\dot{m}_\text{w}=10^{-7}\,M_\odot\,\text{yr}^{-1}$ and $v_\text{w}=50\,\text{km}\,\text{s}^{-1}$ and at $\sim 0.59\,\text{pc}$ for $\dot{m}_\text{w}=10^{-9}\,M_\odot\,\text{yr}^{-1}$ and $v_\text{w}=10\,\text{km}\,\text{s}^{-1}$), see Eqs.~\eqref{eq_Rstag} and \eqref{eq_stellar_wind}, which leads to the matter ablation from the stellar atmosphere. The internal 3D structure of the envelope is approximated using a polytropic radial profile calculated using the Lane-Emden equation (see Subsection~\ref{numone} for details). In addition to the outer layers, we also add the core region to the polytropic profile, whose radius is $\sim 10\,\%$ of the stellar radius and the mass of $\sim 20\,\%$ of the total stellar mass (see Subsection~\ref{numthree} for a detailed description). Unless stated otherwise, we set the stellar mass to $M_{\rm RG}=1\,M_{\odot}$ and its (initial) radius to $R_{\rm RG}=100\,R_{\odot}$. We also implement a spherically symmetric stellar wind which
initially surrounds the star. Its density and temperature are, however, too low to significantly affect
the modelled process. The wind is inserted as an initially static structure with the following parameters: the initial density $\rho_\text{w}$ was set to $\dot{m}_\text{w}/(4\pi r^2v_\text{w})=\rho_\text{0,w}(R_{\rm RG}/r)^2$,
where $\dot{m}_\text{w}=5\times 10^{-11}\,M_\odot \,\text{yr}^{-1}$ and $v_\text{w} = 25\,\text{km}\,{s}^{-1}$ can be regarded as the typical asymptotic wind speed for M-type RGs \citep[while the K-type RGs have the winds even less dense, e.g.][]{2024ApJ...967..120W}. The base density of the wind is thus $\sim 2\times 10^{-18}\,\text{g}\,\text{cm}^{-3}$. We inject the stellar wind with the same parameters during the period when the star is between the jet crossings. At certain time steps (according to the grid cells range), the stellar wind gradually grows to its new expansion front distance. We did not employ dust or line driving nor any other radiative processes in the models. This stellar wind, however, is always completely swept away by the dynamics of the jet as soon as the star enters it.

\begin{figure*}
    \centering
    \includegraphics[width=\textwidth]{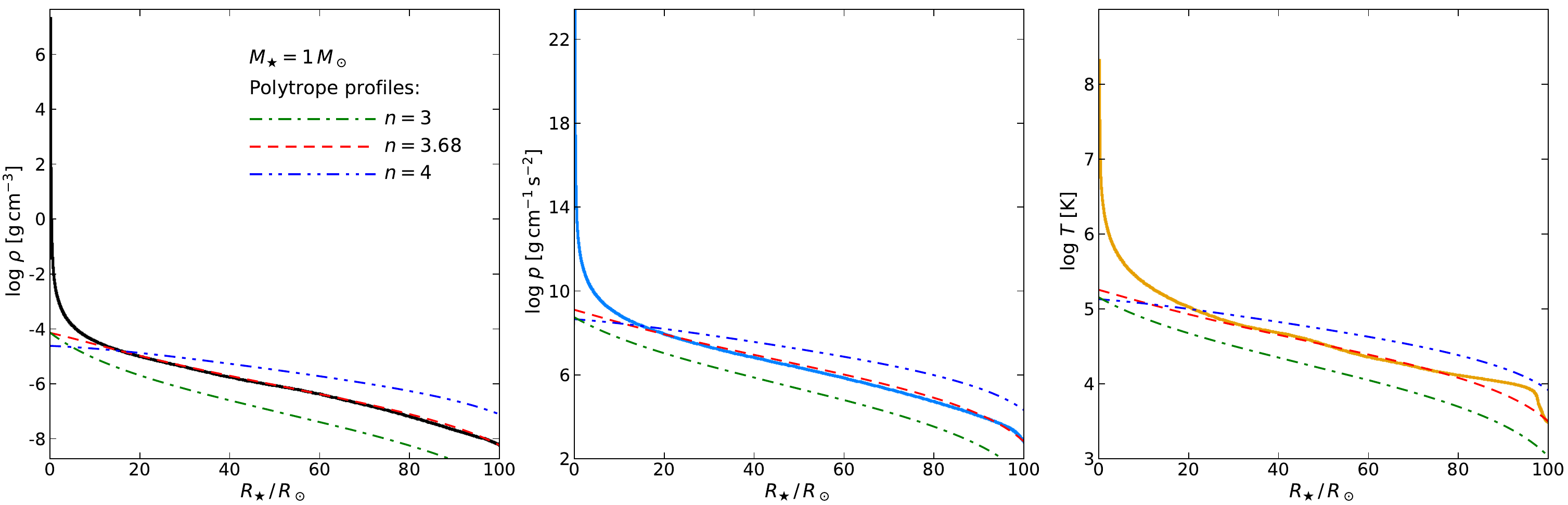}
    \caption{Illustrative comparison of the MESA calculated interior profile of an initial state of a RG with mass $1\,M_\odot$ and radius $100\,R_\odot$ (thick continuous lines) with polytrope profiles of different polytrope indexes $n$ (dashed or dashdotted lines, the values of $n$ that correspond to different linetypes/colours are denoted in the graph).  {\bf Left panel} - density profile, {\bf middle panel} - pressure profile, {\bf right panel} - temperature profile, all of them plotted in a logarithmic scale. The polytrope fits to the MESA model obviously favour the index $n\approx 3.68$ ($\gamma\approx 1.27$) in this case.}
    \label{fig_red_giant_lane_emden}
\end{figure*}

{\color{black}
\section{Numerical setup}
\label{numsetup}
Due to the asymmetry of the studied process in two directions, radial and orbital, which may in the spherical case likely affect the ablation rate in a three-dimensional sense, especially at the entry and exit phase of the star in and from the jet, we have performed full 3D models at least for the first 10 passes through the jet for the star with the lowest assumed orbital distance of $10^{-3}\,\text{pc}$ from the SMBH. In future work, it will probably be possible and appropriate to run some of the models using much less computationally costly 2D simulations and extrapolate the ablation rates from obtained comparative analogies.

For the 3D numerical calculation of the jet-star interactions during multiple orbits around the galactic nucleus, we choose the computational domain as a perfect cube with an edge length of $30\times 10^{12}\,\text{cm}$ or $2\,{\rm au}$, where the position of the star is exactly at its centre; see Figs.~\ref{masslosscurve1e42}\,--\,\ref{masslosscurve1e42_figeight} except for Fig.~\ref{masslosscurve1e42contours_figsecond} whose domain is extended in the jet direction to illustrate the behaviour of the flow downstream behind the star, but only for one passage of the star through the jet.

\subsection{Computational setup and codes}\label{numone}
We first calculated the initial hydrodynamical internal profile of the RG star using the 1D stellar evolutionary code MESA \citep[e.g.,][]{2011ApJS..192....3P}. We keep the mass $1\,M_\odot$ of the star fixed during the whole MESA simulation for simplicity (we add the stellar wind to the model later in the CASTRO code setup - see further in this section).
We stopped the calculation when the radius of the star reached $100\,R_\odot$. The other important parameters of the model were at that moment as follows: age $\approx 1.22\times 10^{10}\,\text{yrs}$, $T_\text{eff}\approx 3400\,\text{K}$, He core mass $\approx 0.4\,M_\odot$, and the Kelvin-Helmholtz (KH) timescale $\approx 1.93\times 10^2\,\text{yrs}$. As the next step, we calculate the stellar internal profile as a set of polytropic profiles in order to get the most similar internal structure to the MESA profile (at least for the most of the outer part of the stellar body). The fundamental advantage of the polytropically determined structure is that it is (unlike the MESA profile) a consequence of the same equation of state employed within the hydrodynamic code. An anomaly occurs near the centre of the star, where the actual distribution of state quantities increases steeply. The polytrope is therefore complemented by a point mass with a non-negligible fraction of the total stellar mass. The complete profile was calculated as spherically symmetric (1D) using a supplementary algorithm and was normalised to the mass of the star. In Figure~\ref{fig_red_giant_lane_emden}, we show the relationships between the different models of the initial state of the stellar interior, showing that the closest approximation of the outer polytropic stellar structure to the MESA model is provided by the polytrope with the index of $n = 3.68$. The further details of the initial-state model will be described later within this section.

During the work on the actual time-evolving multi-dimensional numerical models, we tested several computational codes including our own hydrodynamical code used in several previous works \citep{2014A&A...569A..23K,2018A&A...613A..75K,2020A&A...642A.214K,KKapplmath,2019A&A...625A..24K}.
The best performance was achieved by the widely used code \texttt{CASTRO} \citep{2010ApJ...715.1221A,2011ascl.soft05010A}.
The \texttt{CASTRO} code does not involve the relativistic effects, however, we do not find the necessity for implementing them for the current hydrodynamic models. In particular, we study the star-jet collisions at the distance of $\sim 5230 (x/0.001\,{\rm pc})(M_{\bullet}/4\times 10^6\,M_{\odot})^{-1}$ gravitational radii from the SMBH, where post-Newtonian effects are negligible on the timescale of $\sim 5$ stellar orbits studied in this work, especially for circular orbits considered here.

The main benefits of using the \texttt{CASTRO} code are as follows,
\begin{itemize}
\item  it solves the multi-component compressible hydrodynamics for astrophysical applications in 3D, including self-gravity,
\item  it provides the best resolution and the most reliable results in zones of shocks and contacts of environments mutually moving at very high velocities,
\item  it includes built-in interpolation procedures for embedding the semi-analytically pre-calculated 1D initial states; by using them, we remapped the polytropic stellar internal profiles to the CASTRO initial state (for the initial state of the jet and ambient medium implementation - see the further description).
\end{itemize}
For the hydrodynamic calculations, the CASTRO code employs the fully compressible equations (without reactions and other source terms that we do not apply here) whose conservative form is
\begin{align}\label{conseqns}
\frac{\partial\rho}{\partial t}&=-\vec\nabla\cdot\left(\rho\vec{v}\right),\\
\frac{\partial\rho\vec{v}}{\partial t}&=-\vec\nabla\cdot\left(\rho\vec{v}\vec{v}\right)-\vec\nabla p +\rho\vec{g},\\
\frac{\partial\rho E}{\partial t}&=-\vec\nabla\cdot\left(\rho E+p\right)\vec{v}+\rho\vec{v}\cdot\vec{g},
\end{align}
where $\rho,\vec{v}$, and $p$ are the density, velocity, and pressure, respectively, and $E=e+\vec{v}\cdot\vec{v}/2$ is the total energy density with $e$ representing the internal energy density. The code also complements the above system with an internal energy equation
\begin{align}\label{interneeqns}
\frac{\partial\rho e}{\partial t}=-\vec{v}\cdot\vec\nabla\left(\rho e\right)-\left(\rho e+p\right)\vec\nabla\cdot\vec{v}
\end{align}
which has a benefit namely for high Mach number flows where the kinetic energy can dominate the total gas energy, making the subtraction of the internal energy from the total one numerically unreliable. {We do not include the heat conduction in the energy equation nor the radiative processes in this phase.}
For the dynamical simulation, we employed  an ideal gas equation of state (a gamma law gas)
\begin{align}
p=\left(\gamma-1\right)\rho e,
\end{align}
where the gas is assumed to be monoatomic with $\gamma=5/3$ (which is not inconsistent with the already mentioned polytropic index $n = 3.68$ or $\gamma\approx 1.27$ because we simulate here the dynamical action within the star\,-\,ambient medium interface where no polytropic behaviour occurs).

Besides that, the CASTRO code may potentially employ magnetohydrodynamics (MHD) and radiation hydrodynamics as a basis for possible calculations of observable characteristics which we regard as a  main contribution to the models in the future. The important technical issue is the adaptive mesh refinement (AMR) that provides simultaneous refinement of the grids in both space and time in case of steep gradients. 

\subsection{Initial state and boundary conditions}\label{numtwo}
The physical nature of the initial-state numerical setup of the RG was described in the previous Subsection~\ref{numone}. The initial state of the jet and the ambient medium follows the equations and parameters given in Section~\ref{jetambmod}. For computational simplicity, we reverse the kinematics of the orbital motion and run the model in the reference frame connected with the star. In that frame, the jet moves in the direction against the assumed orbital motion of the star (this applies for all the models). 

The initial settings in the CASTRO code \citep[see][including the nested links]{2011ascl.soft05010A} were as follows: the Cartesian grid cube was divided into $256^3$ cells, we used the classic Colella \& Woodward piecewise parabolic method (PPM) advection scheme in the whole domain \citep{1984JCoPh..54..174C}. The implemented physics is pure hydrodynamics, Poisson gravity outside a certain fraction of stellar radius (see the more detailed description in the next paragraph \ref{numthree}), Newtonian gravity centred to a star's centre inside this radius, and a sponge source term (to damp velocities in buffer regions between the domain of interest and the boundary conditions). We set the Courant-Friedrichs-Lewy (CFL) number to 0.5 for the time step control. We used the AMR maximum level 2 with the refinement ratio 2 between the adjacent levels as a compromise between accuracy and the computational cost.

The choice and arrangement of boundary conditions are of the special importance, either at the outer boundaries of the computational domain or at the ``inner boundary", meaning the interconnection zone between the stellar body and the surroundings. For this reason, we set the external and the ``internal" boundary conditions as follows:
\begin{itemize}
\item  ``Jet inflow boundary" (left-side boundary in Figs~\ref{masslosscurve1e42}\,--\,\ref{masslosscurve1e42_figeight}) is set as an inflow  with the fixed state quantities and velocities that correspond to selected models described in detail in Section~\ref{modsetup}. However, the values repeatedly change over time depending on whether the interface is in the ambient medium, the jet, or in the transition zone between these two environments.
\item  ``Orbital velocity inflow boundary" (bottom-side boundary in Figs~\ref{masslosscurve1e42}\,--\,\ref{masslosscurve1e42_figeight}) is also set as an inflow with the fixed state quantities and velocities that correspond to selected models and also repeatedly change over time depending on the type of environment.
\item  Remaining outer boundaries (right-side and top-side boundaries in Figs~\ref{masslosscurve1e42}\,--\,\ref{masslosscurve1e42_figeight} and the remaining two lateral sides) are set as an outflow (freely evolving).
\end{itemize}
The implementation of the initial states and boundary conditions allows for the proper behaviour of the stellar environment and its interactions with the outer stellar layers. Stellar mass loss due to ablation was calculated separately during postprocessing. We repeatedly integrated the mass within the original volume occupied by the star at certain time steps. We then simply compared the results from these time steps with the original mass of the star. Although this method allows us to obtain an assessment of the approximate ablated mass as a function of time, it is not entirely accurate since not all the mass inside the original volume is bound to the star. However, since the same methodology for the mass-loss estimate is applied to all the models, we can consistently compare the impact of different parameters on the mass-ablation rate. In the future work, we plan to design the algorithm for assessing directly the amount of mass bound to the star.

\begin{figure*}
    \centering
    \includegraphics[width=0.875\textwidth]{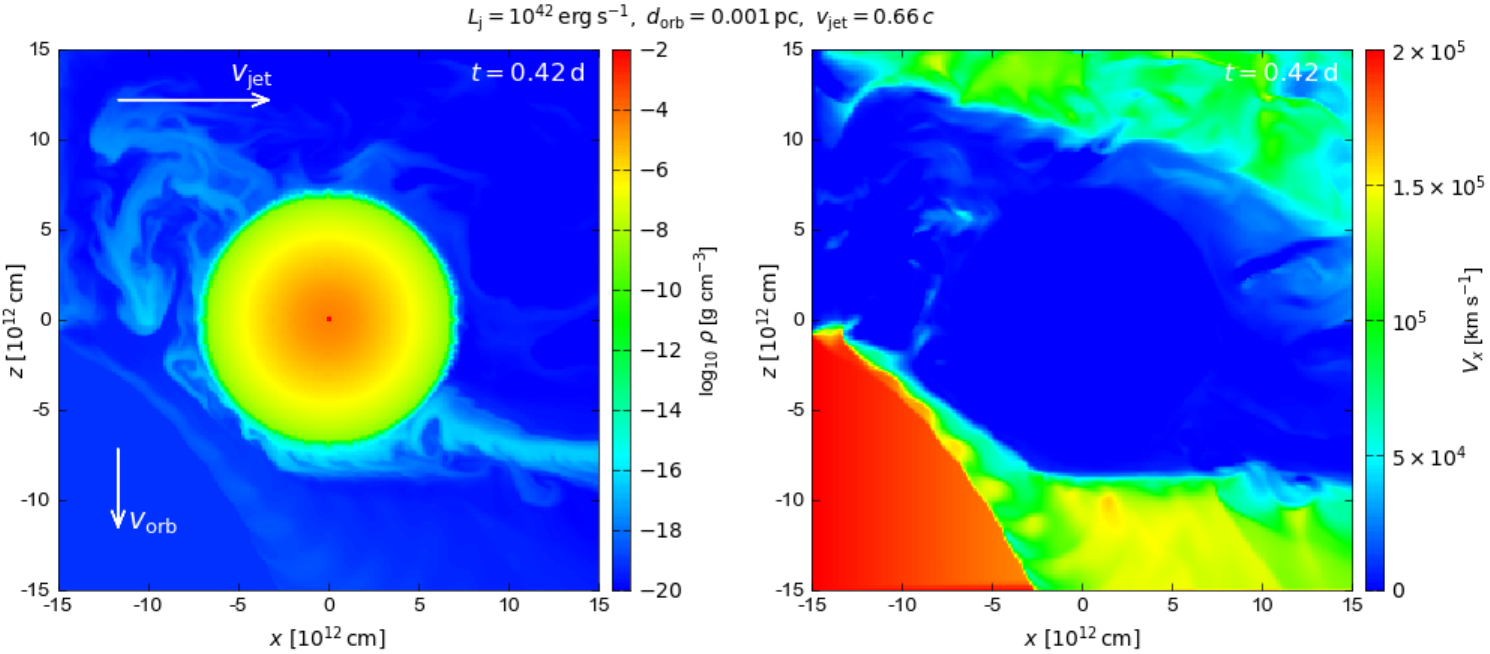}
    \caption{Star-jet interaction model with the jet luminosity $L_\text{j}=10^{42}\,\text{erg}\,\text{s}^{-1}$ and the star's orbital distance $d_\text{orb}=10^{-3}\,\text{pc}$. The simulation time $t\approx 0.42\,\text{d}$ corresponds to an approximate time of the first entry of the star into the jet. The time frame shows a longitudinal-vertical ($x$-$z$ plane, corresponding in this study to the orbital plane of the star) slice of the 3D model running through the centre of the star, with respect to the jet direction which in this case streams from left to right. {\bf Left panel}: Snapshot of the density, the directions of the jet velocity $v_\text{jet}$ and orbital velocity $v_\text{orb}$ are indicated by arrows in the figure. {\bf Right panel}: Same slice of the model as in the left panel, now showing the corresponding longitudinal ($x$-) component of the velocity, streaming from left to right. The incoming jet corresponds to the high-speed region manifested by red color.}
    \label{masslosscurve1e42}
\end{figure*}
\begin{figure*}
    \centering
    \includegraphics[width=0.875\textwidth]{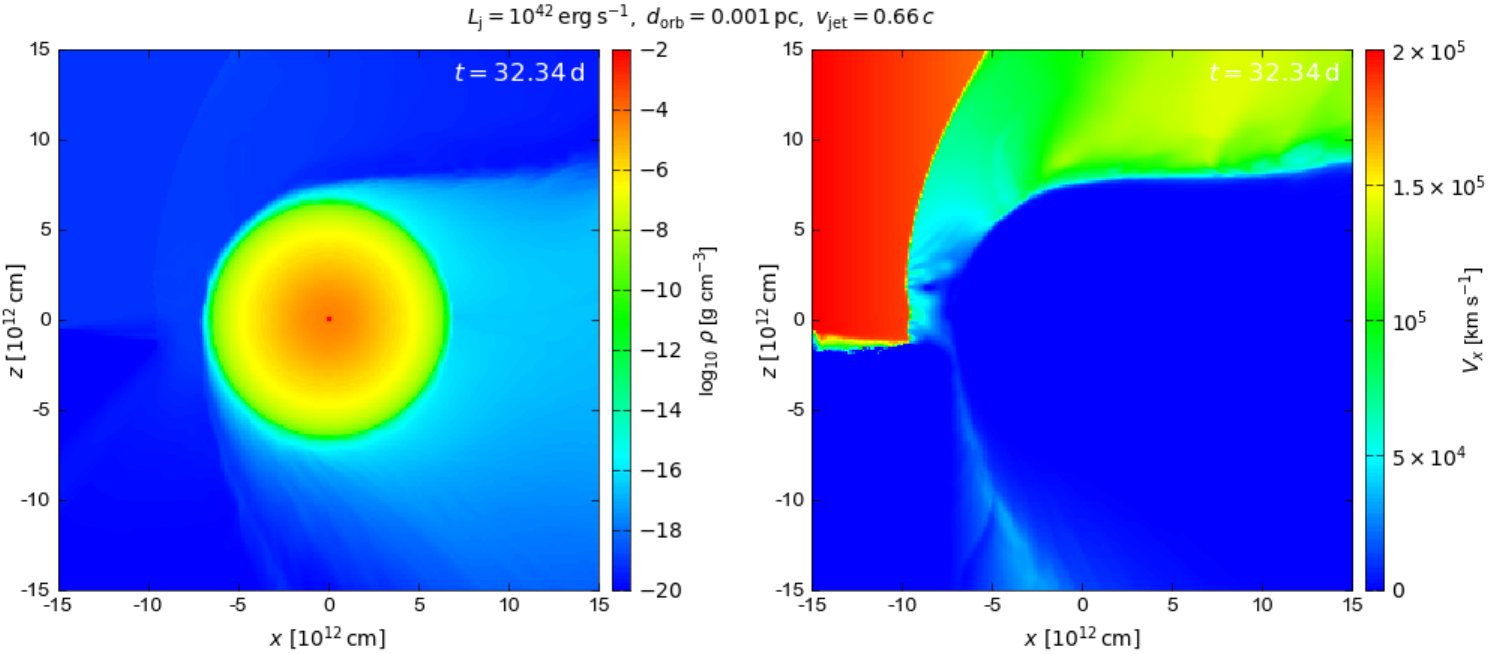}
    \caption{Star-jet interaction model with the jet luminosity $L_\text{j}=10^{42}\,\text{erg}\,\text{s}^{-1}$ and the star's orbital distance $d_\text{orb}=10^{-3}\,\text{pc}$.  The simulation time $t\approx 32.3\,\text{d}$ corresponds to an approximate time of the first exit of the star from the jet. The longitudinal-vertical ($x$-$z$) slice position of the 3D model is the same as in Fig.~\ref{masslosscurve1e42}. {\bf Left panel}: Snapshot of the density. {\bf Right panel}: Same slice of the model as in the left panel, showing the corresponding longitudinal ($x$-) component of the velocity, streaming from left to right. The outgoing jet corresponds to the high-speed region manifested by red color.}
    \label{masslosscurve1e42_figfirst}
\end{figure*}

\subsection{Long-term stability of the RG star's interior structure}\label{numthree}
The main difficulty was to keep the long-term stellar hydrostatic balance in 3D without its artificial stabilisation, which would then not allow the outer layers of the star to evolve freely and it would thus artificially alter the physics of these layers. We followed the approach outlined in \citet{2017A&A...599A...5O}, namely the choice of the core ``point" mass as $0.2\,M_\odot$ (being one-half of their core mass since they study the $2\,M_\odot$ RG star model) and implementing a smoothed Newtonian gravitational force for the core (as given in
Eq.~(10) in the referred paper) inside the sphere of radius $h\approx 0.1\,R_\star$ (the softening length of the interaction).  To avoid too strong discontinuity in density towards the center, we also implemented a smoother increase of density between the radius $0.05\,R_\star$ and the inner ``point" core, to mimic the similarly smoother increase of it in this region in the MESA model (see Fig.~\ref{fig_red_giant_lane_emden}). We have scaled the density transition so that its mass together with the inner core ``point" mass compensates the total core mass ($0.4\,M_\odot$ given by the MESA model), see Sect.~\ref{numone}), which is not accounted for by the polytropic profiles. The gravitation then corresponds to a core mass of $0.4\,M_\odot$ anyway.
The core parameters do not play a role in the actual modelling due to artificial stabilisation of this area (see below in this section).} 

In accordance with the referred article, we have permanently incorporated the damping term for spurious velocity fluctuations
(caused intrinsically by mapping
stellar models to hydrodynamical grids), i.e. $\dot{v}=-v/\tau$, where $\tau$ represents a timescale for damping \citep[see Eqs. (8) and (9) together with the explanatory text in][see also further details in Appendix~\ref{stelinternal}]{2017A&A...599A...5O}.

In addition, since we do not expect the effect of the star's ablation to reach such deep layers (even in the future models where we consider performing a significantly higher number of star-jet passages), we artificially stabilised the very inner part of the star (below $0.05\,R_\star$) by implementing the variation of the ``sponge function'' for velocity \citep[cf.][]{2011ascl.soft05010A}
\begin{align}\label{spongefunc}
v \to v \times 0.5\left\{1-\cos\left[\pi\left(r-0.01\,R_\star\right)/0.04\,R_\star\right]\right\},
\end{align}
where the velocity $v$ is smoothly damped between the radius of $0.01\,R_\star$ and $0.05\,R_\star$ (where $r$ is the radial distance from the stellar centre) so that above the radius $0.05\,R_\star$ it is unaffected by the given function while below $0.01\,R_\star$ it is zero. For this setup, only the complete inner core of the star is artificially absolutely stabilised (see also Fig~\ref{fig_sponges} in Appendix~\ref{stelinternal}).

The combination of these adjustments maintains a stable hydrostatic equilibrium in the star for at least 5 stellar orbits at the distance of $10^{-3}\,\text{pc}$ (approximately $2500\,\text{days}$) or even for a longer time of two star-jet passages at the distance of $10^{-2}\,\text{pc}$ (approximately $9500\,\text{days}$) without any artificial interference with the bulk of the stellar body. In order to achieve the best possible initial stellar model with respect to the long-term stability of the star, we first tested a large number of different internal configurations of the star  very carefully and for a long time with only a static environment, i.e. without the galactic jet, when the simulation time is significantly shorter. We include a plot showing initial stellar density and the temporal stability of the implemented stellar model in Fig.~\ref{fig_steltemporalstability} in Appendix~\ref{stelinternal}.

\begin{figure*}
    \centering
    \includegraphics[width=0.875\textwidth]{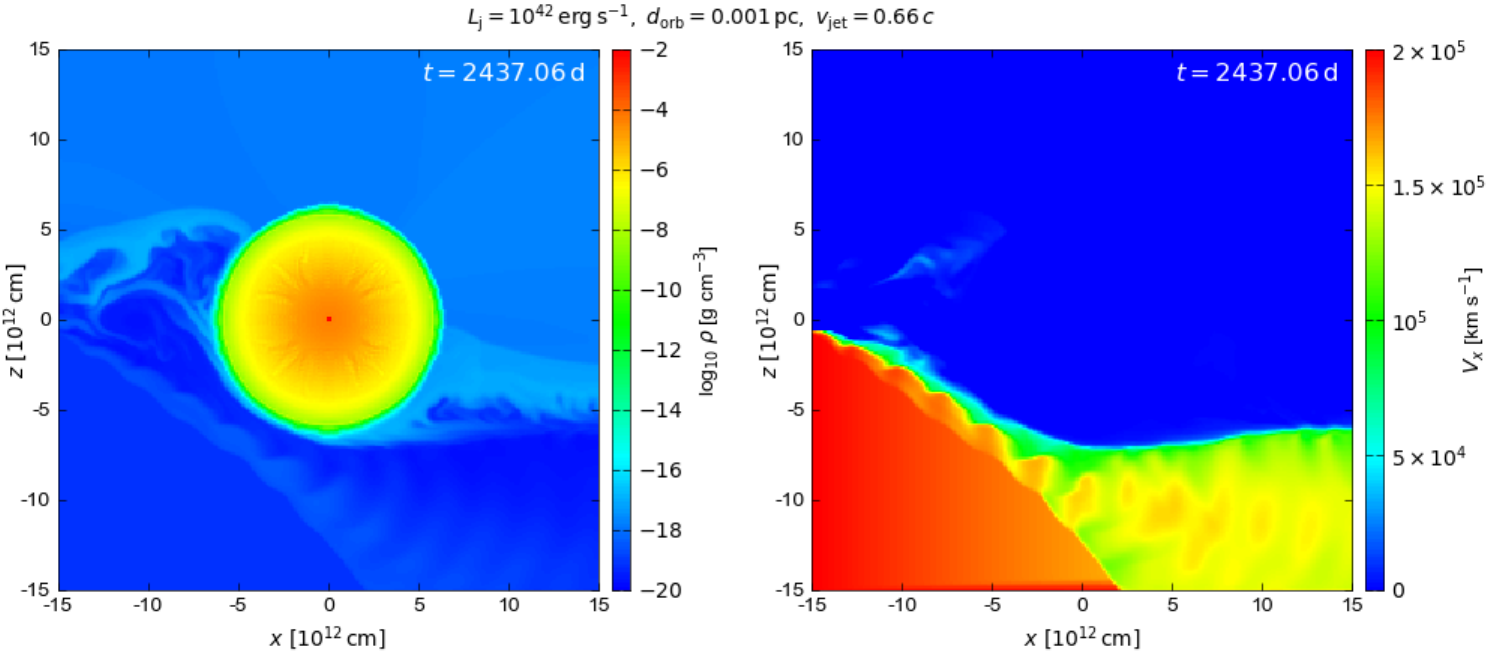}
    \caption{Star-jet interaction model with the jet luminosity $L_\text{j}=10^{42}\,\text{erg}\,\text{s}^{-1}$ and the star's orbital distance $d_\text{orb}=10^{-3}\,\text{pc}$. The simulation time $t\approx 2437\,\text{d}$ corresponds to an approximate time of the 10th entry of the star into the jet (after five stellar orbits around the central SMBH). The longitudinal-vertical ($x$-$z$) slice position of the 3D model is the same as in Fig.~\ref{masslosscurve1e42}. {\bf Left panel}: Snapshot of the density. {\bf Right panel}: Same slice of the model as in the left panel, showing the corresponding longitudinal ($x$-) component of the velocity, streaming from left to right. The incoming jet corresponds to the high-speed region manifested by red color.}
    \label{masslosscurve1e42_figsecond}
\end{figure*}
\begin{table*}
    \centering
     \caption{Overview of set-ups with repetitive red-giant jet collisions. In the third and the fourth columns we list cumulative mass-loss rates (in Solar masses) during the typical TDE phase ($10$\,years) and the AGN phase ($10^5$\,years) estimated using the extrapolation of the best-fit power-law functions (second column) fitted to the temporal evolution of the relative mass loss inferred from 3D runs {for the closer orbital distance $10^{-3}\,\text{pc}$} (see Figs.~\ref{masslosscurve1e42_fignine}, \ref{masslosscurve1e44}, and \ref{masslosscurve1e48}). { In the case of the more distant orbit $10^{-2}\,\text{pc}$, (separated by the grey horizontal line), the third column gives the cumulative mass-loss rate over two passages (about 27 years), and the rate during the complete AGN phase is not extrapolated because the two passages of the star through the jet do not yet seem to provide a sufficient statistical basis for the fitting (cf. Fig.~\ref{masslosscurvetwicedist1e42})}.}
    \begin{tabular}{l|c|c|c}
    \hline
    \hline
    Set-up & Power-law function $\Delta M_{\star}(t)\,[M_{\odot}]$ & $\Delta M_{\star}\,[M_{\odot}]$ (TDE) &  $\Delta M_{\star}\,[M_{\odot}]$ (AGN) \\
    \hline
    $L_{\rm j}=10^{42}\,{\rm erg\,s^{-1}}$, $d_{\rm orb}=10^{-3}\,{\rm pc}$, $v_{\rm jet}=0.33c$     &  $-3.36 \times 10^{-8}(t/1\,{\rm day})^{0.47}$ &   $-1.56 \times 10^{-6}$ & $-1.16 \times 10^{-4}$ \\
      $L_{\rm j}=10^{42}\,{\rm erg\,s^{-1}}$, $d_{\rm orb}=10^{-3}\,{\rm pc}$, $v_{\rm jet}=0.66c$ & $-3.12 \times 10^{-8}(t/1\,{\rm day})^{0.49}$ &   $-1.74 \times 10^{-6}$ & $-1.58 \times 10^{-4}$\\
     $L_{\rm j}=10^{44}\,{\rm erg\,s^{-1}}$, $d_{\rm orb}=10^{-3}\,{\rm pc}$, $v_{\rm jet}=0.33c$ & $-2.88 \times 10^{-6}(t/1466\,\text{days})^{1.09} [0.5(1+(t/1466\,\text{days})^{2.87})]^{-0.39}$  &  $-3.58 \times 10^{-6}$ &   $-2.80 \times 10^{-6}$\\  
       $L_{\rm j}=10^{44}\,{\rm erg\,s^{-1}}$, $d_{\rm orb}=10^{-3}\,{\rm pc}$, $v_{\rm jet}=0.66c$ & $-5.20 \times 10^{-8}(t/1\,{\rm day})^{0.50}$ &  $-3.03 \times 10^{-6}$ & $-2.91 \times 10^{-4}$\\  
      $L_{\rm j}=10^{48}\,{\rm erg\,s^{-1}}$, $d_{\rm orb}=10^{-3}\,{\rm pc}$, $v_{\rm jet}=0.33c$ & $-2.48 \times 10^{-6}(t/1\,{\rm day})^{0.08}$  &   $-4.81 \times 10^{-6}$  & $-1.01 \times 10^{-5}$\\ 
       $L_{\rm j}=10^{48}\,{\rm erg\,s^{-1}}$, $d_{\rm orb}=10^{-3}\,{\rm pc}$, $v_{\rm jet}=0.66c$ & $-6.95 \times 10^{-7}(t/1\,{\rm day})^{0.18}$ &  $-3.03 \times 10^{-6}$ & $-1.59 \times 10^{-5}$\\\grayline $L_{\rm j}=10^{42}\,{\rm erg\,s^{-1}}$, $d_{\rm orb}=10^{-2}\,{\rm pc}$, $v_{\rm jet}=0.66c$ &  &  $-3.65 \times 10^{-8}$ & \\
    \hline     
    \end{tabular}   
    \label{tab:my_label}
\end{table*}
\begin{figure*}
    \centering
    \includegraphics[width=0.805\textwidth]{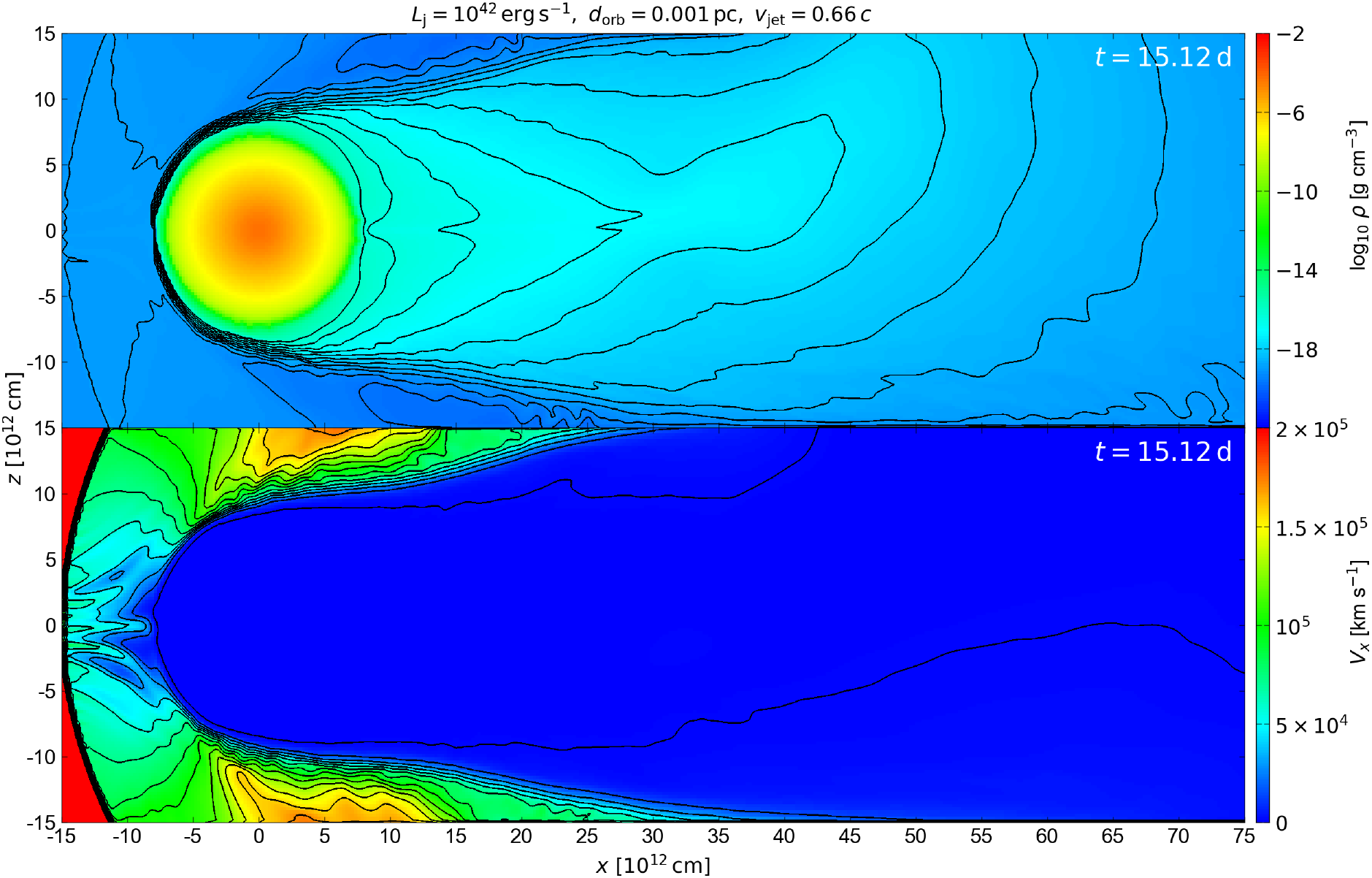}
    \caption{Star-jet interaction model with the jet luminosity $L_\text{j}=10^{42}\,\text{erg}\,\text{s}^{-1}$ and the star's orbital distance $d_\text{orb}=10^{-3}\,\text{pc}$ highlighted with contours. The simulation time $t\approx 15\,\text{d}$ corresponds to an approximate mid-time of the first star-jet passage. The longitudinal-vertical ($x$-$z$, the orbital plane) slice position of the 3D model is the same as in Fig.~\ref{masslosscurve1e42} while the longitudinal range behind the star (in the sense of the jet stream direction) of the snapshot is four-times longer. {\bf Top panel}: Extended snapshot of the density. The significant vertical asymmetry of the denser tail is caused by the orbital motion of the star, in the negative $z$ direction. {\bf Bottom panel}: Same slice of the model as in the top panel, showing the corresponding longitudinal ($x$-) component of the velocity, streaming from left to right, highlighted with contours.}
    \label{masslosscurve1e42contours_figsecond}
\end{figure*}
\begin{figure*}
    \centering
    \includegraphics[width=0.925\textwidth]{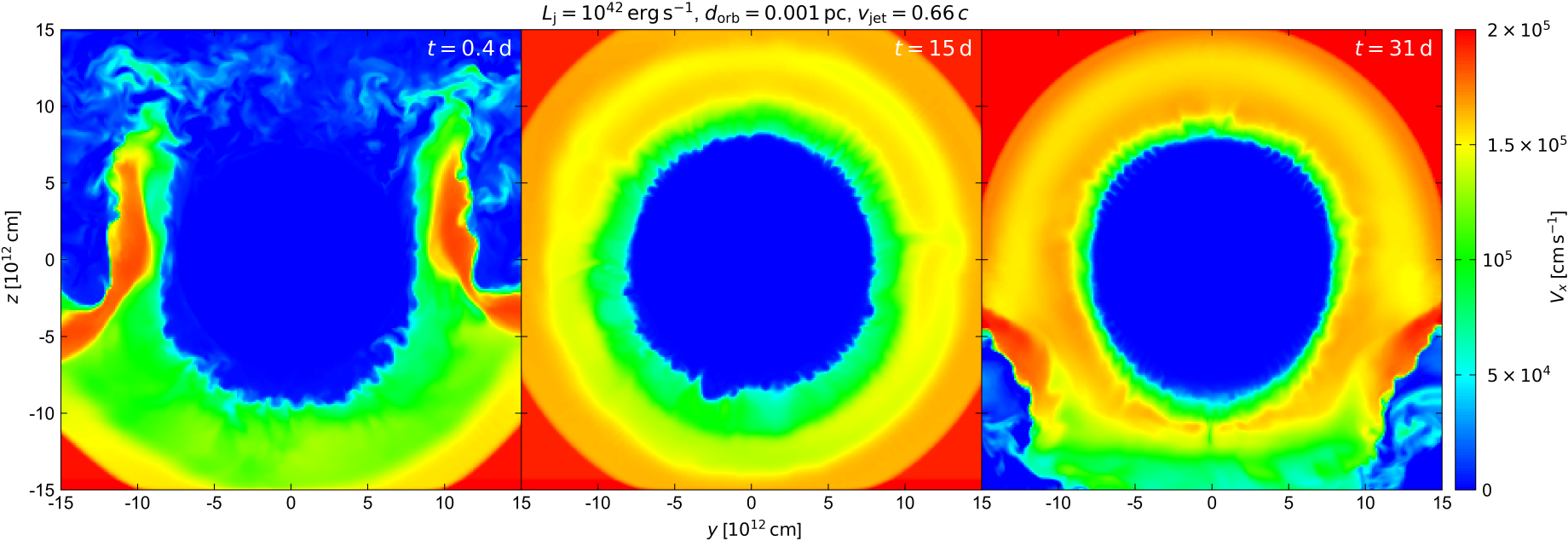}
    \caption{Illustrative snapshots of the longitudinal ($x$-) component of the velocity coloured map in the star-jet interaction model with the jet luminosity $L_\text{j}=10^{42}\,\text{erg}\,\text{s}^{-1}$ and the star's orbital distance $d_\text{orb}=10^{-3}\,\text{pc}$. The section through the 3D model is now in the $y$-$z$ plane, perpendicular to the jet direction, running through the centre of the star, so the jet flow direction is now perpendicular to the image plane, in the direction away from us (while the $y$-axis in the right-hand Cartesian coordinate system now points to the left). The simulation times $t\approx 0.4\,\text{d}$ ({\bf left panel}), $t\approx 15\,\text{d}$ ({\bf middle panel}), and $t\approx 31\,\text{d}$ ({\bf right panel}), roughly correspond to the entry, mid-time, and exit of the first star-jet passage, respectively.}
    \label{masslosscurve1e42_figthird}
\end{figure*}
\begin{figure*}
    \centering
    \includegraphics[width=0.925\textwidth]{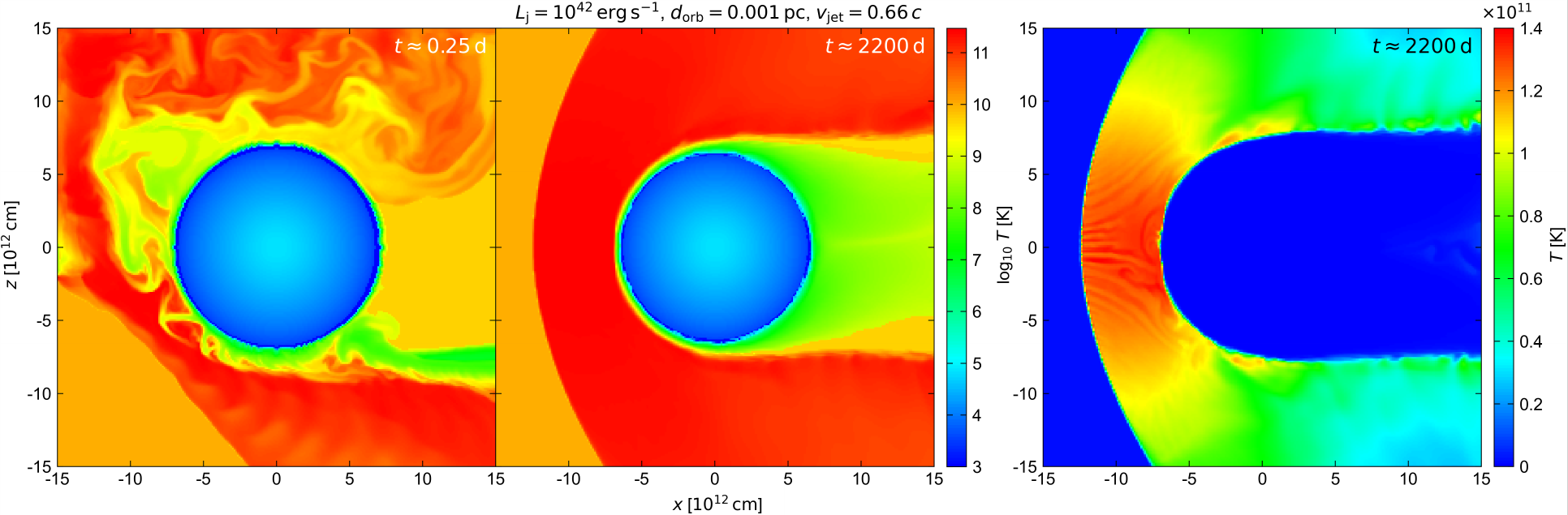}
    \caption{Snapshot of the temperature evolution coloured graph in the star-jet interaction model with the jet luminosity $L_\text{j}=10^{42}\,\text{erg}\,\text{s}^{-1}$ and the star's orbital distance $d_\text{orb}=10^{-3}\,\text{pc}$ in two different times. The longitudinal-vertical ($x$-$z$) slice position of the 3D model is the same as in Fig.~\ref{masslosscurve1e42}. The simulation times $t\approx 0.25\,\text{d}$ ({\bf left panel}) and $t\approx 2200\,\text{d}$ ({\bf middle and right panels}) roughly correspond to the entry of the first star-jet passage and mid-time of the 9th star-jet passage (about four star's orbits around SMBH; the appearance of the temperature structure in other intermediate times is roughly similar to the ones shown here). For illustration, we distinguish the same in the middle and right panel by the plot in log scale (middle panel) or in linear scale (right panel). For the 1D slices of the depicted domain in $x$-direction ($y,z=0$) in various times, see Fig.~\ref{masslosscurve1e42_figfourth_1D}.}
    \label{masslosscurve1e42_figfourth}
\end{figure*}
\begin{figure*}
    \centering
\includegraphics[width=0.675\textwidth]{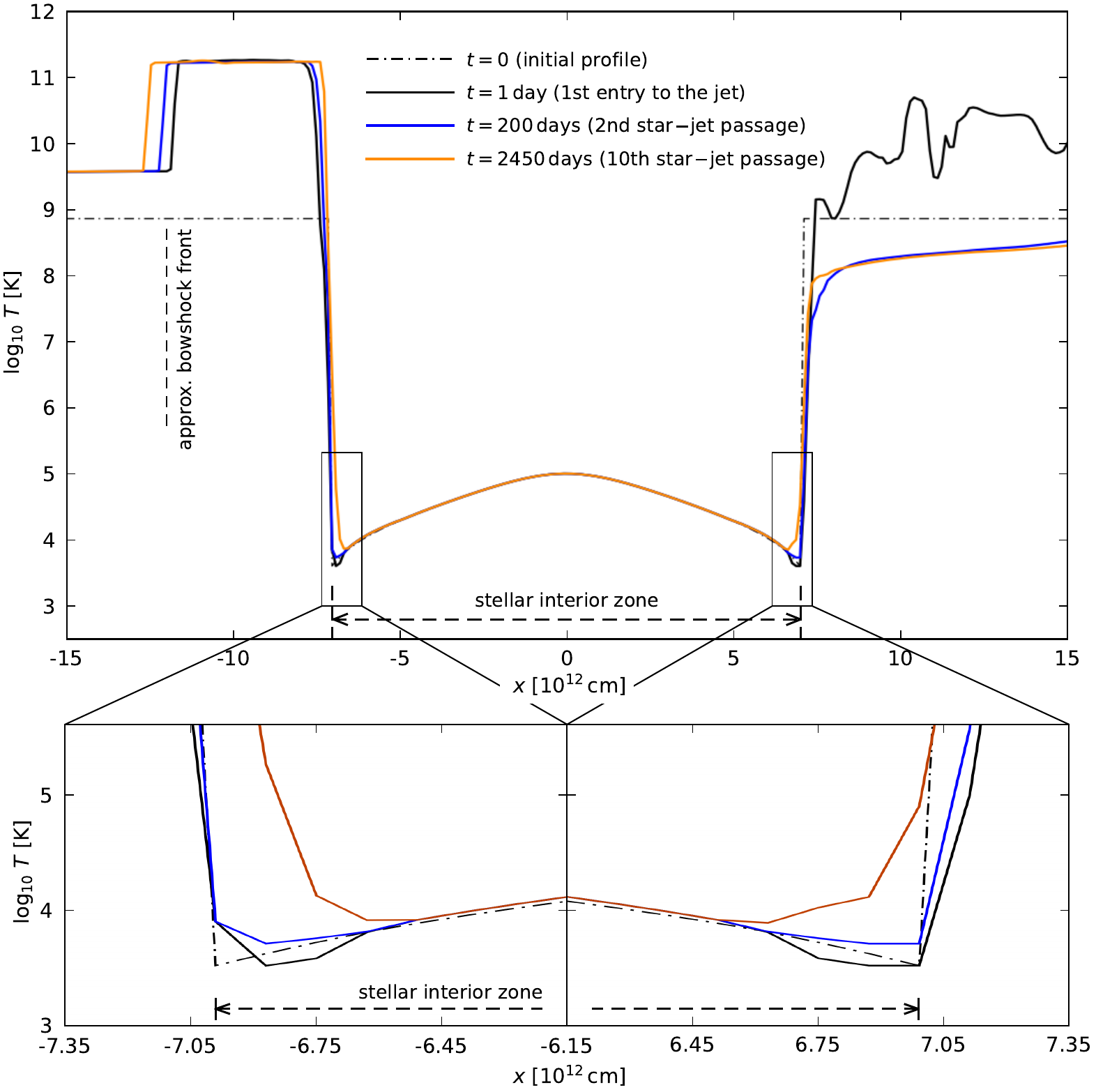}
    \caption{1D longitudinal (coincident with the jet $x$-axis while $y,z=0$, cf.~Figs~\ref{masslosscurve1e42_figfourth}) slice of the temperature structure of the star-jet interaction model with the jet luminosity $L_\text{j}=10^{42}\,\text{erg}\,\text{s}^{-1}$ and the star's orbital distance $d_\text{orb}=10^{-3}\,\text{pc}$ in four different characteristic times (initial state, first star's entry to the jet, star's second passage through the jet, and star's tenth  passage through the jet) when star crosses the jet.
    The two rectangles in the overall picture of the larger domain of $-15$ to $15\times 10^{12}\,\text{cm}$ (centred at the star's core) in the top panel indicate details in the surface regions of the star where the material of the star meets the surroundings and are shown enlarged in the lower panels.}
\label{masslosscurve1e42_figfourth_1D}
\end{figure*}
\begin{figure*}
    \centering
    \includegraphics[width=0.805\textwidth]{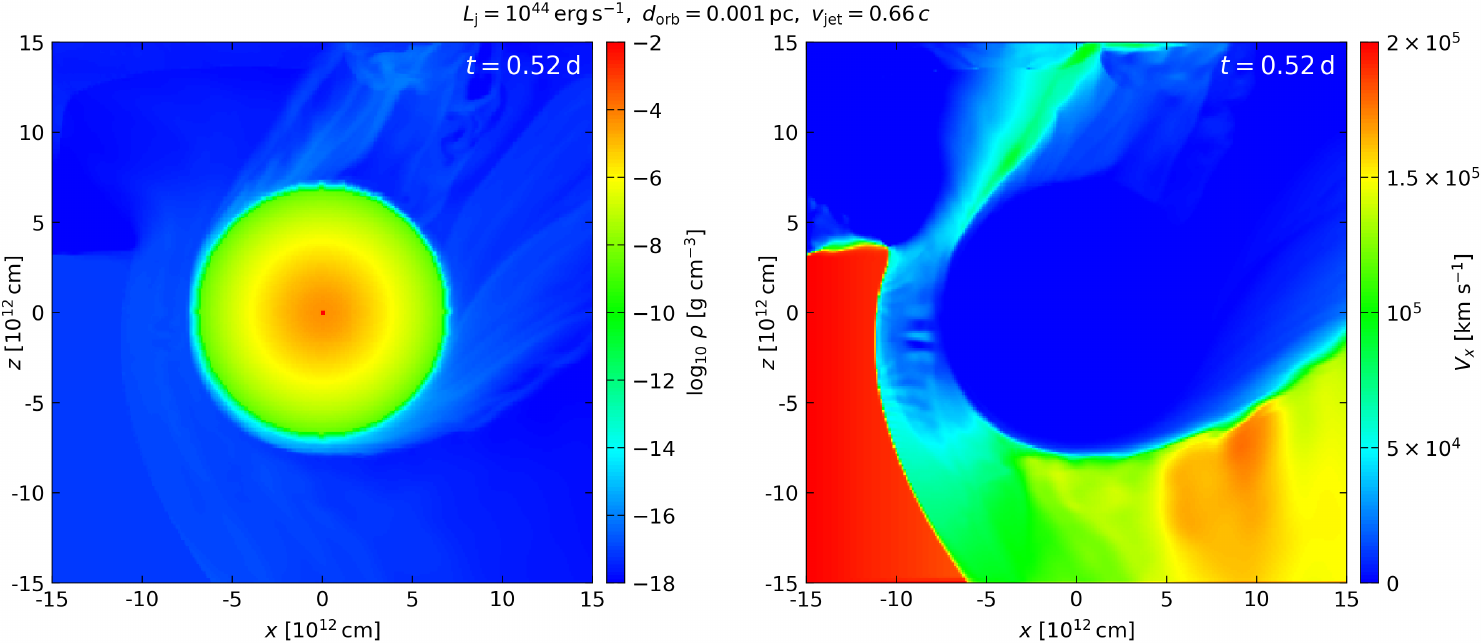}
    \caption{Star-jet interaction model with the jet luminosity $L_\text{j}=10^{44}\,\text{erg}\,\text{s}^{-1}$ and the star's orbital distance $d_\text{orb}=10^{-3}\,\text{pc}$. The simulation time $t\approx 0.52\,\text{d}$ corresponds to an approximate time of the first entry of the star into the jet. The longitudinal-vertical ($x$-$z$) slice position of the 3D model is the same as in Fig.~\ref{masslosscurve1e42}. {\bf Left panel}: Snapshot of the density. {\bf Right panel}: Same slice of the model as in the left panel, showing the corresponding longitudinal ($x$-) component of the velocity, streaming from left to right.}
    \label{masslosscurve1e42_figfifth}
\end{figure*}
\begin{figure*}
    \centering
    \includegraphics[width=0.805\textwidth]{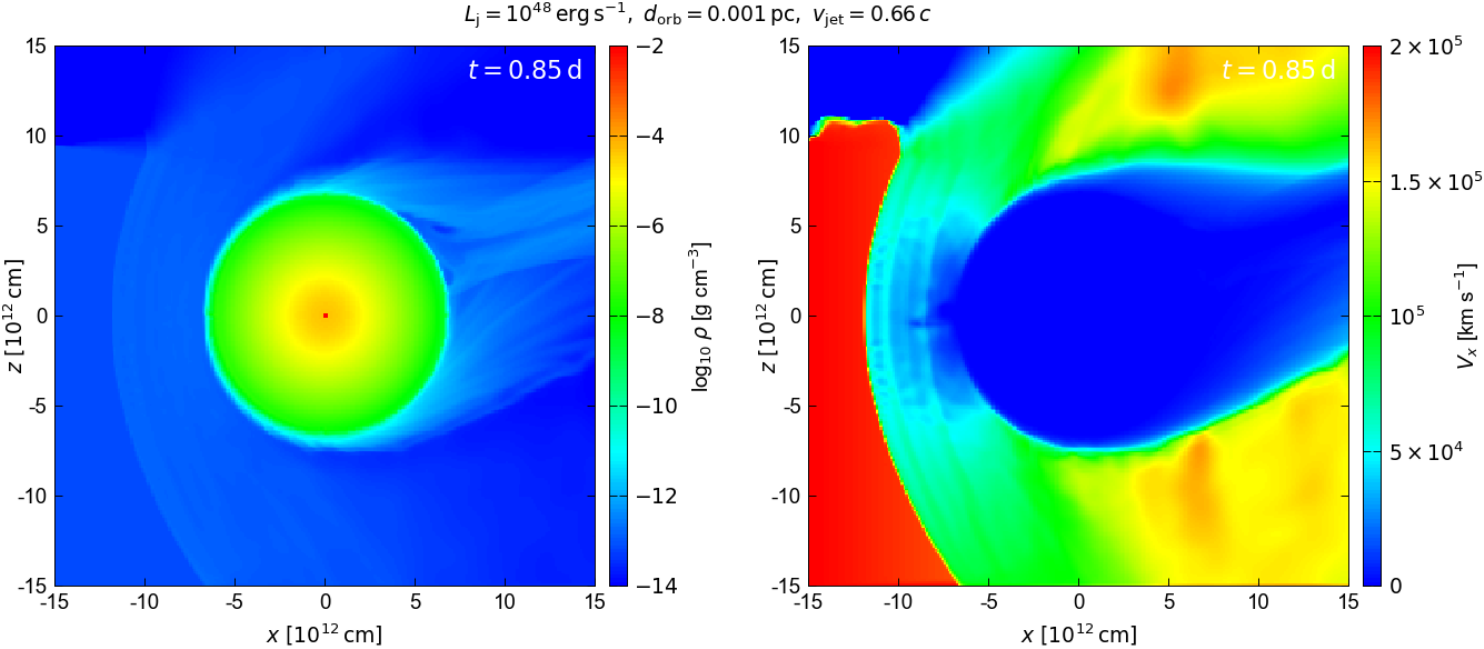}
    \caption{Star-jet interaction model with the rather hypothetical jet luminosity $L_\text{j}=10^{48}\,\text{erg}\,\text{s}^{-1}$ and the star's orbital distance $d_\text{orb}=10^{-3}\,\text{pc}$. The simulation time $t\approx 0.85\,\text{d}$ corresponds to an approximate time (or slightly after) of the first entry of the star into the jet. The longitudinal-vertical ($x$-$z$) slice position of the 3D model is the same as in Fig.~\ref{masslosscurve1e42}. {\bf Left panel}: Snapshot of the density. {\bf Right panel}: Same slice of the model as in the left panel, showing the corresponding longitudinal ($x$-) component of the velocity, streaming from left to right.}
    \label{masslosscurve1e42_figsixth}
\end{figure*}
\begin{figure*}
    \centering
    \includegraphics[width=0.805\textwidth]{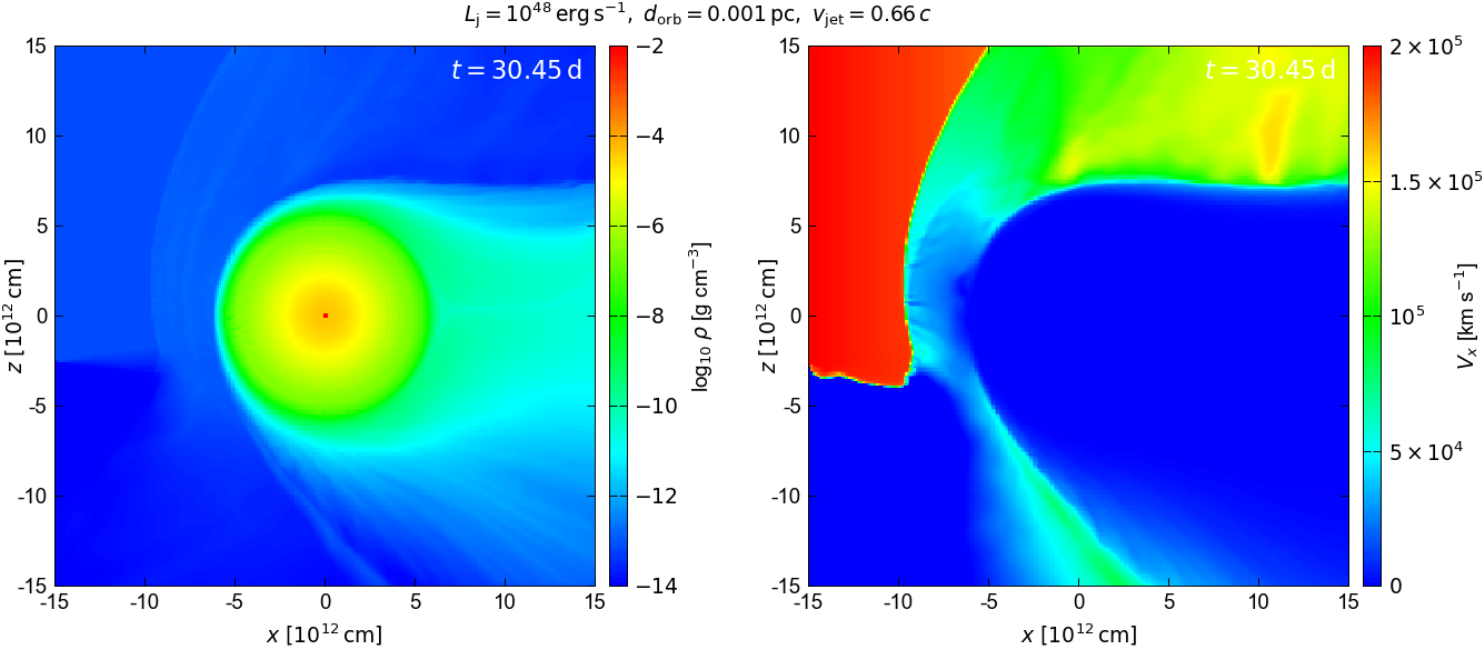}
    \caption{Star-jet interaction model with the rather hypothetical jet luminosity $L_\text{j}=10^{48}\,\text{erg}\,\text{s}^{-1}$ and the star's orbital distance $d_\text{orb}=10^{-3}\,\text{pc}$. The simulation time $t\approx 31\,\text{d}$ corresponds to an approximate time of the first exit of the star from the jet. The longitudinal-vertical ($x$-$z$) slice position of the 3D model is the same as in Fig.~\ref{masslosscurve1e42}. {\bf Left panel}: Snapshot of the density. {\bf Right panel}: Same slice of the model as in the left panel, showing the corresponding longitudinal ($x$-) component of the velocity, streaming from left to right.}
    \label{masslosscurve1e42_figseventh}
\end{figure*}
\begin{figure*}
    \centering
    \includegraphics[width=0.7\textwidth]{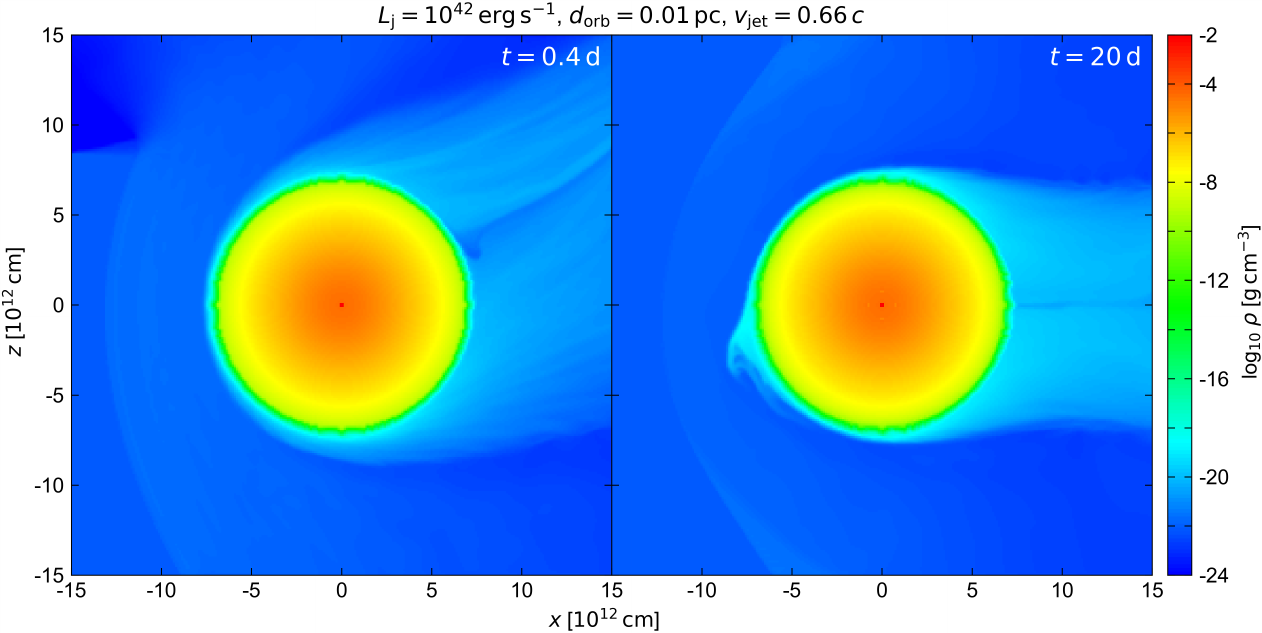}
    \caption{Two snapshots of the 2D model of the density in the star-jet interaction model with the jet luminosity $L_\text{j}=10^{42}\,\text{erg}\,\text{s}^{-1}$ and the star's orbital distance $d_\text{orb}=10^{-2}\,\text{pc}$ (ten times larger than in the previous model) in different times. {\bf Left panel}: First entry of the star into the jet. The longitudinal-vertical ($x$-$z$) 2D plane position is the same as the $x$-$z$ slices of the 3D model in Fig.~\ref{masslosscurve1e42} and other figures. {\bf Right panel}: Snapshot of the identically positioned 2D density configuration of the star-jet interaction at a randomly selected time during the first star's passage through the jet.}
    \label{masslosscurve1e42_figeight}
\end{figure*}
\begin{figure*}
    \centering
    \includegraphics[width=0.6\textwidth]{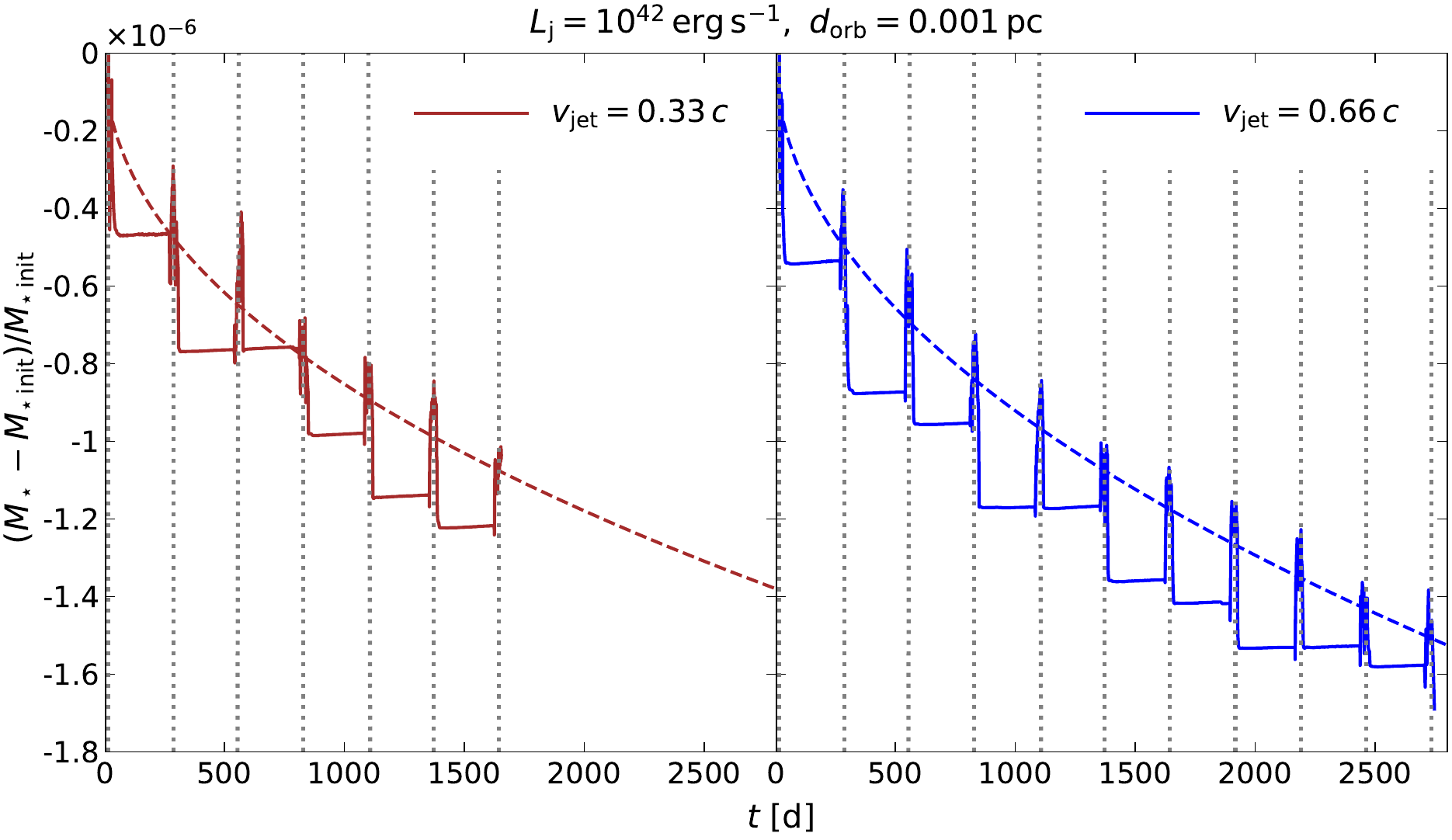}\hspace{0.5cm}
    \includegraphics[width=0.655\columnwidth]{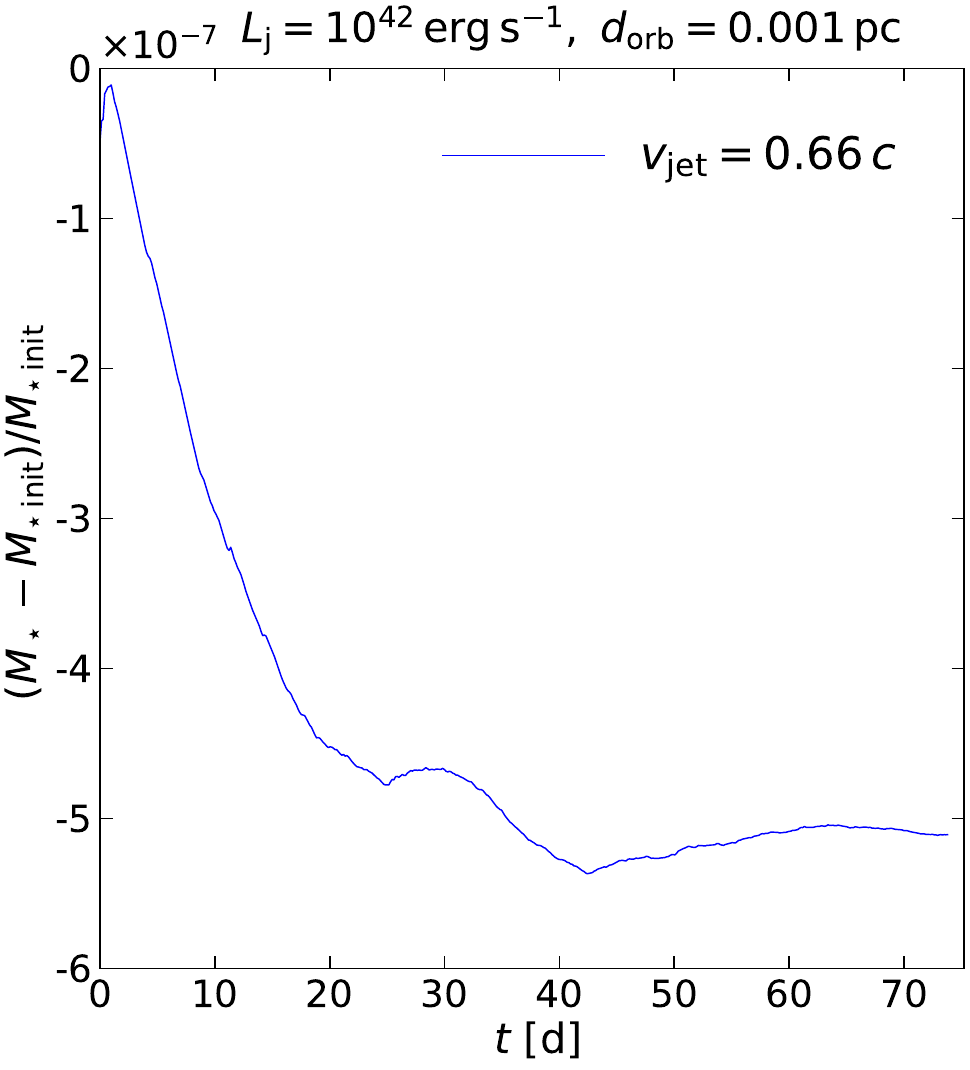}
    \caption{Plot of the relative mass loss due to RG star passages through the galactic jet with the jet luminosity $L_\text{j}=10^{42}\,\text{erg}\,\text{s}^{-1}$ and the star's orbital distance $d_\text{orb}=10^{-3}\,\text{pc}$, calculated directly from the 3D model. {\bf Left panel}: Graph of the relative mass loss for the jet flow velocity $v_\text{jet}=0.33\,c$ (solid brown line), showing 7 passages through the jet including the intermediate periods between passages, corresponding to 3.5 orbits of the star around the central SMBH. The dashed brown line depicts the best-fit power-law relation for the relative mass loss as a function of time in days: $\Delta M_{\star}=-3.36 \times 10^{-8}(t/1\,{\rm day})^{0.47}\,M_{\odot}$. The vertical dotted lines highlight the mean moment of each star's passage through the jet. {\bf Middle panel}: An analogous plot of the relative mass loss for jet velocity $v_\text{jet}=0.66\,c$ (solid blue line), showing 10 passages through the jet, corresponding roughly to 5 orbits of the star around the central SMBH. The dashed blue line depicts the best-fit power-law relation for the relative mass loss as a function of time in days: $\Delta M_{\star}=-3.12 \times 10^{-8}(t/1\,{\rm day})^{0.49}\,M_{\odot}$.
    {\bf Right panel}: A detailed temporal profile of the relative mass loss during the first RG star-jet passage, considering the jet luminosity of $L_\text{j}=10^{42}\,\text{erg}\,\text{s}^{-1}$, the jet flow velocity of $v_\text{jet}=0.66\,c$, the orbital distance of the star of $d_\text{orb}=10^{-3}\,\text{pc}$.}
    \label{masslosscurve1e42_fignine}
\end{figure*}
\begin{figure*}
    \centering
    \includegraphics[width=0.6\textwidth]{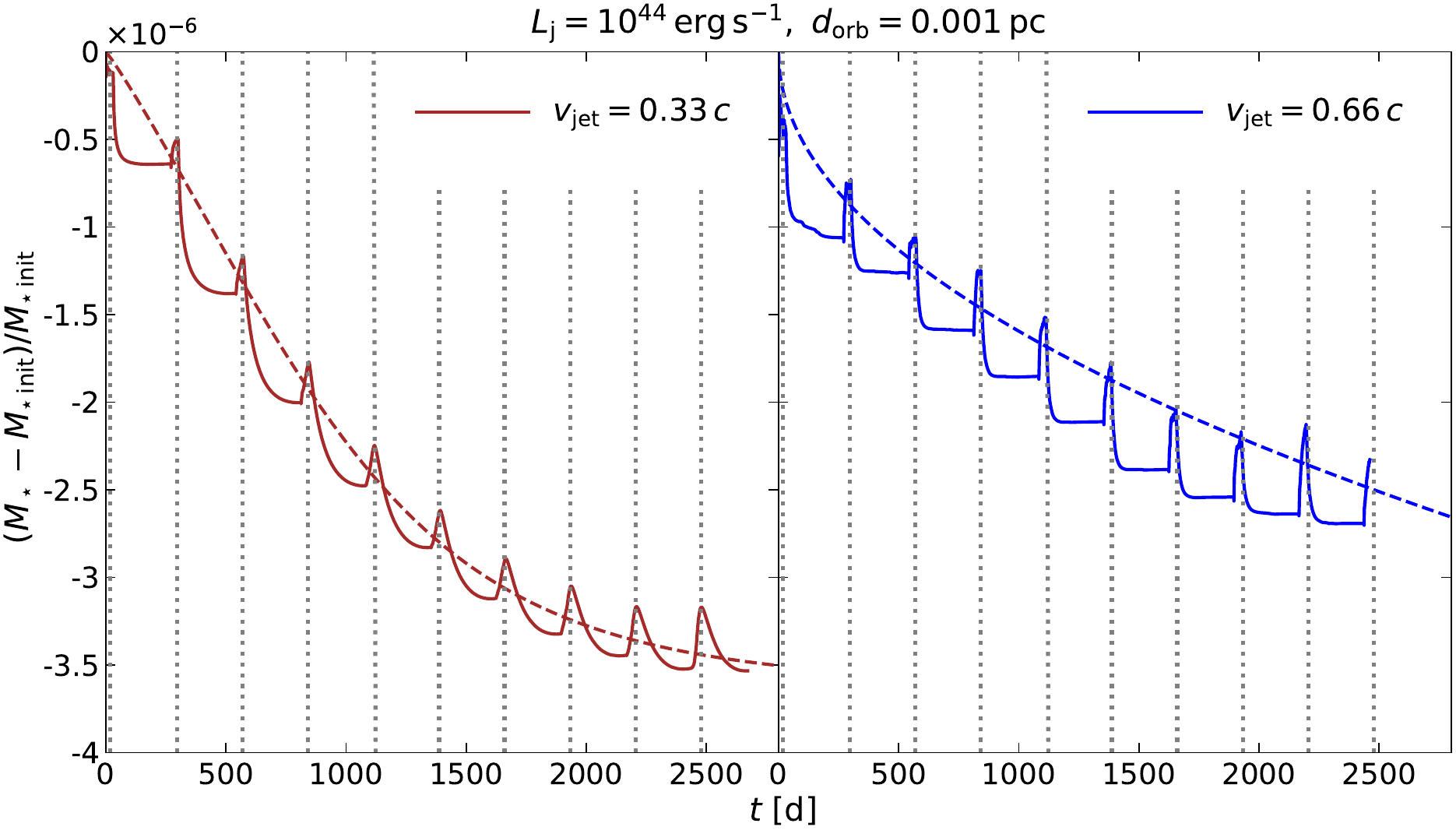}
    \caption{Same as in Fig.~\ref{masslosscurve1e42_fignine} for the jet luminosity of $L_\text{j}=10^{44}\,\text{erg}\,\text{s}^{-1}$ and the orbital distance of the star of $d_\text{orb}=10^{-3}\,\text{pc}$. {\bf Left panel}: A graph of the relative mass loss for the  jet flow velocity of $v_\text{jet}=0.33\,c$, showing 10 passages through the jet (solid brown line). A dashed brown line represents the best broken power-law fit: $\Delta M_{\star}=A(t/t_b)^{\gamma_1} [0.5(1+(t/t_b)^{1/\Delta})]^{(\gamma_1-\gamma_2)\Delta}$, where $A=-2.88\times 10^{-6}\,M_{\odot}$, $t_{\rm b}=1466$ days, $\gamma_1=1.09$, $\gamma_2=2.21$, and $\Delta=0.35$.  {\bf Right panel}: A graph of the relative mass loss for the jet velocity of $v_\text{jet}=0.66\,c$, also showing 10 passages through the jet. A dashed blue line represents the best-fit simple power law function, $\Delta M_{\star}=-5.20\times 10^{-8}(t/1\,\text{days})^{0.50}\,M_{\odot}$.}
    \label{masslosscurve1e44}
\end{figure*}
\begin{figure*}
    \centering
    \includegraphics[width=0.6\textwidth]{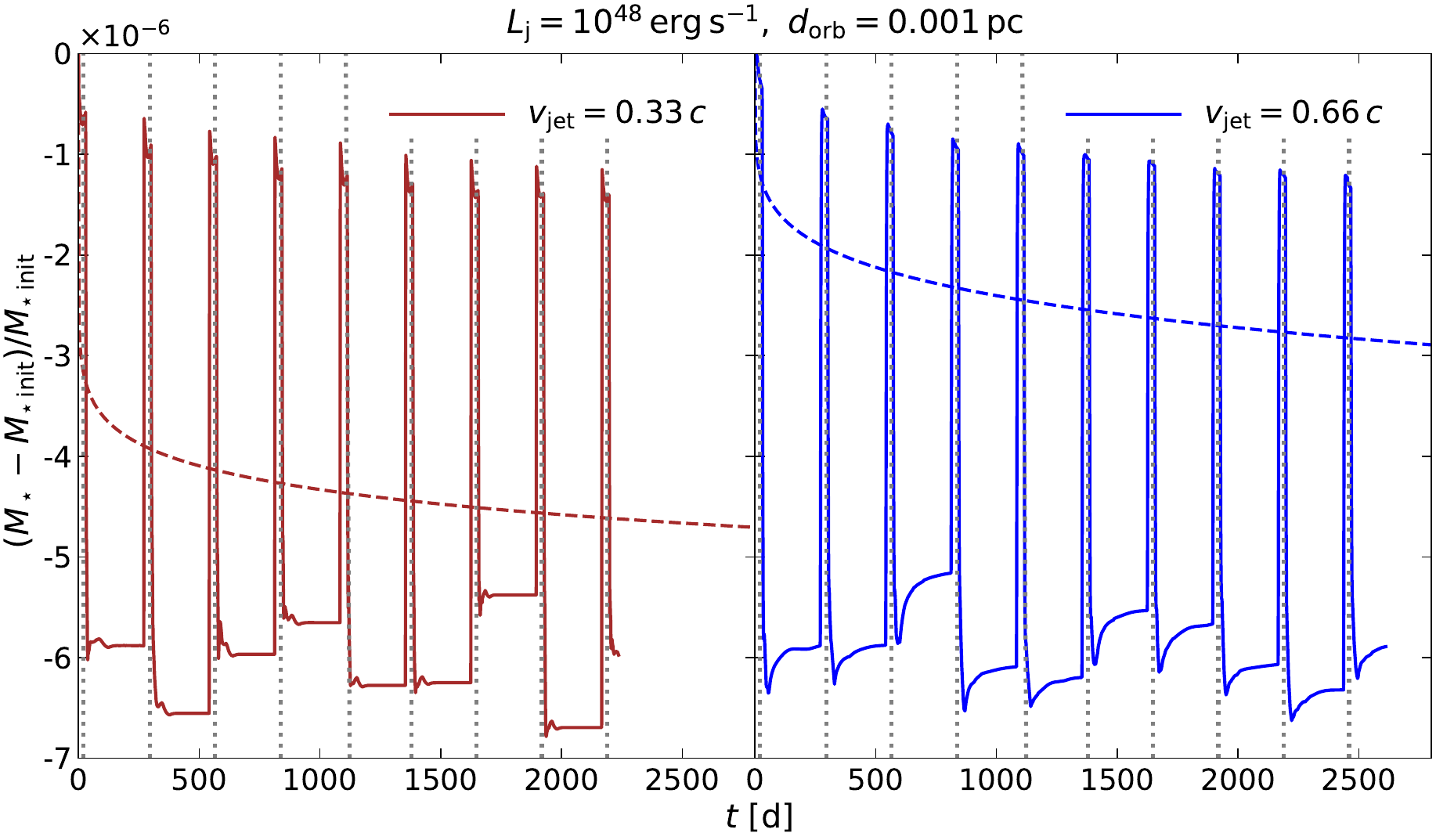}
    \caption{Same as in Fig.~\ref{masslosscurve1e42_fignine} for the jet luminosity of $L_\text{j}=10^{48}\,\text{erg}\,\text{s}^{-1}$ (upper limit) and the orbital distance of the star of $d_\text{orb}=10^{-3}\,\text{pc}$. {\bf Left panel}: Temporal evolution of the relative mass loss for the jet velocity of $v_\text{jet}=0.33\,c$, showing 9 passages through the jet. The brown dashed line represents the best-fit power-law relation, $\Delta M_{\star}=-2.48 \times 10^{-6}(t/1\,\text{day})^{0.08}\,M_{\odot}$. {\bf Right panel}: Temporal evolution of the relative mass loss for the jet velocity of $v_\text{jet}=0.66\,c$, showing 10 star-jet passages. The blue dashed line represents the best-fit power-law relation, $\Delta M_{\star}=-6.95 \times 10^{-7}(t/1\,\text{day})^{0.18}\,M_{\odot}$.}
    \label{masslosscurve1e48}
\end{figure*}
\begin{figure}
    \centering
    \includegraphics[width=0.7\columnwidth]{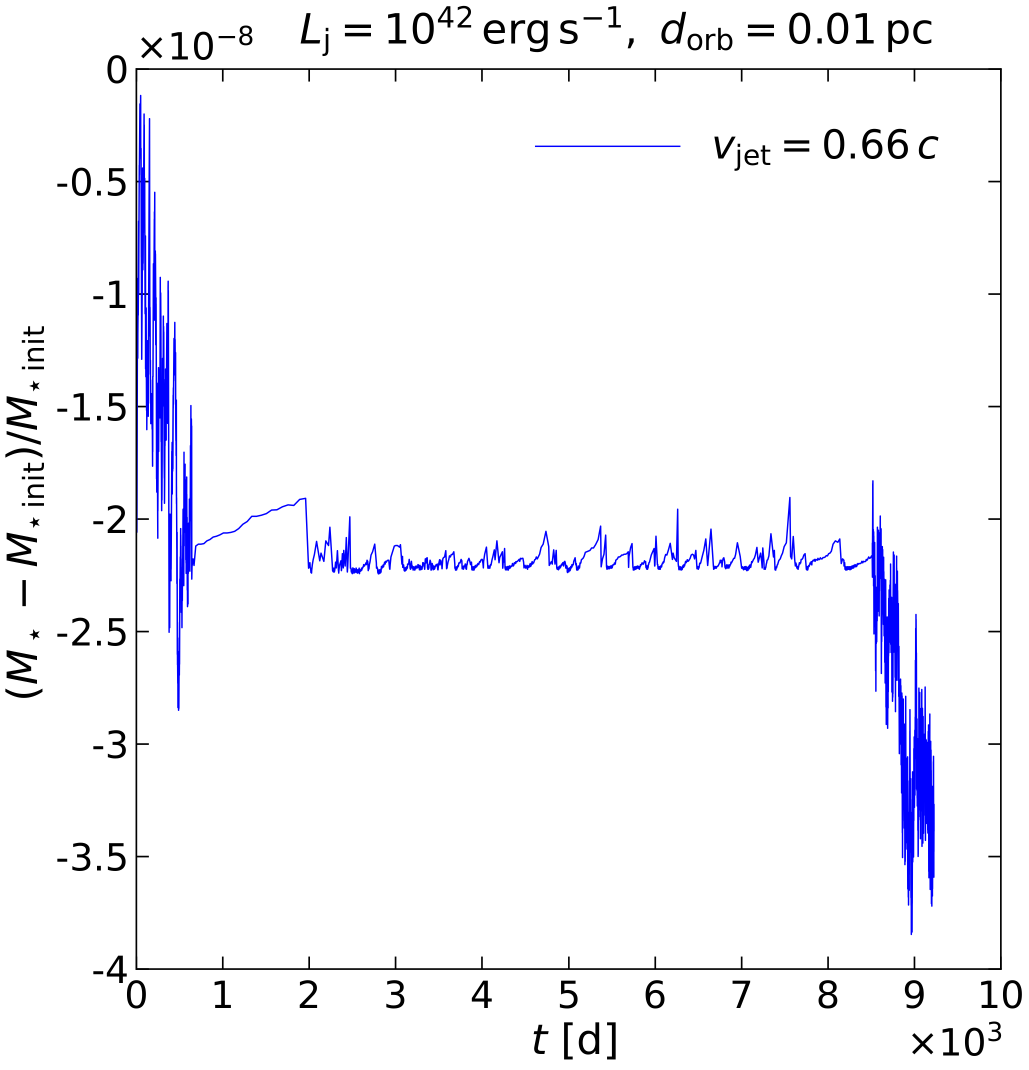}
    \caption{A temporal profile of the relative mass loss during the first two star-jet passages for the jet luminosity of $L_\text{j}=10^{42}\,\text{erg}\,\text{s}^{-1}$ and the jet velocity of $v_\text{jet}=0.66\,c$, calculated from the 2D model in the stellar orbital plane, including the mass loss evolution in the intermediate period between the jet passages. The orbital distance of the star is ten times larger, $d_\text{orb}=10^{-2}\,\text{pc}$.}
    \label{masslosscurvetwicedist1e42}
\end{figure}

\section{Results: Hydrodynamical simulations of star-jet interactions}
\label{sec_results}
In this Section, we describe the results of the hydrodynamical simulations of star-jet interactions in detail.
We will primarily focus on:
\begin{itemize}
    \item the ablation rate of RG envelopes and its evolution during a small number of stellar passages. Based on the power-law fit to the mass-loss evolution, we will estimate the cumulative mass loss during the longer AGN phase ($\sim 0.1$\,Myr) as well as the shorter phase associated with a TDE (10 years; see Subsection~\ref{subsec_mass_ablation}), 
    \item the change in the stellar effective temperature during a few consecutive RG-jet collisions (see Subsection~\ref{subsec_eff_temp}).
\end{itemize}

\subsection{Stellar mass ablation rates from numerical calculations}
\label{subsec_mass_ablation}
3D simulations of the surface ablation within a few orbital periods and subsequent determination of the relative mass loss of a standard RG with the parameters described in Section \ref{rgmod} and an orbital distance of $d_{\rm orb}=10^{-3}\,\text{pc}$ show the following.
  
   In case of the specified jet luminosity $L_\text{j}=10^{42}\,\text{erg}\,\text{s}^{-1}$ (see Figs.~\ref{masslosscurve1e42}\,--\,\ref{masslosscurve1e42_figfourth} together with Fig.~\ref{masslosscurve1e42_fignine} and the first two videos included in Appendix~\ref{videodoc}), which is most consistent with the estimated activity of the past jet in our Galaxy, we calculated the ablation rate for two possible jet velocities - $0.33\,c$ and $0.66\,c$. Figure~\ref{masslosscurve1e42} demonstrates the density and longitudinal ($x-$) velocity snapshots (parallel with the direction of the jet) in the moment of the star's first jet entry at approximately $0.4\,\text{days}$ after the start of the simulation, Fig.~\ref{masslosscurve1e42_figfirst} the first star's  exit of the jet at approx.~$32\,\text{days}$, and Fig.~\ref{masslosscurve1e42_figsecond} the 10th entry of the star into the jet after 5 stellar orbits, at approx.~$2400\,\text{days}$ after the start of the simulation.  

    Figure~\ref{masslosscurve1e42contours_figsecond} snapshots the same model at approx.~15 days after the start of the simulation, i.e., roughly in the middle of the star's first passage through the jet, where the image area is extended longitudinally beyond the star (to the right in the figure) to better demonstrate the behaviour of the gas flow not only around the star but also at a greater distance beyond the star. Due to the added contours, the asymmetric deflection of the stream against the direction of the star's orbital motion (upwards in the image) is clearly visible.

    We also demonstrate in Figures~\ref{masslosscurve1e42_figthird} and \ref{masslosscurve1e42_figfourth} the velocity structure of the star's surroundings seen in the jet direction at three different times, at the beginning, in the middle, and at the end of the star's first jet passage (Fig.~\ref{masslosscurve1e42_figthird}) and the temperature structure of the star and its surroundings at two different times, roughly during the star's first and ninth jet passages (Fig.~\ref{masslosscurve1e42_figfourth}) while at the other either intermediate or later times the temperature structure does not differ much from what is shown here. For a better illustration, the later-time temperature structure is shown in two ways, one on a logarithmic scale and the other on a linear scale, which makes the finer temperature differences around the star beyond the edge of the ambient bow-shock more apparent. Figure~\ref{masslosscurve1e42_figfourth_1D} consequently demonstrates the 1D slice coincident with the $x$-axis centred on the star's centre (where $y,z=0$) at various times denoted in the figure, together with details of the temperature structure evolution at the stellar surface. It shows that the ``left" detail (closer to the bow-shock front region where the jet ``hits" the star) exhibits a slightly higher temperature profile at the stellar surrounding and the surface zone at later times.

    Figure~\ref{masslosscurve1e42_fignine} shows the result of the calculation of the relative mass loss of the star during the first few passages of the star through the jet for the two adopted jet velocities, including the power-law fits of the temporal dependence of this relative mass loss. This, at least as the first estimate, will allow us to extrapolate the evolution of the star's ablation over a longer period of time at this point, until we have relevantly longer simulations with a significantly larger number of passages. The relative mass loss corresponds here to the order of $10^{-6}$ of the original stellar mass, which would imply an overall significant loss of stellar mass, apparently with probabilities in the tens of percentages, for the expected several hundred thousand star-jet passages over the expected duration of Galactic jet activity. In the right panel of Fig.~\ref{masslosscurve1e42_fignine}, we show in more detail the mass loss of the star during the first passage of the star through the jet.

  We demonstrate the same process for the two jet flow velocities for the case of the jet luminosity $L_\text{j}=10^{44}\,\text{erg}\,\text{s}^{-1}$, see Fig.~\ref{masslosscurve1e42_figfifth} together with Fig.~\ref{masslosscurve1e44}, and the third video included in Appendix~\ref{videodoc}. The first of the two figures shows the density and velocity configurations at a selected moment during the star's first entry into the jet. The latter Figure \ref{masslosscurve1e44} shows the result of the calculation of the relative mass loss of the star during the first few (ten) passages of the star through the jet for the two adopted jet velocities, including the power-law fits of the temporal dependence of the relative mass loss. In comparison to the model with $L_\text{j}=10^{42}\,\text{erg}\,\text{s}^{-1}$, the rate of mass loss is roughly double for the same time, for the both considered jet speed variations. Such a model with a given jet luminosity may potentially be useful especially for the case of other galaxies with more massive SMBHs and with presumably stronger central galactic jets.
 
  To investigate these processes more broadly, we have also calculated an extreme case with a rather overestimated jet luminosity of $L_\text{j}=10^{48}\,\text{erg}\,\text{s}^{-1}$. The hydrodynamic effects of the model at the times of the star's first entry into and the exit out of the jet are shown
    in Figures~\ref{masslosscurve1e42_figsixth} and \ref{masslosscurve1e42_figseventh}, and in the last video included in Appendix~\ref{videodoc}. Figure~(\ref{masslosscurve1e48}) shows again the result of the calculation of the relative mass loss of the star during the first few passages (nine for the lower and ten for the higher jet-flow velocity) of the star through the jet, including the power-law fits of the temporal dependence of the relative mass loss. In this case, surprisingly, the mass loss is slower than in the previous model with $L_\text{j}=10^{44}\,\text{erg}\,\text{s}^{-1}$ and is roughly comparable to the model with $L_\text{j}=10^{42}\,\text{erg}\,\text{s}^{-1}$. The exact reason for this mass-ablation flattening for the increasing jet luminosity is not entirely clear at the moment, but we assume that it is the result of the jet temporarily bringing more material into the space originally occupied by the star (cf.~Fig.~\ref{fig_steltemporalstability}), as compared to the inter-jet phase.
    Thus, this additional mass protects the star from being ablated to a greater extent. This would then mean that the rate of ablation would peak at some particular value of $L_\text{j}$ (close to $10^{44}\,\text{erg}\,\text{s}^{-1}$) and decrease towards larger jet luminosities. In any case, the confirmation of such a trend will require further analysis for more values of the jet luminosity within the studied range of ($10^{42},\,10^{48}\,{\rm erg\,s^{-1}}$).

We also calculated the surface ablation for the case of two star-jet passages with $L_\text{j}=10^{42}\,\text{erg}\,\text{s}^{-1}$ and the stellar orbital distance of $d_{\text orb}=10^{-2}\,\text{pc}$. Due to a much longer orbital period, the calculation was performed only in 2D on the analogous Cartesian grid with $256^2$ cells, described in Sect.~\ref{numtwo}. We calculate the subsequent conversion to 3D mass ablation values as the ratio between 3D and 2D relative mass losses as $\left(\delta m/M_\star\right)_{3D}/\left(\delta m/M_\star\right)_{2D}=3/2$. The snapshots of hydrodynamic density evolution at the time of the first entry into the jet and the first exit from the jet are shown in Figure~\ref{masslosscurve1e42_figeight}. Figure~\ref{masslosscurvetwicedist1e42} then shows the relative mass loss within the first two star-jet passages (one orbit) for this case. The noticeable noise in the plot of the relative stellar ablation is probably due to the much longer time interval of the star's passage through the jet, and also the larger relative density and velocity fluctuations of the jet itself at this distance (as opposed to the detail of the single star-jet passage at $d_{\rm orb}=10^{-3}\,{\rm pc}$ and for $L_\text{j}=10^{42}\,\text{erg}\,\text{s}^{-1}$ - see the right panel of Fig.~\ref{masslosscurve1e42_fignine}). Future simulations capturing a significantly larger number of star-jet passages at larger distances from the SMBH are necessary to assess more accurately the evolution of the star's ablation in this configuration. To verify the connection between the 3D and 2D ablation rates mentioned above, we also performed a comparative 2D simulation of the model with $L_\text{j} = 10^{42}\,\text{erg}\,\text{s}^{-1}$ at $d_{\rm orb} = 10^{-3}\,\text{pc}$ for the first ten star-jet passages. The stellar mass ablation was larger by 4.61\,\% for the 2D case, which we consider to be a good agreement.

For a more comprehensive overview and comparison, we present the obtained relative mass loss rates for each model and their extrapolations using the power-law fits of the numerical calculations in Table~\ref{tab:my_label}. As we can see, for most cases with $d_{\rm orb}=10^{-3}\,{\rm pc}$ and $L_{\rm j}\leq 10^{44}\,{\rm erg\,s^{-1}}$, the temporal dependency of the mass loss due to the ablation by jet is consistent with a simple power law and evolves approximately as a square root of time, $\Delta M_{\star}\propto t^{1/2}$. The mass loss temporal evolution is thus slower than linear. This results in the extrapolated cumulative mass loss of the order of $10^{-6}\,M_{\odot}$ during 10 years of jet activity (e.g. post-TDE) and of the order of $10^{-4}\,M_{\odot}$ during the jetted AGN phase lasting $10^5$ years. These estimates do not depend significantly on the adopted jet flow velocity. In contrast, for large jet luminosities of $L_{\rm j}\sim 10^{48}\,{\rm erg\,s^{-1}}$, the extrapolated cumulative mass loss for the AGN phase is one order of magnitude smaller, i.e. $\Delta M_{\star}\sim 10^{-5}\,M_{\odot}$, which is likely, as already explained, due to the additional matter accumulation onto the star during each passage (see the prominent peaks in Figure~\ref{masslosscurve1e48}). In addition, for one case with $d_{\rm orb}=10^{-3}\,{\rm pc}$, $L_{\rm j}=10^{44}\,{\rm erg\,s^{-1}}$, and $v_{\rm jet}=0.33c$, the temporal evolution is more consistent with the broken power-law function, i.e. the slowing down of the mass loss for later times. This results in the comparable cumulative mass loss of the order of $\sim 10^{-6}\,M_{\odot}$ between the short post-TDE and the long AGN phases. A longer simulation capturing at least one order of magnitude larger amount of star-jet passages will help to clarify the mass-loss evolution for later times. In addition, an improved algorithm for calculating the mass difference between individual times needs to be implemented so that it evaluates just the amount of matter bound to the star. This way the temporary mass increases during the passages through the jet can be verified to a much higher accuracy than using the current volume-based mass estimate described in Subsection~\ref{numtwo}.  

\subsection{Change in the stellar effective temperature}
\label{subsec_eff_temp}
As a part of the calculations, we also evaluated the temperature structure of the star and especially of its outermost, surface layers. In Fig.~\ref{masslosscurve1e42_figfourth} and especially using the 1D longitudinal slice of the thermal structure of the star (the slice parallel to the direction of the jet passing through the centre of the star) in Fig.~\ref{masslosscurve1e42_figfourth_1D}, we show the evolution of the surface temperatures of the star during the approximately first 10 passages of the star through the jet. Specifically, the models indicate that after the first ten passages, due to the continuous ablation, the temperature of the surface of the star both on the front and on the far side of the star will increase from $\sim 3600\,\text{K}$ to $\sim 8500\,\text{K}$, which has important consequences both for the appearance of the star (spectral type changes from M to A and the luminosity increases by a factor of $\sim 30$) and especially for near-surface pressure conditions and their impact on the resistance to further jet ablation (see the discussion in Section~\ref{sec_discussion} for more details). Since the KH time should be in this case $1.93\times 10^2\,\text{yrs}$ (see Sect.~\ref{numone}), the interval $\approx 240\,\text{d}$ between two star-jet passages for $d_\text{orb}=10^{-3}\,\text{pc}$ is more
than two orders of magnitude less. Even if the star is surrounded by very hot, albeit rarefied material in this region, it is probably not enough for significant thermal relaxation (further refinement in this matter would come from the inclusion of radiative processes). Depending on the number of ablated RGs, this can also affect the radiation density and the spectrum of the diffusive light of the inner nuclear stellar cluster since the spectrum should shift towards the ultraviolet domain. However, we would need to run longer simulations to assess how long the increased surface temperature remains after the turn-off of the jet and how the overall stellar evolution of the ablated RGs is impacted.

\section{Discussion}
\label{sec_discussion}
{
\begin{table*}
    \centering
     {\caption{Comparison of the analytically estimated mass loss during the first star-jet passage $\Delta M_1^\text{max}/M_\odot$ from the equation 14 in \citet{2020ApJ...903..140Z} with the values $\Delta M_1/M_\odot$ calculated in the current phase of numerical modelling, for the individual set-ups. The analytical relation does not take into account the jet velocity, so for the different velocities within the same jet luminosity $L_{\rm j}$ the analytical mass loss rate is always the same.}
    \begin{tabular}{l|c|c}
    \hline
    \hline
    Set-up & analytical prediction of $\Delta M_1^\text{max}/M_\odot$ & numerical output for $\Delta M_1/M_\odot$ \\
    \hline
    $L_{\rm j}=10^{42}\,{\rm erg\,s^{-1}}$, $d_{\rm orb}=10^{-3}\,{\rm pc}$, $v_{\rm jet}=0.33c$     &  $-2.5 \times 10^{-7}$ &   $-4.64 \times 10^{-7}$\\
      $L_{\rm j}=10^{42}\,{\rm erg\,s^{-1}}$, $d_{\rm orb}=10^{-3}\,{\rm pc}$, $v_{\rm jet}=0.66c$ & $-2.5 \times 10^{-7}$ &   $-5.31 \times 10^{-7}$\\
     $L_{\rm j}=10^{44}\,{\rm erg\,s^{-1}}$, $d_{\rm orb}=10^{-3}\,{\rm pc}$, $v_{\rm jet}=0.33c$ & $-2.5 \times 10^{-5}$  &  $-6.23 \times 10^{-7}$\\  
       $L_{\rm j}=10^{44}\,{\rm erg\,s^{-1}}$, $d_{\rm orb}=10^{-3}\,{\rm pc}$, $v_{\rm jet}=0.66c$ & $-2.5 \times 10^{-5}$ &  $-9.55 \times 10^{-7}$\\  
      $L_{\rm j}=10^{48}\,{\rm erg\,s^{-1}}$, $d_{\rm orb}=10^{-3}\,{\rm pc}$, $v_{\rm jet}=0.33c$ & $-2.5 \times 10^{-1}$  &   $-5.81 \times 10^{-6}$ \\ 
       $L_{\rm j}=10^{48}\,{\rm erg\,s^{-1}}$, $d_{\rm orb}=10^{-3}\,{\rm pc}$, $v_{\rm jet}=0.66c$ & $-2.5 \times 10^{-1}$ &  $-6.32 \times 10^{-6}$\\\grayline $L_{\rm j}=10^{42}\,{\rm erg\,s^{-1}}$, $d_{\rm orb}=10^{-2}\,{\rm pc}$, $v_{\rm jet}=0.66c$ & $-3 \times 10^{-9}$ &  $-2.23 \times 10^{-8}$ \\
    \hline     
    \end{tabular} 
    \label{tab2:my_label}}  
\end{table*}
According to the performed first numerical models of the interaction of standard RGs with the galactic jet, we evaluated the mass loss of these stars via the jet ablation for roughly the first ten passages of the star through the galactic jet with all the above described parameters. Consequently, we  extrapolated the trends of this ablation, i.e. the mass removal from the outermost stellar layers, for the much longer time intervals as well (AGN and post-TDE phases). Naturally, the question arises as to how relevant such an extrapolation is. To increase the credibility of these models, more statistically relevant calculations of these processes should be performed over a much longer time interval. This will require an additional effort and the involvement of more powerful computational resources. For these reasons, we plan to continue working on this project and evaluate the impact of the stellar passages through the jet for at least a ten-times longer period of time. In this section, we compare the mass loss during one passage as inferred from numerical runs with the analytical predictions (Subsection~\ref{subsec_comparison_mass_loss}). Then we discuss the potential effect of magnetic field (Subsection~\ref{subsec_mag_field}). Subsequently, we look at observational signatures of the interaction of stars with a nuclear outflow in the Galactic centre nowadays (Subsection~\ref{subsec_star_outflow}). We also discuss the effect of the jet half-opening angle in Subsection~\ref{subsec_narrower_jet}. Finally, we revisit the likelihood assessment of the red-giant/jet interaction in the Milky Way-like nuclear star cluster in Subsection~\ref{subsec_statistics}. 

\subsection{Comparison of mass loss with analytical predictions}
\label{subsec_comparison_mass_loss}

If we evaluate the mass loss as predicted analytically for one passage of the star through the jet using Eq. (14) in \citet{2020ApJ...903..140Z}, we obtain the values (for different set-ups with $L_\text{j}$ of the jet and other parameters) listed in the second column of Table~\ref{tab2:my_label}. The third column of Table~\ref{tab2:my_label} lists the values calculated numerically based on the current models presented in this paper.

This comparison shows a fairly good agreement of the amount of mass removed from the star only for the orbital distance of $d_{\rm orb}=10^{-3}\,{\rm pc}$ and the jet luminosity of $L_{\rm j}=10^{42}\,{\rm erg\,s^{-1}}$ (this value could be consistent with the power of the previous jet activity in our Galaxy). We stress that only one passage of the star through the jet is taken into account, specifically the first one in the numerical calculations, which appears to be mostly above the average. For jets with a higher kinetic luminosity, however, the mass loss differs significantly by several orders of magnitude. This discrepancy seems to be caused by a combination of the following factors. The higher jet luminosity implies a higher density of the hadronic jet which brings more material during the first moment after the entry of the star into the jet that is temporarily accumulated around the star in its original volume, i.e. in the space originally filled by the star}. To some extent, it prevents the ablation of the actual stellar body. This temporary material accumulation is fairly visible in the graphs of mass loss in Figs.~\ref{masslosscurve1e42_fignine}\,--\,\ref{masslosscurve1e48}. It is not considered in the analytical approach. Another reason for the discrepancy could be the increased temperature of the surface layers of the star after their partial ablation. This leads to higher thermodynamic pressure resisting further ablation of the respective layers (see the detailed increase of the stellar surface temperature with time for the case of $L_{\rm j}=10^{42}\,{\rm erg\,s^{-1}}$ in Fig.~\ref{masslosscurve1e42_figfourth_1D}), which is also a phenomenon not taken in account in the simple analytical relation. We also note here that the current simulations take about 4\,\% of the KH time in total and the inter-jet times of the inter-jet motion of the star at $d_\text{orb}=10^{-3}\,\text{pc}$ are about $240$ days each, so the fundamental thermal relaxation does not seem to occur (see Sect.~\ref{subsec_eff_temp} for a more specific description of this context).

At this point, we also note that we did not include the radiative cooling within the study due to the extensive complexity of the problem. The significance of the radiative cooling on the stellar surface temperature evolution will be studied separately. Due to the high computational cost of the global radiative hydrodynamic simulations, this will likely have to be done as a particular comparative study within the limited spatial and temporal range and then it should be extrapolated to the overall evolution of the studied process. For the same reason, we also neglect the heat conduction in the energy equation (cf.~Section~\ref{numone}).

We also note that the numerical study does not yet take into account the stellar rotation.
Thus, the significance of the rotational effect is still questionable and can be evaluated by a detailed sub-study, for example, by comparing the degree of surface ablation of a non-rotating and a rotating star during one stellar orbit.

\subsection{Effect of magnetic field}
\label{subsec_mag_field}

The inclusion of the magnetic field of the jet can in principle influence the behaviour of the ablation process under investigation. On one hand, the magnetisation of the surface layers of the star should probably result in the slowing down of the ablation due to some degree of magnetically induced confinement of the stellar surface layers and/or of the surrounding regions of the star. The quantification of the influence of magnetic fields on mass loss rates from stars is given, e.g. in \citet{2021A&A...656A.131D}, however, in this case it is only a qualitative estimate and a more detailed study would be needed to determine the actual impact of this effect.
 On the other hand, it is possible that the matter in the jet will be more strongly bound within a given opening angle due to the helical magnetic field, so the effect caused by the reinforced ram pressure of the jet onto the star will be stronger. Using the first principles, it can be estimated that at the orbital distance of $d_{\rm orb}=10^{-3}\,{\rm pc}$, the magnetic field becomes dynamically relevant for the magnitude of $B\sim 30\,{\rm G}$ considering the jet luminosity of $L_{\rm j}= 10^{42}\,{\rm erg\,s^{-1}}$, see Appendix~\ref{app_mag_field}. In any case, the effects of the magnetic field should be explored within the future simulations.

The processes described above desire a more detailed hydrodynamic modelling, probably within a limited box containing only a part (or multiple parts) of the contact between the stellar surface and the jet stream to achieve a more detailed description of the physics along the interaction zone. However, for such a set-up, the question arises how much such a limited (albeit detailed and high-resolution) sample can capture a phenomenon such as the total temporary mass packing on a star requiring a computational domain containing the whole star and its surroundings. In any case, this is a relevant problem and a major challenge for subsequent more advanced models covering a significantly longer time period of the process.

\subsection{Observational motivation and current signs of star--nuclear outflow interaction}
\label{subsec_star_outflow}

Observations of the Galactic centre SMBH (Sgr~A*) do not show a clear sign of an active radio jet in its current state of a low activity \citep{2022ApJ...930L..12E,2024ApJ...964L..25E}, yet the existence of the weaker jet emanating from Sgr~A* is expected based on the nature of its surrounding hot magnetised accretion flow with the dominant poloidal component of the magnetic field \citep{2000A&A...362..113F,2014ARA&A..52..529Y,2018A&A...618L..10G}. There have been claims of the detection of linear spatial structures resembling jets in the inner parsec in the radio, mm \citep{2012ApJ...758L..11Y,2020MNRAS.499.3909Y} as well as X-ray domains \citep{2013ApJ...779..154L} that remain to be revisited in the future. Regardless of these putative jet structures, it appears that the jet associated with Sgr~A* could have been more powerful a few million years ago. This is hinted by smaller $\gamma-$ray Fermi and larger X-ray eROSITA bubbles \citep{2010ApJ...724.1044S,2020Natur.588..227P} that point towards one or a series of energetic events (related to accretion onto the SMBH or star-formation) in the nuclear region a few million years ago. Numerical cosmic-ray/MHD models of the increased jet activity in the Galactic centre a few million years ago can reproduce several properties of Fermi and eROSITA bubbles \citep{2012ApJ...756..181G,2022NatAs...6..584Y} when the jet kinetic power is in the range of $L_{\rm j}\simeq 10^{42}-10^{44}\,{\rm erg\,s^{-1}}$ and the jet activity lasted for $\sim 0.1 -1.0$ Myr. Since the total injected energy was $\sim 10^{56}-10^{57}\,{\rm erg}$ \citep{2019ApJ...886...45B}, the jet-phase duration and the total luminosity are mutually degenerate. That is also why we consider a range of jet kinetic luminositities in our hydrodynamical models. 

Even in the current low state of Sgr~A*, there are several examples of the interaction of stars with their surrounding ambient medium. Since these stars generally move supersonically, they drive bow shocks into the surrounding medium that are symmetric with respect to the relative velocity vector. There are several prominent examples of bow-shock sources in the NIR and MIR domains, especially along the ionised dusty streamers in the Minispiral region \citep{2014A&A...567A..21S,2022ApJ...929..178B}; see also \citet{2017bhns.work..237Z} and \citet{2018acps.confE..49Z} for overviews.

This supersonic interaction can mimic the jet-star interaction in the limit of low-Mach numbers. For evolved stars, a prominent example is the pulsating M1/M2 red supergiant IRS 7 star (5.5'' north of Sgr~A*), which exhibits a shell-like structure and a long tail with a few trailing components at 1''-3.6'' north of IRS 7 \citep{1991ApJ...371L..59Y,1992ApJ...385L..41Y,2020PASJ...72...36T,2022ApJ...929..178B}. Another prominent bow-shock source associated with a late-type star is the AGB star IRS 3 \citep{2005A&A...433..117V}, which is generally the brightest and the most extended stellar source in the central parsec. The infrared emission associated with IRS 3 has a radius of $\sim 2\times 10^6\,R_{\odot}$, which is in excess of the standard evolution of red supergiants or AGB stars. A possible interpretation involves expansion via shock heating induced by a fast nuclear outflow in combination with tidal prolongation as the star orbits around Sgr~A* \citep{2017ApJ...837...93Y,2022ApJ...929..178B}.

Furthermore, in the central cavity region, there were two sources identified -- X3 and X7 -- that are clearly comet-shaped and oriented away from Sgr~A* \citep{2010A&A...521A..13M,2021ApJ...909...62P,2023ApJ...944..231P}. Such features are consistent with wind-blowing stars interacting with the fast nuclear outflow with the velocity of a few 1000 km/s. The X7 source has been undergoing a rapid evolution \citep{2021ApJ...909...62P}, which could be associated with the detachment of the bow-shock shell that is further undergoing tidal stretching under the influence of Sgr~A* \citep{2023ApJ...944..136C}. Another extended bow-shock source oriented away from Sgr~A* is X8 \citep{2019A&A...624A..97P}. All of these sources deserve further analysis as they could manifest the interactions of stars with the accretion-flow nuclear outflow or potentially a weaker jet. In this regard, they can serve as ideal test-beds for 3D computational models of star-nuclear outflow interactions.

\subsection{Effect of a jet opening angle}
\label{subsec_narrower_jet}
The best observations of the gradually varying collimation of the galactic jet are available for the galaxies M87 \citep{1999Natur.401..891J} and for Centaurus A \citep{2021NatAs...5.1017J}. Of course, these are special cases and the jet in Sgr~A* could have had different properties. For distances of a few tens of $r_\text{g}$ (gravitational radii), a half-opening angle could be as much as $\sim 30^\circ$. If we consider distances comparable to or greater than $2000\,r_\text{g}$, it is likely reasonable to consider a half-opening angle of $5^\circ$. In this sense, for the past Galactic jet (noting that the distance $10^{-3}\,\text{pc}$ corresponds to approx. $5231\,\,r_\text{g}$ in our Galaxy), we should rather consider a half-opening angle of $~5^\circ$, when the star orbited at a distance $\lesssim 0.01\,\text{pc}$. For the comparable orbital distance around, e.g., M\,87, the half-opening angle could be, however, larger. Since our original idea was to investigate more versatile set-ups in galactic nuclei, we first chose a larger half-opening angle.

\begin{figure}
    \centering
    \includegraphics[width=0.7\columnwidth]{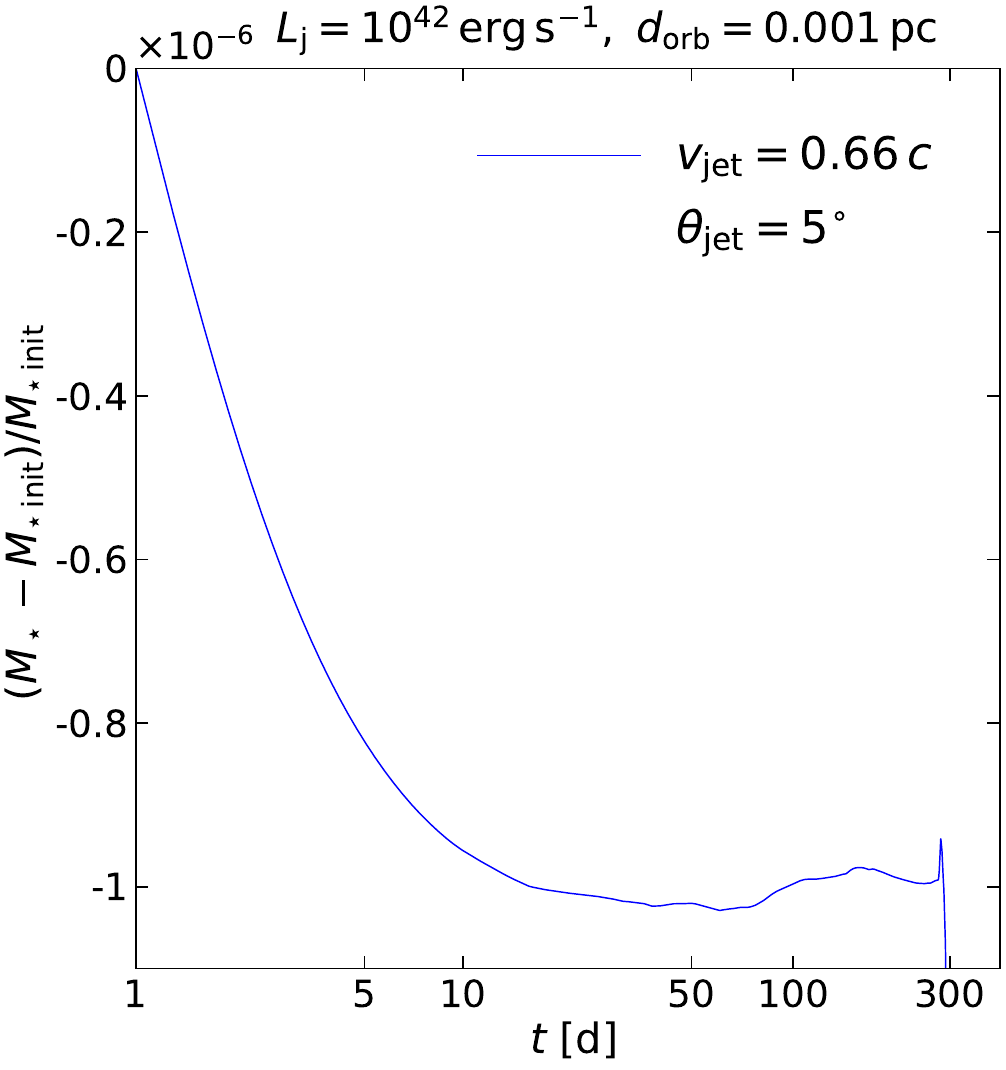}
    \caption{Detailed temporal profile of the relative mass loss for the case of a narrower jet with a half-opening angle of $\theta=5^\circ$, during the first RG-jet passage up to the beginning of the second passage. We consider the jet luminosity of $L_\text{j} = 10^{42}\,\text{erg}\,\text{s}^{-1}$, the jet flow velocity
    of $v_\text{jet} = 0.66\,c$, and the orbital distance of the star of $d_\text{orb} = 10^{-3}\,\text{pc}$. The horizontal (time) axis is plotted using a logarithmic scale.}
    \label{masslosscurve1e42_narrower}
\end{figure}

Here we present an initial 3D simulation of a RG-jet interaction with a half-opening angle of $\theta = 5^\circ$, i.e. half the value of $\theta$ introduced in the previous models. We will not demonstrate here the detailed hydrodynamics, which does not differ significantly from the one already reported. Due to the significant cost of 3D models, we have so far only performed an initial stage of such a simulation up to 300 days. This covers one passage of the star through the jet and one intermediate phase, up to the beginning of the second passage. We quantified the mass that might be ablated within the first star-jet passage, the result is shown in Fig.~\ref{masslosscurve1e42_narrower}. The relative mass loss is now almost twice as large as for the model with a broader jet (see Fig.~\ref{masslosscurve1e42_fignine}) which may correspond to the fact that the jet density is four times higher than for the case with $\theta=10^\circ$, see Eq.~\eqref{eq_jet_number_density}, whereas the star spends half the time in the jet. However, since such simple linear relations do not translate proportionally to the mass-loss rate in the models mentioned above and the simulation time is too short in this case, we can claim no unambiguous conclusions from this initial investigation, except that the ablation should be higher than for a broader jet. Future simulations that will last longer are necessary for more precise conclusions.

\subsection{Statistical limitations of the red giant-jet interaction model}
\label{subsec_statistics}

We focused on the scenario of a RG interacting repeatedly with the jet. Here we address the basic statistical incidence of this setup. We approximate the number density distribution of late-type stars using,
\begin{equation}
    n_{\rm LT}=n_{\rm LT,m}\left(\frac{r}{r_{\rm m}}\right)^{-\gamma}\,
    \label{eq_bahcall_wolf}
\end{equation}
where we assume a relaxed cusp of late-type stars approximately following the Bahcall-Wolf profile for a nuclear star cluster with unequal stellar masses with $\gamma=3/2$ \citep{1977ApJ...216..883B}. The density profile is scaled to the gravitational influence radius of the SMBH, $r_{\rm m}\sim 2.5\,{\rm pc}$, which coincides approximately with the influence radius of the jet, $\mathcal{R}_{\rm jet-inf}\sim r_{\rm m}$, where the aligned orbiting star can experience at least one interaction with the jet. The number density of late-type stars at $r_{\rm m}$ is estimated using,
\begin{equation}
    n_{\rm LT,m}\simeq \frac{3f_{\rm LT}M_{\bullet}}{2 \pi r_{\rm m}^3 <m_{\star}>}\sim 10^5\,{\rm pc^{-3}}\,,
\end{equation}
where we considered $M_{\star}(r<r_{\rm m})=2M_{\bullet}$ from the definition of the sphere of influence, the fraction of late-type stars $f_{\rm LT}\sim 0.8$ \citep{2020A&A...641A.102S}, and the mean mass of late-type stars $<m_{\star}>=1\,M_{\odot}$.

Using the power-law relation for $n_{\rm LT}(r)$ and the biconical approximation for the jet, we arrive at the mean number of stars $N_{\rm LT, jet}(<r)$ that are interacting with the jet at any time,
\begin{equation}
    N_{\rm LT, jet}(<r)=\frac{2\pi n_{\rm LT, m} r_{\rm m}^{\gamma} r^{3-\gamma}\tan^2{\theta}}{3-\gamma}\,,
\end{equation}
which is $N_{\rm LT, jet}(<10^{-3}\,{\rm pc})\sim 1.6$ and $N_{\rm LT, jet}(<10^{-2}\,{\rm pc})\sim 51.5$, assuming $\theta=10^{\circ}$. Hence, for the adopted stellar density distribution, the number of interacting late-type stars is of the order of unity on milliparsec scales. 

A physically motivated measure of the RG-jet interactions is the total interaction rate per unit time at a given distance from the SMBH, i.e. $\dot{N}_{\rm LT}\approx n_{\rm LT}(r)\sigma_{\star}(r)S_{\rm jet}$, where $\sigma_{\star}$ is the stellar velocity dispersion that we approximate by the Keplerian orbital velocity and $S_{\rm jet}\sim 2r^2 \tan{\theta}$ is the total surface area of the jet. For the interaction rate we obtain,
\begin{equation}
    \dot{N}_{\rm LT}(r)\approx 2 n_{\rm LT,m}r_{\rm m}^{\gamma} (GM_{\bullet})^{1/2} r^{3/2-\gamma} \tan{\theta}\,.
\end{equation}
The number of RG-jet interactions per orbital period at a given radius is
\begin{equation}
    N_{\rm LT}(r)=\dot{N}_{\rm LT}(r)P_{\star}=4 \pi n_{\rm LT,m}r_{\rm m}^{\gamma} r^{3-\gamma} \tan{\theta}\,,
\end{equation}
and the mean number of interactions in the sphere around the SMBH with the outer radius $r_{\rm out}$ is
\begin{align}
    <N_{\rm LT}>(r_{\rm out})&=\frac{\int_0^{r_{\rm out}} N_{\rm LT}(r')4\pi r'^2\mathrm{d}r'}{\int_0^{r_{\rm out}} 4\pi r'^2\mathrm{d}r'}\,\notag\\
    &=\frac{12 \pi n_{\rm LT,m}r_{\rm m}^{\gamma}r_{\rm out}^{3-\gamma}\tan{\theta}}{6-\gamma}\,,
\end{align}
which is $<N_{\rm LT}>(10^{-3}\,{\rm pc})\sim 18.5$ and $<N_{\rm LT}>(10^{-2}\,{\rm pc})\sim 583.9$.

These order-of-magnitude estimates show that once a cusp of late-type stars exists around the SMBH, several tens to hundreds of stars can be affected by repetitive passages through the jet on subparsec scales. Larger late-type giants are expected to be ablated more because of the larger cross-sectional area. On the other hand, their number is small in comparison with the whole late-type population, hence the mean number $<N_{\rm LT}>$ is an upper limit for the ablation of large red giants.   The estimate of $<N_{\rm LT}>$ is still consistent with the observationally inferred number of ``missing'' bright late-type stars:  $\sim 4-5$ within $\sim 0.04$ pc \citep{2019ApJ...872L..15H} and about 100 missing giants within $\sim 0.3$ pc \citep{2018A&A...609A..26G}.  Although the models presented here were focused on a special, fine-tuned RG evolutionary stage of a Solar-type star ($R_{\star}=100\,R_{\odot}$, $m_{\star}=1\,M_{\odot}$), which is rather short-lived ($\sim 2$ million years), the mechanism is applicable to all late-type stars with both smaller and larger stellar radii. The only condition for the photosphere ablation is the terminal stellar wind velocity that should be smaller than the critical value provided by Eq.~\eqref{eq_stellar_wind}. A hydrodynamical study of the ablation of late-type stars of different masses and ages will be necessary to assess the impact of the jet-induced ablation on the whole population of the nuclear star cluster.

\section{Conclusions}
\label{sec_conclusions}
In this study, we present the results of numerical hydrodynamical models that aim to verify the analytical scenario presented in \citet{2020ApJ...903..140Z}, which suggested that the lack of bright RGs in the inner regions of the Galactic NSC could partially be caused by repetitive star-jet interactions during the phase when Sgr~A* was more active a few million years ago. Although this process likely occurred simultaneously with other previously proposed mechanisms, such as star-disc interactions, star-star collisions, or tidal disruptions due to the Galactic SMBH, the ablation of RGs by an active jet could alter their appearance, i.e. mainly increase their surface temperature. For the smallest orbital distance of $10^{-3}\,\text{pc}$, we find a rapid growth in temperature during the first 10 passages from $\sim$3600\,K to $\sim$8500\,K. Hence,
the RG may at least temporarily change the stellar type from M to A within the relatively short time period. This change is limited to the phase when the RG repeatedly interacts with the jet. After the interaction ceases, the temperature will decrease due to radiative losses on the Kelvin-Helmholtz timescale of $\sim 10-100$ years.

We performed 3D hydrodynamical simulations of about 10 passages of the star through the jet for the case of the orbital distance of $10^{-3}\,\text{pc}$ from the SMBH and 2D simulations of 2 passages for the case of the orbital distance of $10^{-2}\,\text{pc}$. We calculate the star's ablation rate for three values of the jet luminosity, $10^{42}\,\text{erg}\,\text{s}^{-1}$, $10^{44}\,\text{erg}\,\text{s}^{-1}$, both of which fall into the likely range of luminosities of the putative past jet of Sgr~A*, and $10^{48}\,\text{erg}\,\text{s}^{-1}$, which is generally an upper limit for the jet kinetic luminosity. We also account for the subsequent evolution of the stellar surface layers during the period between each star-jet interaction. The ablated mass behaves with time approximately as $\Delta M_{\star}\propto \sqrt{t}$ for $L_{\rm j} \leq 10^{44}\,{\rm erg\,s^{-1}}$. At the present stage of the simulations, we extrapolate the mass loss for the longer duration of the jet activity. For the case of $\sim$10 stellar passages through jet at the orbital distance of $10^{-3}\,\text{pc}$, we estimate the cumulative mass loss of $\Delta M_{\star}\sim 10^{-6}\,M_{\odot}$ after 10 years of the jet activity (e.g. during the post-TDE phase) and $\Delta M_{\star}\sim 10^{-4}\,M_{\odot}$ for the AGN phase of $10^5$ years considering the jet luminosity of $L_{\rm j} \leq 10^{44}\,{\rm erg\,s^{-1}}$. For the larger jet luminosity of $L_{\rm j}=10^{48}\,{\rm erg\,s^{-1}}$, the mass loss tends to be significantly flatter with time, resulting in the total ablated mass of $\Delta M_{\star}\sim 10^{-5}\,M_{\odot}$ at the end of the AGN phase. Overall, the repeated RG-jet interaction leads to the ablated mass that is comparable to and amplifies the mass loss due to the stellar wind.

The current stage of the numerical simulations shows a relatively good agreement with the analytical predictions for the orbital distance of $10^{-3}\,\text{pc}$ and the jet luminosity of $10^{42}\,\text{erg}\,\text{s}^{-1}$, while for the other cases, the results significantly differ -- the ablation rates in the numerical models are generally lower. The 3D models do not imply significant changes in the stellar size. Instead they point towards an increase in the surface temperature of RGs repeatedly interacting with the jet. An important contribution of this first stage of the numerical simulations and solutions of the jet--star interaction phenomenon is that we have verified in detail and technically gone through the methods of calculating the interaction of the upper layers of the RG with the surrounding matter on the global scale of the stellar body and thus overcame numerous caveats that appear during such simulations. This gives us a solid basis for the future work, where we will aim to perform longer and thus statistically more relevant simulations, involving at least a few hundred star-jet passages. This will allow us to verify the extrapolated values of the total mass ablated from RGs and link the numerical results more closely to the observed properties of stars and their subsequent stellar evolution in the inner regions of the Galactic centre as well as other nuclei.

\section*{Acknowledgements}

We are grateful to the anonymous referee for providing constructive comments that helped to improve the manuscript.
We enjoyed fruitful discussions with Vladimír Karas, Matúš Labaj, and Asif ud-Doula for helpful comments that improved the paper.
 NW and PK have received support from the GA\v{C}R EXPRO grant No. 21-13491X (``Exploring the Hot Universe and Understanding Cosmic Feedback'').  MZ acknowledges the GA\v{C}R JUNIOR STAR grant no. GM24-10599M (``Stars in galactic nuclei: interrelation with massive black holes'') for support.

\section*{Data Availability}

The radiative/MHD hydrodynamics code \texttt{CASTRO}, which was used to obtain the results presented in this study, is available on \url{https://amrex-astro.github.io/Castro/}. The data concerning simulation runs of red giant-jet interactions presented in this paper can be made available to the reader upon request.




\begin{thebibliography}{}
\makeatletter
\relax
\def\mn@urlcharsother{\let\do\@makeother \do\$\do\&\do\#\do\^\do\_\do\%\do\~}
\def\mn@doi{\begingroup\mn@urlcharsother \@ifnextchar [ {\mn@doi@}
  {\mn@doi@[]}}
\def\mn@doi@[#1]#2{\def\@tempa{#1}\ifx\@tempa\@empty \href
  {http://dx.doi.org/#2} {doi:#2}\else \href {http://dx.doi.org/#2} {#1}\fi
  \endgroup}
\def\mn@eprint#1#2{\mn@eprint@#1:#2::\@nil}
\def\mn@eprint@arXiv#1{\href {http://arxiv.org/abs/#1} {{\tt arXiv:#1}}}
\def\mn@eprint@dblp#1{\href {http://dblp.uni-trier.de/rec/bibtex/#1.xml}
  {dblp:#1}}
\def\mn@eprint@#1:#2:#3:#4\@nil{\def\@tempa {#1}\def\@tempb {#2}\def\@tempc
  {#3}\ifx \@tempc \@empty \let \@tempc \@tempb \let \@tempb \@tempa \fi \ifx
  \@tempb \@empty \def\@tempb {arXiv}\fi \@ifundefined
  {mn@eprint@\@tempb}{\@tempb:\@tempc}{\expandafter \expandafter \csname
  mn@eprint@\@tempb\endcsname \expandafter{\@tempc}}}

\bibitem[\protect\citeauthoryear{{Ali} et~al.,}{{Ali}
  et~al.}{2020}]{2020ApJ...896..100A}
{Ali} B.,  et~al., 2020, \mn@doi [\apj] {10.3847/1538-4357/ab93ae}, \href
  {https://ui.adsabs.harvard.edu/abs/2020ApJ...896..100A} {896, 100}

\bibitem[\protect\citeauthoryear{{Almgren} et~al.,}{{Almgren}
  et~al.}{2010}]{2010ApJ...715.1221A}
{Almgren} A.~S.,  et~al., 2010, \mn@doi [\apj] {10.1088/0004-637X/715/2/1221},
  \href {https://ui.adsabs.harvard.edu/abs/2010ApJ...715.1221A} {715, 1221}

\bibitem[\protect\citeauthoryear{{Almgren} et~al.,}{{Almgren}
  et~al.}{2011}]{2011ascl.soft05010A}
{Almgren} A.~S.,  et~al., 2011, {CASTRO: Multi-dimensional Eulerian AMR
  Radiation-hydrodynamics Code}, Astrophysics Source Code Library, record
  ascl:1105.010

\bibitem[\protect\citeauthoryear{{Amaro-Seoane} \& {Chen}}{{Amaro-Seoane} \&
  {Chen}}{2014}]{2014ApJ...781L..18A}
{Amaro-Seoane} P.,  {Chen} X.,  2014, \mn@doi [\apjl]
  {10.1088/2041-8205/781/1/L18}, \href
  {https://ui.adsabs.harvard.edu/abs/2014ApJ...781L..18A} {781, L18}

\bibitem[\protect\citeauthoryear{{Amaro-Seoane}, {Chen}, {Sch{\"o}del}  \&
  {Casanellas}}{{Amaro-Seoane} et~al.}{2020}]{2020MNRAS.492..250A}
{Amaro-Seoane} P.,  {Chen} X.,  {Sch{\"o}del} R.,   {Casanellas} J.,  2020,
  \mn@doi [\mnras] {10.1093/mnras/stz3507}, \href
  {https://ui.adsabs.harvard.edu/abs/2020MNRAS.492..250A} {492, 250}

\bibitem[\protect\citeauthoryear{{Antonucci} \& {Miller}}{{Antonucci} \&
  {Miller}}{1985}]{1985ApJ...297..621A}
{Antonucci} R.~R.~J.,  {Miller} J.~S.,  1985, \mn@doi [\apj] {10.1086/163559},
  \href {https://ui.adsabs.harvard.edu/abs/1985ApJ...297..621A} {297, 621}

\bibitem[\protect\citeauthoryear{{Araudo}, {Bosch-Ramon}  \& {Romero}}{{Araudo}
  et~al.}{2013}]{2013MNRAS.436.3626A}
{Araudo} A.~T.,  {Bosch-Ramon} V.,   {Romero} G.~E.,  2013, \mn@doi [\mnras]
  {10.1093/mnras/stt1840}, \href
  {https://ui.adsabs.harvard.edu/abs/2013MNRAS.436.3626A} {436, 3626}

\bibitem[\protect\citeauthoryear{{Baganoff} et~al.,}{{Baganoff}
  et~al.}{2003}]{2003ApJ...591..891B}
{Baganoff} F.~K.,  et~al., 2003, \mn@doi [\apj] {10.1086/375145}, \href
  {https://ui.adsabs.harvard.edu/abs/2003ApJ...591..891B} {591, 891}

\bibitem[\protect\citeauthoryear{{Bahcall} \& {Wolf}}{{Bahcall} \&
  {Wolf}}{1977}]{1977ApJ...216..883B}
{Bahcall} J.~N.,  {Wolf} R.~A.,  1977, \mn@doi [\apj] {10.1086/155534}, \href
  {https://ui.adsabs.harvard.edu/abs/1977ApJ...216..883B} {216, 883}

\bibitem[\protect\citeauthoryear{{Barkov}, {Aharonian}, {Bogovalov}, {Kelner}
  \& {Khangulyan}}{{Barkov} et~al.}{2012}]{2012ApJ...749..119B}
{Barkov} M.~V.,  {Aharonian} F.~A.,  {Bogovalov} S.~V.,  {Kelner} S.~R.,
  {Khangulyan} D.,  2012, \mn@doi [\apj] {10.1088/0004-637X/749/2/119}, \href
  {https://ui.adsabs.harvard.edu/abs/2012ApJ...749..119B} {749, 119}

\bibitem[\protect\citeauthoryear{{Bednarek} \& {Banasi{\'n}ski}}{{Bednarek} \&
  {Banasi{\'n}ski}}{2015}]{2015ApJ...807..168B}
{Bednarek} W.,  {Banasi{\'n}ski} P.,  2015, \mn@doi [\apj]
  {10.1088/0004-637X/807/2/168}, \href
  {https://ui.adsabs.harvard.edu/abs/2015ApJ...807..168B} {807, 168}

\bibitem[\protect\citeauthoryear{{Bhat}, {Sabha}, {Zaja{\v{c}}ek}, {Eckart},
  {Sch{\"o}del}, {Hosseini}, {Pei{\ss}ker}  \& {Zensus}}{{Bhat}
  et~al.}{2022}]{2022ApJ...929..178B}
{Bhat} H.~K.,  {Sabha} N.~B.,  {Zaja{\v{c}}ek} M.,  {Eckart} A.,  {Sch{\"o}del}
  R.,  {Hosseini} S.~E.,  {Pei{\ss}ker} F.,   {Zensus} A.,  2022, \mn@doi
  [\apj] {10.3847/1538-4357/ac6106}, \href
  {https://ui.adsabs.harvard.edu/abs/2022ApJ...929..178B} {929, 178}

\bibitem[\protect\citeauthoryear{{Bland-Hawthorn} et~al.,}{{Bland-Hawthorn}
  et~al.}{2019}]{2019ApJ...886...45B}
{Bland-Hawthorn} J.,  et~al., 2019, \mn@doi [\apj] {10.3847/1538-4357/ab44c8},
  \href {https://ui.adsabs.harvard.edu/abs/2019ApJ...886...45B} {886, 45}

\bibitem[\protect\citeauthoryear{{Bosch-Ramon}, {Perucho}  \&
  {Barkov}}{{Bosch-Ramon} et~al.}{2012}]{2012A&A...539A..69B}
{Bosch-Ramon} V.,  {Perucho} M.,   {Barkov} M.~V.,  2012, \mn@doi [\aap]
  {10.1051/0004-6361/201118622}, \href
  {https://ui.adsabs.harvard.edu/abs/2012A&A...539A..69B} {539, A69}

\bibitem[\protect\citeauthoryear{{Burrows} et~al.,}{{Burrows}
  et~al.}{2011}]{2011Natur.476..421B}
{Burrows} D.~N.,  et~al., 2011, \mn@doi [\nat] {10.1038/nature10374}, \href
  {https://ui.adsabs.harvard.edu/abs/2011Natur.476..421B} {476, 421}

\bibitem[\protect\citeauthoryear{{Cantiello}, {Jermyn}  \& {Lin}}{{Cantiello}
  et~al.}{2021}]{2021ApJ...910...94C}
{Cantiello} M.,  {Jermyn} A.~S.,   {Lin} D. N.~C.,  2021, \mn@doi [\apj]
  {10.3847/1538-4357/abdf4f}, \href
  {https://ui.adsabs.harvard.edu/abs/2021ApJ...910...94C} {910, 94}

\bibitem[\protect\citeauthoryear{{Ciurlo} et~al.,}{{Ciurlo}
  et~al.}{2023}]{2023ApJ...944..136C}
{Ciurlo} A.,  et~al., 2023, \mn@doi [\apj] {10.3847/1538-4357/acb344}, \href
  {https://ui.adsabs.harvard.edu/abs/2023ApJ...944..136C} {944, 136}

\bibitem[\protect\citeauthoryear{{Colella} \& {Woodward}}{{Colella} \&
  {Woodward}}{1984}]{1984JCoPh..54..174C}
{Colella} P.,  {Woodward} P.~R.,  1984, \mn@doi [Journal of Computational
  Physics] {10.1016/0021-9991(84)90143-8}, \href
  {http://adsabs.harvard.edu/abs/1984JCoPh..54..174C} {54, 174}

\bibitem[\protect\citeauthoryear{{Czerny} et~al.,}{{Czerny}
  et~al.}{2023}]{2023Ap&SS.368....8C}
{Czerny} B.,  et~al., 2023, \mn@doi [\apss] {10.1007/s10509-023-04165-7}, \href
  {https://ui.adsabs.harvard.edu/abs/2023Ap&SS.368....8C} {368, 8}

\bibitem[\protect\citeauthoryear{{Dale}, {Davies}, {Church}  \&
  {Freitag}}{{Dale} et~al.}{2009}]{2009MNRAS.393.1016D}
{Dale} J.~E.,  {Davies} M.~B.,  {Church} R.~P.,   {Freitag} M.,  2009, \mn@doi
  [\mnras] {10.1111/j.1365-2966.2008.14254.x}, \href
  {https://ui.adsabs.harvard.edu/abs/2009MNRAS.393.1016D} {393, 1016}

\bibitem[\protect\citeauthoryear{{Davies} \& {Lin}}{{Davies} \&
  {Lin}}{2020}]{2020MNRAS.498.3452D}
{Davies} M.~B.,  {Lin} D. N.~C.,  2020, \mn@doi [\mnras]
  {10.1093/mnras/staa2590}, \href
  {https://ui.adsabs.harvard.edu/abs/2020MNRAS.498.3452D} {498, 3452}

\bibitem[\protect\citeauthoryear{{Do} et~al.,}{{Do}
  et~al.}{2019}]{2019Sci...365..664D}
{Do} T.,  et~al., 2019, \mn@doi [Science] {10.1126/science.aav8137}, \href
  {https://ui.adsabs.harvard.edu/abs/2019Sci...365..664D} {365, 664}

\bibitem[\protect\citeauthoryear{{Driessen}, {Kee}  \& {Sundqvist}}{{Driessen}
  et~al.}{2021}]{2021A&A...656A.131D}
{Driessen} F.~A.,  {Kee} N.~D.,   {Sundqvist} J.~O.,  2021, \mn@doi [\aap]
  {10.1051/0004-6361/202142175}, \href
  {https://ui.adsabs.harvard.edu/abs/2021A&A...656A.131D} {656, A131}

\bibitem[\protect\citeauthoryear{{Eckart} \& {Genzel}}{{Eckart} \&
  {Genzel}}{1996}]{1996Natur.383..415E}
{Eckart} A.,  {Genzel} R.,  1996, \mn@doi [\nat] {10.1038/383415a0}, \href
  {https://ui.adsabs.harvard.edu/abs/1996Natur.383..415E} {383, 415}

\bibitem[\protect\citeauthoryear{{Eckart} \& {Genzel}}{{Eckart} \&
  {Genzel}}{1997}]{1997MNRAS.284..576E}
{Eckart} A.,  {Genzel} R.,  1997, \mn@doi [\mnras] {10.1093/mnras/284.3.576},
  \href {https://ui.adsabs.harvard.edu/abs/1997MNRAS.284..576E} {284, 576}

\bibitem[\protect\citeauthoryear{{Eckart} et~al.,}{{Eckart}
  et~al.}{2017}]{2017FoPh...47..553E}
{Eckart} A.,  et~al., 2017, \mn@doi [Foundations of Physics]
  {10.1007/s10701-017-0079-2}, \href
  {https://ui.adsabs.harvard.edu/abs/2017FoPh...47..553E} {47, 553}

\bibitem[\protect\citeauthoryear{{Event Horizon Telescope Collaboration}
  et~al.,}{{Event Horizon Telescope Collaboration}
  et~al.}{2022}]{2022ApJ...930L..12E}
{Event Horizon Telescope Collaboration} et~al., 2022, \mn@doi [\apjl]
  {10.3847/2041-8213/ac6674}, \href
  {https://ui.adsabs.harvard.edu/abs/2022ApJ...930L..12E} {930, L12}

\bibitem[\protect\citeauthoryear{{Event Horizon Telescope Collaboration}
  et~al.,}{{Event Horizon Telescope Collaboration}
  et~al.}{2024}]{2024ApJ...964L..25E}
{Event Horizon Telescope Collaboration} et~al., 2024, \mn@doi [\apjl]
  {10.3847/2041-8213/ad2df0}, \href
  {https://ui.adsabs.harvard.edu/abs/2024ApJ...964L..25E} {964, L25}

\bibitem[\protect\citeauthoryear{{Falcke} \& {Markoff}}{{Falcke} \&
  {Markoff}}{2000}]{2000A&A...362..113F}
{Falcke} H.,  {Markoff} S.,  2000, \mn@doi [\aap]
  {10.48550/arXiv.astro-ph/0102186}, \href
  {https://ui.adsabs.harvard.edu/abs/2000A&A...362..113F} {362, 113}

\bibitem[\protect\citeauthoryear{{Feldmeier-Krause} et~al.,}{{Feldmeier-Krause}
  et~al.}{2025}]{2025arXiv250311856F}
{Feldmeier-Krause} A.,  et~al., 2025, \mn@doi [\aap]
  {10.1051/0004-6361/202453414}, \href
  {https://ui.adsabs.harvard.edu/abs/2025A&A...696A.213F} {696, A213}

\bibitem[\protect\citeauthoryear{{Frank}, {King}  \& {Raine}}{{Frank}
  et~al.}{2002}]{2002apa..book.....F}
{Frank} J.,  {King} A.,   {Raine} D.~J.,  2002, {Accretion Power in
  Astrophysics: Third Edition (Cambridge: Cambridge University Press)}

\bibitem[\protect\citeauthoryear{{GRAVITY Collaboration} et~al.,}{{GRAVITY
  Collaboration} et~al.}{2018}]{2018A&A...618L..10G}
{GRAVITY Collaboration} et~al., 2018, \mn@doi [\aap]
  {10.1051/0004-6361/201834294}, \href
  {https://ui.adsabs.harvard.edu/abs/2018A&A...618L..10G} {618, L10}

\bibitem[\protect\citeauthoryear{{Gallego-Cano}, {Sch{\"o}del}, {Dong},
  {Nogueras-Lara}, {Gallego-Calvente}, {Amaro-Seoane}  \&
  {Baumgardt}}{{Gallego-Cano} et~al.}{2018}]{2018A&A...609A..26G}
{Gallego-Cano} E.,  {Sch{\"o}del} R.,  {Dong} H.,  {Nogueras-Lara} F.,
  {Gallego-Calvente} A.~T.,  {Amaro-Seoane} P.,   {Baumgardt} H.,  2018,
  \mn@doi [\aap] {10.1051/0004-6361/201730451}, \href
  {https://ui.adsabs.harvard.edu/abs/2018A&A...609A..26G} {609, A26}

\bibitem[\protect\citeauthoryear{{Genzel}, {Thatte}, {Krabbe}, {Kroker}  \&
  {Tacconi-Garman}}{{Genzel} et~al.}{1996}]{1996ApJ...472..153G}
{Genzel} R.,  {Thatte} N.,  {Krabbe} A.,  {Kroker} H.,   {Tacconi-Garman}
  L.~E.,  1996, \mn@doi [\apj] {10.1086/178051}, \href
  {https://ui.adsabs.harvard.edu/abs/1996ApJ...472..153G} {472, 153}

\bibitem[\protect\citeauthoryear{{Genzel}, {Eisenhauer}  \&
  {Gillessen}}{{Genzel} et~al.}{2010}]{2010RvMP...82.3121G}
{Genzel} R.,  {Eisenhauer} F.,   {Gillessen} S.,  2010, \mn@doi [Reviews of
  Modern Physics] {10.1103/RevModPhys.82.3121}, \href
  {https://ui.adsabs.harvard.edu/abs/2010RvMP...82.3121G} {82, 3121}

\bibitem[\protect\citeauthoryear{{Ghez}, {Klein}, {Morris}  \&
  {Becklin}}{{Ghez} et~al.}{1998}]{1998ApJ...509..678G}
{Ghez} A.~M.,  {Klein} B.~L.,  {Morris} M.,   {Becklin} E.~E.,  1998, \mn@doi
  [\apj] {10.1086/306528}, \href
  {https://ui.adsabs.harvard.edu/abs/1998ApJ...509..678G} {509, 678}

\bibitem[\protect\citeauthoryear{{Gillessen}, {Eisenhauer}, {Trippe},
  {Alexander}, {Genzel}, {Martins}  \& {Ott}}{{Gillessen}
  et~al.}{2009}]{2009ApJ...692.1075G}
{Gillessen} S.,  {Eisenhauer} F.,  {Trippe} S.,  {Alexander} T.,  {Genzel} R.,
  {Martins} F.,   {Ott} T.,  2009, \mn@doi [\apj]
  {10.1088/0004-637X/692/2/1075}, \href
  {https://ui.adsabs.harvard.edu/abs/2009ApJ...692.1075G} {692, 1075}

\bibitem[\protect\citeauthoryear{{G{\"o}tberg}, {de Mink}, {Groh}, {Kupfer},
  {Crowther}, {Zapartas}  \& {Renzo}}{{G{\"o}tberg}
  et~al.}{2018}]{2018A&A...615A..78G}
{G{\"o}tberg} Y.,  {de Mink} S.~E.,  {Groh} J.~H.,  {Kupfer} T.,  {Crowther}
  P.~A.,  {Zapartas} E.,   {Renzo} M.,  2018, \mn@doi [\aap]
  {10.1051/0004-6361/201732274}, \href
  {https://ui.adsabs.harvard.edu/abs/2018A&A...615A..78G} {615, A78}

\bibitem[\protect\citeauthoryear{{Grossov{\'a}} et~al.,}{{Grossov{\'a}}
  et~al.}{2022}]{2022ApJS..258...30G}
{Grossov{\'a}} R.,  et~al., 2022, \mn@doi [\apjs] {10.3847/1538-4365/ac366c},
  \href {https://ui.adsabs.harvard.edu/abs/2022ApJS..258...30G} {258, 30}

\bibitem[\protect\citeauthoryear{{Guo} \& {Mathews}}{{Guo} \&
  {Mathews}}{2012}]{2012ApJ...756..181G}
{Guo} F.,  {Mathews} W.~G.,  2012, \mn@doi [\apj]
  {10.1088/0004-637X/756/2/181}, \href
  {https://ui.adsabs.harvard.edu/abs/2012ApJ...756..181G} {756, 181}

\bibitem[\protect\citeauthoryear{{Habibi} et~al.,}{{Habibi}
  et~al.}{2019}]{2019ApJ...872L..15H}
{Habibi} M.,  et~al., 2019, \mn@doi [\apjl] {10.3847/2041-8213/ab03cf}, \href
  {https://ui.adsabs.harvard.edu/abs/2019ApJ...872L..15H} {872, L15}

\bibitem[\protect\citeauthoryear{{Heber}}{{Heber}}{2016}]{2016PASP..128h2001H}
{Heber} U.,  2016, \mn@doi [\pasp] {10.1088/1538-3873/128/966/082001}, \href
  {https://ui.adsabs.harvard.edu/abs/2016PASP..128h2001H} {128, 082001}

\bibitem[\protect\citeauthoryear{{Heywood} et~al.,}{{Heywood}
  et~al.}{2019}]{2019Natur.573..235H}
{Heywood} I.,  et~al., 2019, \mn@doi [\nat] {10.1038/s41586-019-1532-5}, \href
  {https://ui.adsabs.harvard.edu/abs/2019Natur.573..235H} {573, 235}

\bibitem[\protect\citeauthoryear{{Hopman} \& {Alexander}}{{Hopman} \&
  {Alexander}}{2006}]{2006ApJ...645.1152H}
{Hopman} C.,  {Alexander} T.,  2006, \mn@doi [\apj] {10.1086/504400}, \href
  {https://ui.adsabs.harvard.edu/abs/2006ApJ...645.1152H} {645, 1152}

\bibitem[\protect\citeauthoryear{{Hure}, {Collin-Souffrin}, {Le Bourlot}  \&
  {Pineau des Forets}}{{Hure} et~al.}{1994}]{1994A&A...290...19H}
{Hure} J.~M.,  {Collin-Souffrin} S.,  {Le Bourlot} J.,   {Pineau des Forets}
  G.,  1994, \aap, \href
  {https://ui.adsabs.harvard.edu/abs/1994A&A...290...19H} {290, 19}

\bibitem[\protect\citeauthoryear{{Janssen} et~al.,}{{Janssen}
  et~al.}{2021}]{2021NatAs...5.1017J}
{Janssen} M.,  et~al., 2021, \mn@doi [Nature Astronomy]
  {10.1038/s41550-021-01417-w}, \href
  {https://ui.adsabs.harvard.edu/abs/2021NatAs...5.1017J} {5, 1017}

\bibitem[\protect\citeauthoryear{{Junor}, {Biretta}  \& {Livio}}{{Junor}
  et~al.}{1999}]{1999Natur.401..891J}
{Junor} W.,  {Biretta} J.~A.,   {Livio} M.,  1999, \mn@doi [\nat]
  {10.1038/44780}, \href
  {https://ui.adsabs.harvard.edu/abs/1999Natur.401..891J} {401, 891}

\bibitem[\protect\citeauthoryear{{Karas}, {Svoboda}  \&
  {Zaja{\v{c}}ek}}{{Karas} et~al.}{2021}]{2021bhns.confE...1K}
{Karas} V.,  {Svoboda} J.,   {Zaja{\v{c}}ek} M.,  2021, in RAGtime: Workshops
  on black holes and netron stars. p.~E1 (\mn@eprint {arXiv} {1901.06507}),
  \mn@doi{10.48550/arXiv.1901.06507}

\bibitem[\protect\citeauthoryear{{Kejriwal}, {Witzany}, {Zaja{\v{c}}ek},
  {Pasham}  \& {Chua}}{{Kejriwal} et~al.}{2024}]{2024MNRAS.532.2143K}
{Kejriwal} S.,  {Witzany} V.,  {Zaja{\v{c}}ek} M.,  {Pasham} D.~R.,   {Chua} A.
  J.~K.,  2024, \mn@doi [\mnras] {10.1093/mnras/stae1599}, \href
  {https://ui.adsabs.harvard.edu/abs/2024MNRAS.532.2143K} {532, 2143}

\bibitem[\protect\citeauthoryear{{Kieffer} \& {Bogdanovi{\'c}}}{{Kieffer} \&
  {Bogdanovi{\'c}}}{2016}]{2016ApJ...823..155K}
{Kieffer} T.~F.,  {Bogdanovi{\'c}} T.,  2016, \mn@doi [\apj]
  {10.3847/0004-637X/823/2/155}, \href
  {https://ui.adsabs.harvard.edu/abs/2016ApJ...823..155K} {823, 155}

\bibitem[\protect\citeauthoryear{{Komissarov}}{{Komissarov}}{1994}]{1994MNRAS.269..394K}
{Komissarov} S.~S.,  1994, \mn@doi [\mnras] {10.1093/mnras/269.2.394}, \href
  {https://ui.adsabs.harvard.edu/abs/1994MNRAS.269..394K} {269, 394}

\bibitem[\protect\citeauthoryear{{Kurf{\"u}rst} \& {Krti{\v
  c}ka}}{{Kurf{\"u}rst} \& {Krti{\v c}ka}}{2017}]{KKapplmath}
{Kurf{\"u}rst} P.,  {Krti{\v c}ka} J.,  2017, Applications of Mathematics, 62,
  633

\bibitem[\protect\citeauthoryear{{Kurf{\"u}rst} \&
  {Krti{\v{c}}ka}}{{Kurf{\"u}rst} \&
  {Krti{\v{c}}ka}}{2019}]{2019A&A...625A..24K}
{Kurf{\"u}rst} P.,  {Krti{\v{c}}ka} J.,  2019, \mn@doi [\aap]
  {10.1051/0004-6361/201833429}, \href
  {https://ui.adsabs.harvard.edu/abs/2019A&A...625A..24K} {625, A24}

\bibitem[\protect\citeauthoryear{{Kurf{\"u}rst}, {Feldmeier}  \& {Krti{\v
  c}ka}}{{Kurf{\"u}rst} et~al.}{2014}]{2014A&A...569A..23K}
{Kurf{\"u}rst} P.,  {Feldmeier} A.,   {Krti{\v c}ka} J.,  2014, \mn@doi [\aap]
  {10.1051/0004-6361/201424272}, \href
  {http://adsabs.harvard.edu/abs/2014A%26A...569A..23K} {569, A23}

\bibitem[\protect\citeauthoryear{{Kurf{\"u}rst}, {Feldmeier}  \& {Krti{\v
  c}ka}}{{Kurf{\"u}rst} et~al.}{2018}]{2018A&A...613A..75K}
{Kurf{\"u}rst} P.,  {Feldmeier} A.,   {Krti{\v c}ka} J.,  2018, \mn@doi [\aap]
  {10.1051/0004-6361/201731300}, \href
  {http://adsabs.harvard.edu/abs/2018A%26A...613A..75K} {613, A75}

\bibitem[\protect\citeauthoryear{{Kurf{\"u}rst}, {Pejcha}  \&
  {Krti{\v{c}}ka}}{{Kurf{\"u}rst} et~al.}{2020}]{2020A&A...642A.214K}
{Kurf{\"u}rst} P.,  {Pejcha} O.,   {Krti{\v{c}}ka} J.,  2020, \mn@doi [\aap]
  {10.1051/0004-6361/202039073}, \href
  {https://ui.adsabs.harvard.edu/abs/2020A&A...642A.214K} {642, A214}

\bibitem[\protect\citeauthoryear{{Li}, {Morris}  \& {Baganoff}}{{Li}
  et~al.}{2013}]{2013ApJ...779..154L}
{Li} Z.,  {Morris} M.~R.,   {Baganoff} F.~K.,  2013, \mn@doi [\apj]
  {10.1088/0004-637X/779/2/154}, \href
  {https://ui.adsabs.harvard.edu/abs/2013ApJ...779..154L} {779, 154}

\bibitem[\protect\citeauthoryear{{Lu} \& {Quataert}}{{Lu} \&
  {Quataert}}{2023}]{2023MNRAS.524.6247L}
{Lu} W.,  {Quataert} E.,  2023, \mn@doi [\mnras] {10.1093/mnras/stad2203},
  \href {https://ui.adsabs.harvard.edu/abs/2023MNRAS.524.6247L} {524, 6247}

\bibitem[\protect\citeauthoryear{{Merritt}}{{Merritt}}{2013}]{2013degn.book.....M}
{Merritt} D.,  2013, {Dynamics and Evolution of Galactic Nuclei (Princeton:
  Princeton University Press)}

\bibitem[\protect\citeauthoryear{{Meyer} et~al.,}{{Meyer}
  et~al.}{2012}]{2012Sci...338...84M}
{Meyer} L.,  et~al., 2012, \mn@doi [Science] {10.1126/science.1225506}, \href
  {https://ui.adsabs.harvard.edu/abs/2012Sci...338...84M} {338, 84}

\bibitem[\protect\citeauthoryear{{Milosavljevi{\'c}} \&
  {Loeb}}{{Milosavljevi{\'c}} \& {Loeb}}{2004}]{2004ApJ...604L..45M}
{Milosavljevi{\'c}} M.,  {Loeb} A.,  2004, \mn@doi [\apjl] {10.1086/383467},
  \href {https://ui.adsabs.harvard.edu/abs/2004ApJ...604L..45M} {604, L45}

\bibitem[\protect\citeauthoryear{{Mu{\v{z}}i{\'c}}, {Eckart}, {Sch{\"o}del},
  {Buchholz}, {Zamaninasab}  \& {Witzel}}{{Mu{\v{z}}i{\'c}}
  et~al.}{2010}]{2010A&A...521A..13M}
{Mu{\v{z}}i{\'c}} K.,  {Eckart} A.,  {Sch{\"o}del} R.,  {Buchholz} R.,
  {Zamaninasab} M.,   {Witzel} G.,  2010, \mn@doi [\aap]
  {10.1051/0004-6361/200913087}, \href
  {https://ui.adsabs.harvard.edu/abs/2010A&A...521A..13M} {521, A13}

\bibitem[\protect\citeauthoryear{{Netzer}}{{Netzer}}{2015}]{2015ARA&A..53..365N}
{Netzer} H.,  2015, \mn@doi [\araa] {10.1146/annurev-astro-082214-122302},
  \href {https://ui.adsabs.harvard.edu/abs/2015ARA&A..53..365N} {53, 365}

\bibitem[\protect\citeauthoryear{{Neumayer}, {Seth}  \& {B{\"o}ker}}{{Neumayer}
  et~al.}{2020}]{2020A&ARv..28....4N}
{Neumayer} N.,  {Seth} A.,   {B{\"o}ker} T.,  2020, \mn@doi [\aapr]
  {10.1007/s00159-020-00125-0}, \href
  {https://ui.adsabs.harvard.edu/abs/2020A&ARv..28....4N} {28, 4}

\bibitem[\protect\citeauthoryear{{Ohlmann}, {R{\"o}pke}, {Pakmor}  \&
  {Springel}}{{Ohlmann} et~al.}{2017}]{2017A&A...599A...5O}
{Ohlmann} S.~T.,  {R{\"o}pke} F.~K.,  {Pakmor} R.,   {Springel} V.,  2017,
  \mn@doi [\aap] {10.1051/0004-6361/201629692}, \href
  {https://ui.adsabs.harvard.edu/abs/2017A&A...599A...5O} {599, A5}

\bibitem[\protect\citeauthoryear{{Paumard} et~al.,}{{Paumard}
  et~al.}{2006}]{paumard2006}
{Paumard} T.,  et~al., 2006, \mn@doi [\apj] {10.1086/503273}, \href
  {http://adsabs.harvard.edu/abs/2006ApJ...643.1011P} {643, 1011}

\bibitem[\protect\citeauthoryear{{Paxton}, {Bildsten}, {Dotter}, {Herwig},
  {Lesaffre}  \& {Timmes}}{{Paxton} et~al.}{2011}]{2011ApJS..192....3P}
{Paxton} B.,  {Bildsten} L.,  {Dotter} A.,  {Herwig} F.,  {Lesaffre} P.,
  {Timmes} F.,  2011, \mn@doi [\apjs] {10.1088/0067-0049/192/1/3}, \href
  {https://ui.adsabs.harvard.edu/abs/2011ApJS..192....3P} {192, 3}

\bibitem[\protect\citeauthoryear{{Pei{\ss}ker}, {Zaja{\v{c}}ek}, {Eckart},
  {Sabha}, {Shahzamanian}  \& {Parsa}}{{Pei{\ss}ker}
  et~al.}{2019}]{2019A&A...624A..97P}
{Pei{\ss}ker} F.,  {Zaja{\v{c}}ek} M.,  {Eckart} A.,  {Sabha} N.~B.,
  {Shahzamanian} B.,   {Parsa} M.,  2019, \mn@doi [\aap]
  {10.1051/0004-6361/201834947}, \href
  {https://ui.adsabs.harvard.edu/abs/2019A&A...624A..97P} {624, A97}

\bibitem[\protect\citeauthoryear{{Pei{\ss}ker}, {Eckart}  \&
  {Parsa}}{{Pei{\ss}ker} et~al.}{2020a}]{2020ApJ...889...61P}
{Pei{\ss}ker} F.,  {Eckart} A.,   {Parsa} M.,  2020a, \mn@doi [\apj]
  {10.3847/1538-4357/ab5afd}, \href
  {https://ui.adsabs.harvard.edu/abs/2020ApJ...889...61P} {889, 61}

\bibitem[\protect\citeauthoryear{{Pei{\ss}ker}, {Eckart}, {Zaja{\v{c}}ek},
  {Ali}  \& {Parsa}}{{Pei{\ss}ker} et~al.}{2020b}]{2020ApJ...899...50P}
{Pei{\ss}ker} F.,  {Eckart} A.,  {Zaja{\v{c}}ek} M.,  {Ali} B.,   {Parsa} M.,
  2020b, \mn@doi [\apj] {10.3847/1538-4357/ab9c1c}, \href
  {https://ui.adsabs.harvard.edu/abs/2020ApJ...899...50P} {899, 50}

\bibitem[\protect\citeauthoryear{{Pei{\ss}ker} et~al.,}{{Pei{\ss}ker}
  et~al.}{2021}]{2021ApJ...909...62P}
{Pei{\ss}ker} F.,  et~al., 2021, \mn@doi [\apj] {10.3847/1538-4357/abd9c6},
  \href {https://ui.adsabs.harvard.edu/abs/2021ApJ...909...62P} {909, 62}

\bibitem[\protect\citeauthoryear{{Pei{\ss}ker}, {Eckart}, {Zaja{\v{c}}ek}  \&
  {Britzen}}{{Pei{\ss}ker} et~al.}{2022}]{2022ApJ...933...49P}
{Pei{\ss}ker} F.,  {Eckart} A.,  {Zaja{\v{c}}ek} M.,   {Britzen} S.,  2022,
  \mn@doi [\apj] {10.3847/1538-4357/ac752f}, \href
  {https://ui.adsabs.harvard.edu/abs/2022ApJ...933...49P} {933, 49}

\bibitem[\protect\citeauthoryear{{Pei{\ss}ker} et~al.,}{{Pei{\ss}ker}
  et~al.}{2023}]{2023ApJ...944..231P}
{Pei{\ss}ker} F.,  et~al., 2023, \mn@doi [\apj] {10.3847/1538-4357/aca977},
  \href {https://ui.adsabs.harvard.edu/abs/2023ApJ...944..231P} {944, 231}

\bibitem[\protect\citeauthoryear{{Pl{\v{s}}ek}, {Werner}, {Grossov{\'a}},
  {Topinka}, {Simionescu}  \& {Allen}}{{Pl{\v{s}}ek}
  et~al.}{2022}]{2022MNRAS.517.3682P}
{Pl{\v{s}}ek} T.,  {Werner} N.,  {Grossov{\'a}} R.,  {Topinka} M.,
  {Simionescu} A.,   {Allen} S.~W.,  2022, \mn@doi [\mnras]
  {10.1093/mnras/stac2770}, \href
  {https://ui.adsabs.harvard.edu/abs/2022MNRAS.517.3682P} {517, 3682}

\bibitem[\protect\citeauthoryear{{Predehl} et~al.,}{{Predehl}
  et~al.}{2020}]{2020Natur.588..227P}
{Predehl} P.,  et~al., 2020, \mn@doi [\nat] {10.1038/s41586-020-2979-0}, \href
  {https://ui.adsabs.harvard.edu/abs/2020Natur.588..227P} {588, 227}

\bibitem[\protect\citeauthoryear{{Rauch} \& {Tremaine}}{{Rauch} \&
  {Tremaine}}{1996}]{1996NewA....1..149R}
{Rauch} K.~P.,  {Tremaine} S.,  1996, \mn@doi [\na]
  {10.1016/S1384-1076(96)00012-7}, \href
  {https://ui.adsabs.harvard.edu/abs/1996NewA....1..149R} {1, 149}

\bibitem[\protect\citeauthoryear{{R{\'o}{\.z}a{\'n}ska}, {Mr{\'o}z},
  {Mo{\'s}cibrodzka}, {Sobolewska}  \& {Adhikari}}{{R{\'o}{\.z}a{\'n}ska}
  et~al.}{2015}]{2015A&A...581A..64R}
{R{\'o}{\.z}a{\'n}ska} A.,  {Mr{\'o}z} P.,  {Mo{\'s}cibrodzka} M.,
  {Sobolewska} M.,   {Adhikari} T.~P.,  2015, \mn@doi [\aap]
  {10.1051/0004-6361/201424386}, \href
  {https://ui.adsabs.harvard.edu/abs/2015A&A...581A..64R} {581, A64}

\bibitem[\protect\citeauthoryear{{Sanchez-Bermudez}, {Sch{\"o}del}, {Alberdi},
  {Muzi{\'c}}, {Hummel}  \& {Pott}}{{Sanchez-Bermudez}
  et~al.}{2014}]{2014A&A...567A..21S}
{Sanchez-Bermudez} J.,  {Sch{\"o}del} R.,  {Alberdi} A.,  {Muzi{\'c}} K.,
  {Hummel} C.~A.,   {Pott} J.~U.,  2014, \mn@doi [\aap]
  {10.1051/0004-6361/201423657}, \href
  {https://ui.adsabs.harvard.edu/abs/2014A&A...567A..21S} {567, A21}

\bibitem[\protect\citeauthoryear{{Schawinski}, {Koss}, {Berney}  \&
  {Sartori}}{{Schawinski} et~al.}{2015}]{2015MNRAS.451.2517S}
{Schawinski} K.,  {Koss} M.,  {Berney} S.,   {Sartori} L.~F.,  2015, \mn@doi
  [\mnras] {10.1093/mnras/stv1136}, \href
  {https://ui.adsabs.harvard.edu/abs/2015MNRAS.451.2517S} {451, 2517}

\bibitem[\protect\citeauthoryear{{Sch{\"o}del} et~al.,}{{Sch{\"o}del}
  et~al.}{2002}]{2002Natur.419..694S}
{Sch{\"o}del} R.,  et~al., 2002, \mn@doi [\nat] {10.1038/nature01121}, \href
  {https://ui.adsabs.harvard.edu/abs/2002Natur.419..694S} {419, 694}

\bibitem[\protect\citeauthoryear{{Sch{\"o}del}, {Feldmeier}, {Neumayer},
  {Meyer}  \& {Yelda}}{{Sch{\"o}del} et~al.}{2014}]{2014CQGra..31x4007S}
{Sch{\"o}del} R.,  {Feldmeier} A.,  {Neumayer} N.,  {Meyer} L.,   {Yelda} S.,
  2014, \mn@doi [Classical and Quantum Gravity]
  {10.1088/0264-9381/31/24/244007}, \href
  {https://ui.adsabs.harvard.edu/abs/2014CQGra..31x4007S} {31, 244007}

\bibitem[\protect\citeauthoryear{{Sch{\"o}del}, {Nogueras-Lara},
  {Gallego-Cano}, {Shahzamanian}, {Gallego-Calvente}  \&
  {Gardini}}{{Sch{\"o}del} et~al.}{2020}]{2020A&A...641A.102S}
{Sch{\"o}del} R.,  {Nogueras-Lara} F.,  {Gallego-Cano} E.,  {Shahzamanian} B.,
  {Gallego-Calvente} A.~T.,   {Gardini} A.,  2020, \mn@doi [\aap]
  {10.1051/0004-6361/201936688}, \href
  {https://ui.adsabs.harvard.edu/abs/2020A&A...641A.102S} {641, A102}

\bibitem[\protect\citeauthoryear{{Sellgren}, {McGinn}, {Becklin}  \&
  {Hall}}{{Sellgren} et~al.}{1990}]{1990ApJ...359..112S}
{Sellgren} K.,  {McGinn} M.~T.,  {Becklin} E.~E.,   {Hall} D.~N.,  1990,
  \mn@doi [\apj] {10.1086/169039}, \href
  {https://ui.adsabs.harvard.edu/abs/1990ApJ...359..112S} {359, 112}

\bibitem[\protect\citeauthoryear{{Su}, {Slatyer}  \& {Finkbeiner}}{{Su}
  et~al.}{2010}]{2010ApJ...724.1044S}
{Su} M.,  {Slatyer} T.~R.,   {Finkbeiner} D.~P.,  2010, \mn@doi [\apj]
  {10.1088/0004-637X/724/2/1044}, \href
  {https://ui.adsabs.harvard.edu/abs/2010ApJ...724.1044S} {724, 1044}

\bibitem[\protect\citeauthoryear{{Sukov{\'a}}, {Zaja{\v{c}}ek}, {Witzany}  \&
  {Karas}}{{Sukov{\'a}} et~al.}{2021}]{2021ApJ...917...43S}
{Sukov{\'a}} P.,  {Zaja{\v{c}}ek} M.,  {Witzany} V.,   {Karas} V.,  2021,
  \mn@doi [\apj] {10.3847/1538-4357/ac05c6}, \href
  {https://ui.adsabs.harvard.edu/abs/2021ApJ...917...43S} {917, 43}

\bibitem[\protect\citeauthoryear{{Tsuboi}, {Kitamura}, {Tsutsumi}, {Miyawaki},
  {Miyoshi}  \& {Miyazaki}}{{Tsuboi} et~al.}{2020}]{2020PASJ...72...36T}
{Tsuboi} M.,  {Kitamura} Y.,  {Tsutsumi} T.,  {Miyawaki} R.,  {Miyoshi} M.,
  {Miyazaki} A.,  2020, \mn@doi [\pasj] {10.1093/pasj/psaa013}, \href
  {https://ui.adsabs.harvard.edu/abs/2020PASJ...72...36T} {72, 36}

\bibitem[\protect\citeauthoryear{{Urry} \& {Padovani}}{{Urry} \&
  {Padovani}}{1995}]{1995PASP..107..803U}
{Urry} C.~M.,  {Padovani} P.,  1995, \mn@doi [\pasp] {10.1086/133630}, \href
  {https://ui.adsabs.harvard.edu/abs/1995PASP..107..803U} {107, 803}

\bibitem[\protect\citeauthoryear{{Viehmann}, {Eckart}, {Sch{\"o}del},
  {Moultaka}, {Straubmeier}  \& {Pott}}{{Viehmann}
  et~al.}{2005}]{2005A&A...433..117V}
{Viehmann} T.,  {Eckart} A.,  {Sch{\"o}del} R.,  {Moultaka} J.,  {Straubmeier}
  C.,   {Pott} J.~U.,  2005, \mn@doi [\aap] {10.1051/0004-6361:20041748}, \href
  {https://ui.adsabs.harvard.edu/abs/2005A&A...433..117V} {433, 117}

\bibitem[\protect\citeauthoryear{{Vilkoviskij} \& {Czerny}}{{Vilkoviskij} \&
  {Czerny}}{2002}]{2002A&A...387..804V}
{Vilkoviskij} E.~Y.,  {Czerny} B.,  2002, \mn@doi [\aap]
  {10.1051/0004-6361:20020255}, \href
  {https://ui.adsabs.harvard.edu/abs/2002A&A...387..804V} {387, 804}

\bibitem[\protect\citeauthoryear{{Wang} et~al.,}{{Wang}
  et~al.}{2013}]{2013Sci...341..981W}
{Wang} Q.~D.,  et~al., 2013, \mn@doi [Science] {10.1126/science.1240755}, \href
  {https://ui.adsabs.harvard.edu/abs/2013Sci...341..981W} {341, 981}

\bibitem[\protect\citeauthoryear{{Wardle} \& {Yusef-Zadeh}}{{Wardle} \&
  {Yusef-Zadeh}}{2008}]{2008ApJ...683L..37W}
{Wardle} M.,  {Yusef-Zadeh} F.,  2008, \mn@doi [\apjl] {10.1086/591471}, \href
  {https://ui.adsabs.harvard.edu/abs/2008ApJ...683L..37W} {683, L37}

\bibitem[\protect\citeauthoryear{{Wardle} \& {Yusef-Zadeh}}{{Wardle} \&
  {Yusef-Zadeh}}{2014}]{2014ApJ...787L..14W}
{Wardle} M.,  {Yusef-Zadeh} F.,  2014, \mn@doi [\apjl]
  {10.1088/2041-8205/787/1/L14}, \href
  {https://ui.adsabs.harvard.edu/abs/2014ApJ...787L..14W} {787, L14}

\bibitem[\protect\citeauthoryear{{Werner}, {McNamara}, {Churazov}  \&
  {Scannapieco}}{{Werner} et~al.}{2019}]{2019SSRv..215....5W}
{Werner} N.,  {McNamara} B.~R.,  {Churazov} E.,   {Scannapieco} E.,  2019,
  \mn@doi [\ssr] {10.1007/s11214-018-0571-9}, \href
  {https://ui.adsabs.harvard.edu/abs/2019SSRv..215....5W} {215, 5}

\bibitem[\protect\citeauthoryear{{Wood}, {Harper}  \& {M{\"u}ller}}{{Wood}
  et~al.}{2024}]{2024ApJ...967..120W}
{Wood} B.~E.,  {Harper} G.~M.,   {M{\"u}ller} H.-R.,  2024, \mn@doi [\apj]
  {10.3847/1538-4357/ad401f}, \href
  {https://ui.adsabs.harvard.edu/abs/2024ApJ...967..120W} {967, 120}

\bibitem[\protect\citeauthoryear{{Yang}, {Ruszkowski}  \& {Zweibel}}{{Yang}
  et~al.}{2022}]{2022NatAs...6..584Y}
{Yang} H. Y.~K.,  {Ruszkowski} M.,   {Zweibel} E.~G.,  2022, \mn@doi [Nature
  Astronomy] {10.1038/s41550-022-01618-x}, \href
  {https://ui.adsabs.harvard.edu/abs/2022NatAs...6..584Y} {6, 584}

\bibitem[\protect\citeauthoryear{{Yuan} \& {Narayan}}{{Yuan} \&
  {Narayan}}{2014}]{2014ARA&A..52..529Y}
{Yuan} F.,  {Narayan} R.,  2014, \mn@doi [\araa]
  {10.1146/annurev-astro-082812-141003}, \href
  {https://ui.adsabs.harvard.edu/abs/2014ARA&A..52..529Y} {52, 529}

\bibitem[\protect\citeauthoryear{{Yusef-Zadeh} \& {Melia}}{{Yusef-Zadeh} \&
  {Melia}}{1992}]{1992ApJ...385L..41Y}
{Yusef-Zadeh} F.,  {Melia} F.,  1992, \mn@doi [\apjl] {10.1086/186273}, \href
  {https://ui.adsabs.harvard.edu/abs/1992ApJ...385L..41Y} {385, L41}

\bibitem[\protect\citeauthoryear{{Yusef-Zadeh} \& {Morris}}{{Yusef-Zadeh} \&
  {Morris}}{1991}]{1991ApJ...371L..59Y}
{Yusef-Zadeh} F.,  {Morris} M.,  1991, \mn@doi [\apjl] {10.1086/186002}, \href
  {https://ui.adsabs.harvard.edu/abs/1991ApJ...371L..59Y} {371, L59}

\bibitem[\protect\citeauthoryear{{Yusef-Zadeh} et~al.,}{{Yusef-Zadeh}
  et~al.}{2012}]{2012ApJ...758L..11Y}
{Yusef-Zadeh} F.,  et~al., 2012, \mn@doi [\apjl] {10.1088/2041-8205/758/1/L11},
  \href {https://ui.adsabs.harvard.edu/abs/2012ApJ...758L..11Y} {758, L11}

\bibitem[\protect\citeauthoryear{{Yusef-Zadeh}, {Wardle}, {Cotton},
  {Sch{\"o}del}, {Royster}, {Roberts}  \& {Kunneriath}}{{Yusef-Zadeh}
  et~al.}{2017}]{2017ApJ...837...93Y}
{Yusef-Zadeh} F.,  {Wardle} M.,  {Cotton} W.,  {Sch{\"o}del} R.,  {Royster}
  M.~J.,  {Roberts} D.~A.,   {Kunneriath} D.,  2017, \mn@doi [\apj]
  {10.3847/1538-4357/aa5ea2}, \href
  {https://ui.adsabs.harvard.edu/abs/2017ApJ...837...93Y} {837, 93}

\bibitem[\protect\citeauthoryear{{Yusef-Zadeh}, {Royster}, {Wardle}, {Cotton},
  {Kunneriath}, {Heywood}  \& {Michail}}{{Yusef-Zadeh}
  et~al.}{2020}]{2020MNRAS.499.3909Y}
{Yusef-Zadeh} F.,  {Royster} M.,  {Wardle} M.,  {Cotton} W.,  {Kunneriath} D.,
  {Heywood} I.,   {Michail} J.,  2020, \mn@doi [\mnras]
  {10.1093/mnras/staa2399}, \href
  {https://ui.adsabs.harvard.edu/abs/2020MNRAS.499.3909Y} {499, 3909}

\bibitem[\protect\citeauthoryear{{Zajacek}, {Eckart}  \& {Hosseini}}{{Zajacek}
  et~al.}{2018}]{2018acps.confE..49Z}
{Zajacek} M.,  {Eckart} A.,   {Hosseini} S.~E.,  2018, in Accretion Processes
  in Cosmic Sources - II. p.~49, \mn@doi{10.22323/1.342.0049}

\bibitem[\protect\citeauthoryear{{Zajacek}, {Araudo}, {Karas}, {Czerny},
  {Eckart}, {Sukova}, {Stolc}  \& {Witzany}}{{Zajacek}
  et~al.}{2020}]{2020bhns.work..357Z}
{Zajacek} M.,  {Araudo} A.,  {Karas} V.,  {Czerny} B.,  {Eckart} A.,  {Sukova}
  P.,  {Stolc} M.,   {Witzany} V.,  2020, in {Stuchl{\'\i}k} Z.,
  {T{\"o}r{\"o}k} G.,   {Karas} V.,  eds, RAGtime 20-22: Workshops on Black
  Holes and Neutron Stars. Proceedings of RAGtime 20-22. Edited by Z.
  Stuchl{\'\i}k. pp 357--374

\bibitem[\protect\citeauthoryear{{Zaja{\v{c}}ek} et~al.,}{{Zaja{\v{c}}ek}
  et~al.}{2017}]{2017bhns.work..237Z}
{Zaja{\v{c}}ek} M.,  et~al., 2017, in RAGtime 17-19: Workshops on Black Holes
  and Neutron Stars. pp 237--252

\bibitem[\protect\citeauthoryear{{Zaja{\v{c}}ek}, {Araudo}, {Karas}, {Czerny}
  \& {Eckart}}{{Zaja{\v{c}}ek} et~al.}{2020}]{2020ApJ...903..140Z}
{Zaja{\v{c}}ek} M.,  {Araudo} A.,  {Karas} V.,  {Czerny} B.,   {Eckart} A.,
  2020, \mn@doi [\apj] {10.3847/1538-4357/abbd94}, \href
  {https://ui.adsabs.harvard.edu/abs/2020ApJ...903..140Z} {903, 140}

\bibitem[\protect\citeauthoryear{{Zaja{\v{c}}ek}, {Czerny}, {Sch{\"o}del},
  {Werner}  \& {Karas}}{{Zaja{\v{c}}ek} et~al.}{2022}]{2022NatAs...6.1008Z}
{Zaja{\v{c}}ek} M.,  {Czerny} B.,  {Sch{\"o}del} R.,  {Werner} N.,   {Karas}
  V.,  2022, \mn@doi [Nature Astronomy] {10.1038/s41550-022-01785-x}, \href
  {https://ui.adsabs.harvard.edu/abs/2022NatAs...6.1008Z} {6, 1008}

\bibitem[\protect\citeauthoryear{{Zaja{\v{c}}ek} et~al.,}{{Zaja{\v{c}}ek}
  et~al.}{2024a}]{2024arXiv241012090Z}
{Zaja{\v{c}}ek} M.,  et~al., 2024a, \mn@doi [arXiv e-prints]
  {10.48550/arXiv.2410.12090}, \href
  {https://ui.adsabs.harvard.edu/abs/2024arXiv241012090Z} {p. arXiv:2410.12090}

\bibitem[\protect\citeauthoryear{{Zaja{\v{c}}ek} et~al.,}{{Zaja{\v{c}}ek}
  et~al.}{2024b}]{2024SSRv..220...29Z}
{Zaja{\v{c}}ek} M.,  et~al., 2024b, \mn@doi [\ssr]
  {10.1007/s11214-024-01062-5}, \href
  {https://ui.adsabs.harvard.edu/abs/2024SSRv..220...29Z} {220, 29}

\bibitem[\protect\citeauthoryear{{Zaja{\v{c}}ek} et~al.,}{{Zaja{\v{c}}ek}
  et~al.}{2025}]{2025arXiv250119365Z}
{Zaja{\v{c}}ek} M.,  et~al., 2025, \mn@doi [arXiv e-prints]
  {10.48550/arXiv.2501.19365}, \href
  {https://ui.adsabs.harvard.edu/abs/2025arXiv250119365Z} {p. arXiv:2501.19365}

\bibitem[\protect\citeauthoryear{{de la Cita}, {Bosch-Ramon},
  {Paredes-Fortuny}, {Khangulyan}  \& {Perucho}}{{de la Cita}
  et~al.}{2016}]{2016A&A...591A..15D}
{de la Cita} V.~M.,  {Bosch-Ramon} V.,  {Paredes-Fortuny} X.,  {Khangulyan} D.,
    {Perucho} M.,  2016, \mn@doi [\aap] {10.1051/0004-6361/201527084}, \href
  {https://ui.adsabs.harvard.edu/abs/2016A&A...591A..15D} {591, A15}

\bibitem[\protect\citeauthoryear{{{\v{D}}urov{\v{c}}{\'\i}kov{\'a}}
  et~al.,}{{{\v{D}}urov{\v{c}}{\'\i}kov{\'a}}
  et~al.}{2025}]{2025arXiv250500080D}
{{\v{D}}urov{\v{c}}{\'\i}kov{\'a}} D.,  et~al., 2025, \mn@doi [arXiv e-prints]
  {10.48550/arXiv.2505.00080}, \href
  {https://ui.adsabs.harvard.edu/abs/2025arXiv250500080D} {p. arXiv:2505.00080}

\bibitem[\protect\citeauthoryear{{von Fellenberg} et~al.,}{{von Fellenberg}
  et~al.}{2022}]{2022ApJ...932L...6V}
{von Fellenberg} S.~D.,  et~al., 2022, \mn@doi [\apjl]
  {10.3847/2041-8213/ac68ef}, \href
  {https://ui.adsabs.harvard.edu/abs/2022ApJ...932L...6V} {932, L6}

\makeatother
\end{thebibliography}




\appendix

\section{Animation documentation}
\label{videodoc}
\begin{itemize}
\item Animation of the density and $x-$velocity component evolution of the star-jet interaction; the jet luminosity $L_\text{j}=10^{42}\,\text{erg}\,\text{s}^{-1}$. The time domain of the simulation includes 10 stellar passages through the jet:
        \href{https://www.youtube.com/watch?v=vZFDMLpGXdY}{\color{blue}{\underline{star-jet\_10\_passages\_0.001pc\_Lj\_42.mp4}}}
\vspace{0.2cm}
\item Animation of the density and x-velocity evolution within the extended spatial domain of the first star-jet passage, with the jet luminosity $L_\text{j}=10^{42}\,\text{erg}\,\text{s}^{-1}$; the spatial domain is $\approx$ five-times stretched in the direction of the jet so the longer tail of the jet is shown highlighted by contours of the density and velocity:
\href{https://www.youtube.com/watch?v=gYjtf9LwtPY}{\color{blue}{\underline{extended\_0.001pc\_Lj\_42.mp4}}}
\vspace{0.2cm}
\item Animation of the density and x-velocity evolution of the star-jet passages where the jet luminosity $L_\text{j}=10^{44}\,\text{erg}\,\text{s}^{-1}$; the time domain of the simulation includes 10 stellar passages through the jet:
        \href{https://www.youtube.com/watch?v=Sjq02YnOCt4}{\color{blue}{\underline{star-jet\_10\_passages\_0.001pc\_Lj\_44.mp4}}} 
\vspace{0.2cm}
\item Animation of the density and x-velocity evolution of the star-jet passages where the jet luminosity $L_\text{j}=10^{48}\,\text{erg}\,\text{s}^{-1}$; the time domain of the simulation includes 10 stellar passages through the jet:
     \href{https://www.youtube.com/watch?v=Gusa4Axk8V8}{\color{blue}{\underline{star-jet\_10\_passages\_0.001pc\_Lj\_48.mp4}}}
\end{itemize}

\section{Effect of RG - ADAF interaction}
\label{app_ADAF}

\begin{figure}
    \centering
    \includegraphics[width=0.5\textwidth]{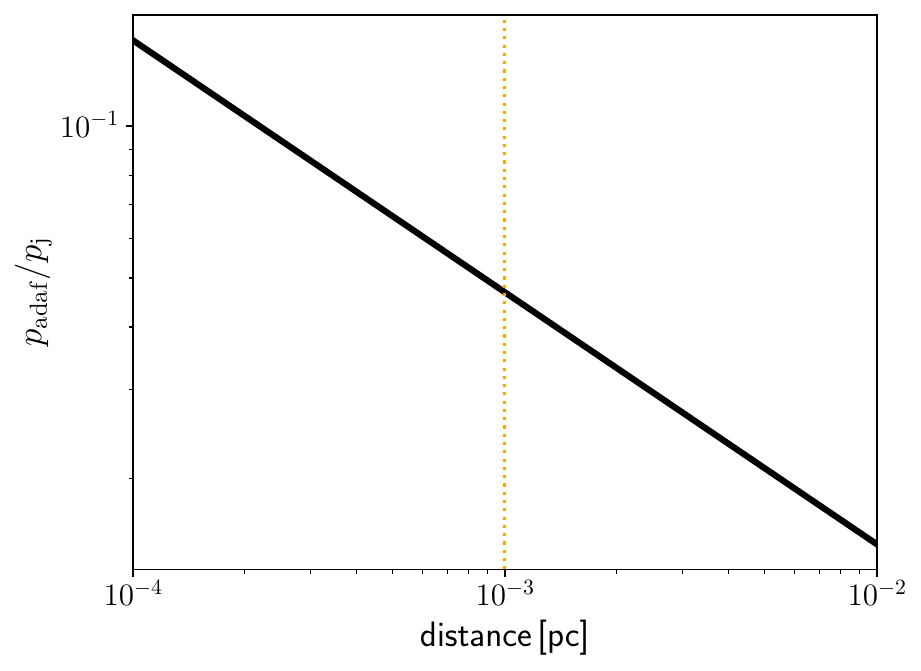}
    \caption{The ratio of the ADAF ram pressure to the jet ram pressure, $p_{\rm adaf}/p_{\rm j}$, as a function of the distance of a RG ($M_{\star}=1\,M_{\odot}$, $R_{\star}=100\,R_{\odot}$) from the SMBH. The vertical dotted line represents the RG distance in most of our simulation runs.}
    \label{fig_ratio_ram}
\end{figure}

Here we consider the ram pressure due to the passages of the star through the ADAF, $p_{\rm adaf}=1/2 \rho_{\rm adaf} v_{\star}^2$, where $\rho_{\rm adaf}$ is the mass density adopted from \citet{2014ARA&A..52..529Y} and $v_{\star}$ is the orbital velocity of the star. We adopt the Eddington ratio of $\dot{m}=10^{-2}$, which is roughly consistent with the jet luminosity of $L_{\rm j}=10^{42}\,{\rm erg\,s^{-1}}$ for $M_{\bullet}=4 \times 10^6\,M_{\odot}$. The jet ram pressure is estimated using $p_{\rm j}=L_{\rm j}/(c\pi x^2\tan{\theta}^2)$, where we adopt $L_{\rm j}=10^{42}\,{\rm erg\,s^{-1}}$. The calculated ratio of the ADAF ram pressure to the jet ram pressure is plotted in Fig.~\ref{fig_ratio_ram} and it is generally lower than unity for the distance range studied in this work. The ratio depends on the radial distance as $p_{\rm adaf}/p_{\rm j} \propto d_{\rm orb}^{-1/2}$, hence it tends to decrease with distance from the SMBH. Therefore, the jet ram pressure is expected to be more profound in terms of the ablated mass from a RG than the ADAF.

\begin{figure}
    \centering
    \includegraphics[width=0.5\textwidth]{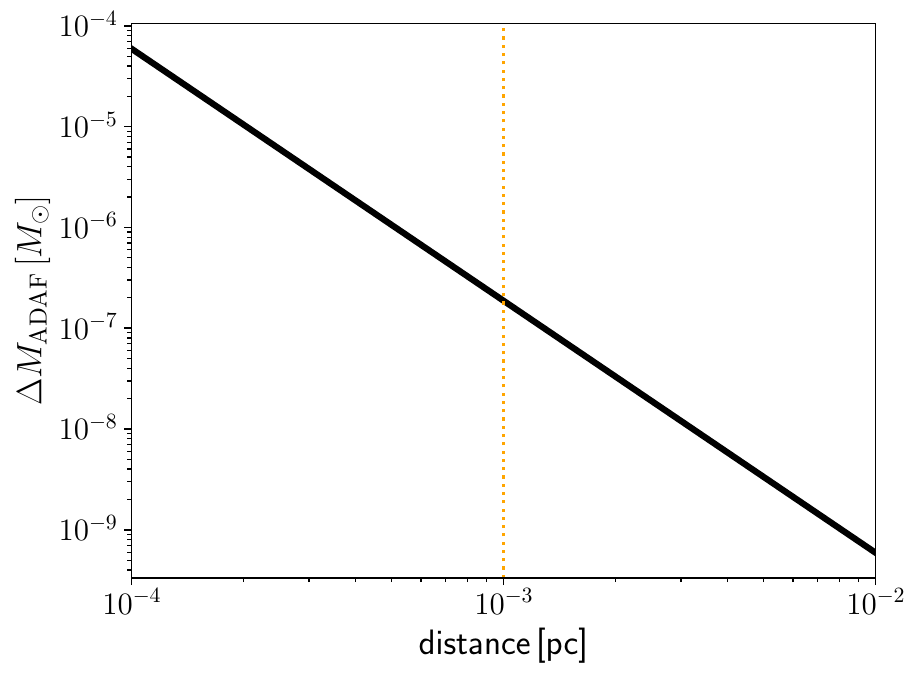}
    \caption{An estimate of the stripped mass from a RG ($M_{\star}=1\,M_{\odot}$, $R_{\star}=100\,R_{\odot}$) during one passage of the star through an ADAF as a function of its distance from the SMBH. The vertical dotted line represents the RG distance in most of our simulation runs.}
    \label{fig_delta_m_adaf}
\end{figure}

We also estimate the stripped mass from the envelope of a RG interacting with the hot ADAF. For the estimate, we follow \citet{2023MNRAS.524.6247L}, who derived a simple relation for the amount of ablated mass due to the collision of a star with the disc, $\Delta M_{\rm ADAF}\simeq M_{\star} (p_{\rm adaf}/\overline{p})$, where $p_{\rm adaf}$ is again the ADAF ram pressure and $\overline{p}=GM_{\star}^2/(4 \pi R_{\star}^4)$ is the mean pressure inside the red giant star. In Figure~\ref{fig_delta_m_adaf}, we plot the estimated $\Delta M_{\rm ADAF}$ as a function of distance of a red-giant star from the SMBH.

Overall, at $d_{\rm orb}=10^{-3}\,{\rm pc}$, we obtain $\Delta M_{\rm ADAF}\sim  1.9 \times 10^{-7}\,M_{\odot} $ and $p_{\rm adaf}/p_{\rm j} \sim 4.7 \times 10^{-2}$, which demonstrates that the interaction of a RG with an ADAF can slighly amplify the mass loss due to the RG-jet collision.

\section{Effect of magnetic field}
\label{app_mag_field}

When the star with the mass-loss rate of $\dot{m}_{\rm w}$ and the terminal wind speed of $v_{\rm w}$ passes through the magnetised jet with the luminosity of $L_{\rm j}$ and the magnetic field poloidal strength $B(x)$, the stagnation radius position can be solved using the equation,
\begin{equation}
    \frac{B(x)^2}{8 \pi}+\frac{L_{\rm j}}{\pi c x^2 \tan^2{\theta} }=\frac{\dot{m}_{\rm w} v_{\rm w}}{4 \pi R_{\rm stag}^2}\,,
\end{equation}
from which one can derive the magnetic field strength at the distance of $x$, for which the magnetic field pressure is comparable to the jet ram pressure,
\begin{align}
    B(x)&=\sqrt{\frac{8L_{\rm j}}{c}} (x\tan{\theta})^{-1}\,\notag\\
    &\simeq 30 \left(\frac{L_{\rm j}}{10^{42}\,{\rm erg\,s^{-1}}} \right)^{1/2} \left(\frac{x}{10^{-3}\,{\rm pc}} \right)^{-1}\,{\rm G}\,,
\end{align}
where we adopted $\theta=10^{\circ}$ for the jet half-opening angle.

\section{Initial stellar model and its temporal stability}
\label{stelinternal}
\begin{figure}
    \centering
\includegraphics[width=\columnwidth]{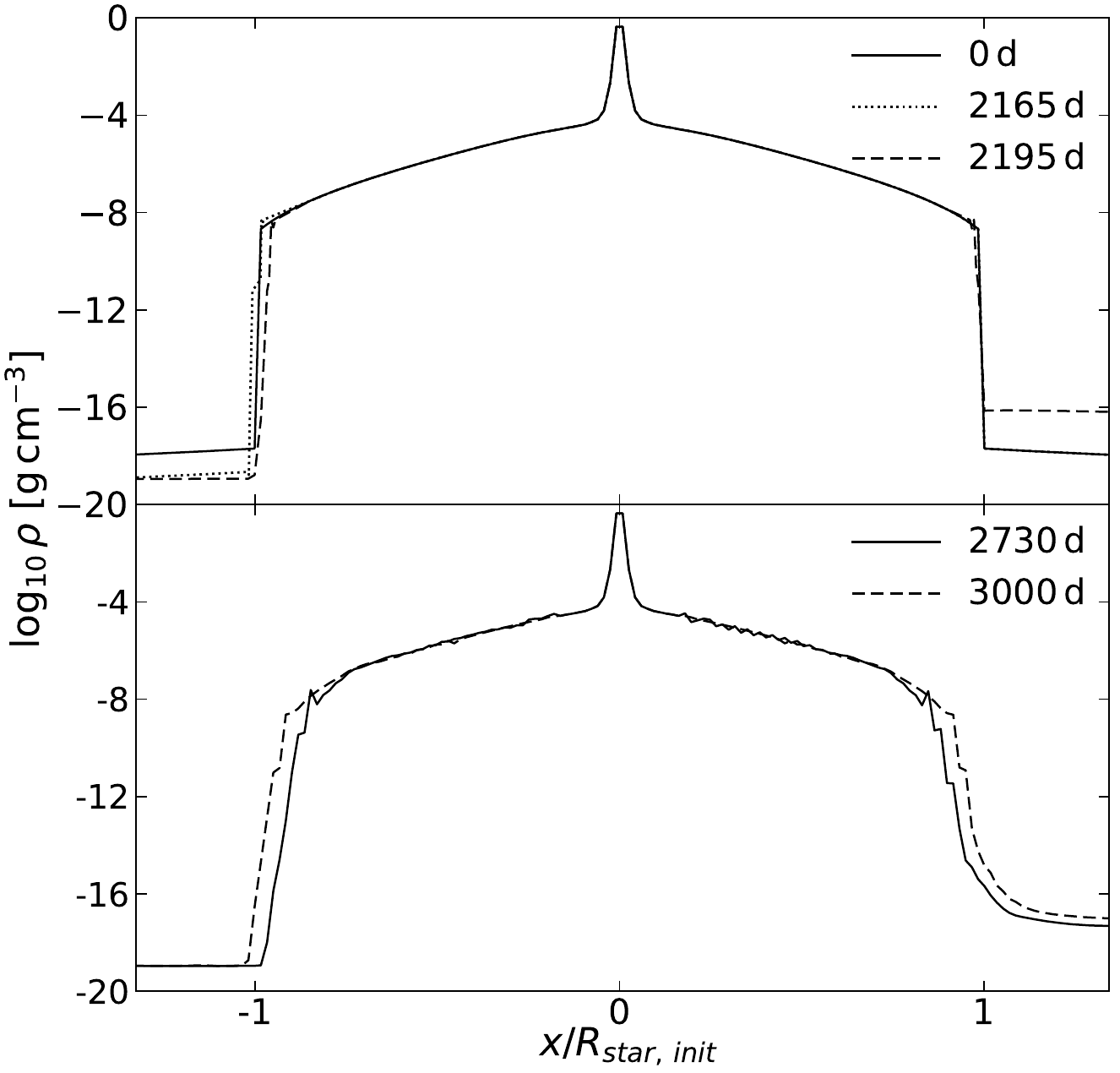}
    \caption{Plots of the temporal stability of the star plotted as 1D slices of mass density through the stellar center along the jet $x$-direction. {\bf Top panel:} Snapshots of the star's density structure during the 9th star-jet passage. The solid line shows the initial stellar density profile. The thin dotted line shows the density soon after the entry of the star into the jet while the dashed line shows the same at the moment the star exits the jet. {\bf Bottom panel:} Same as in the top panel, now in the very late times of simulation. The solid line shows the density profile right before the star's exit from the jet during the 11th star-jet passage while the dashed line shows the same at the ultimate time of simulation. Both slices in this bottom panel show that the structure of the star starts to break down at this late time of the simulation.}
\label{fig_steltemporalstability}
\end{figure}
\begin{figure}
\includegraphics[width=\columnwidth]{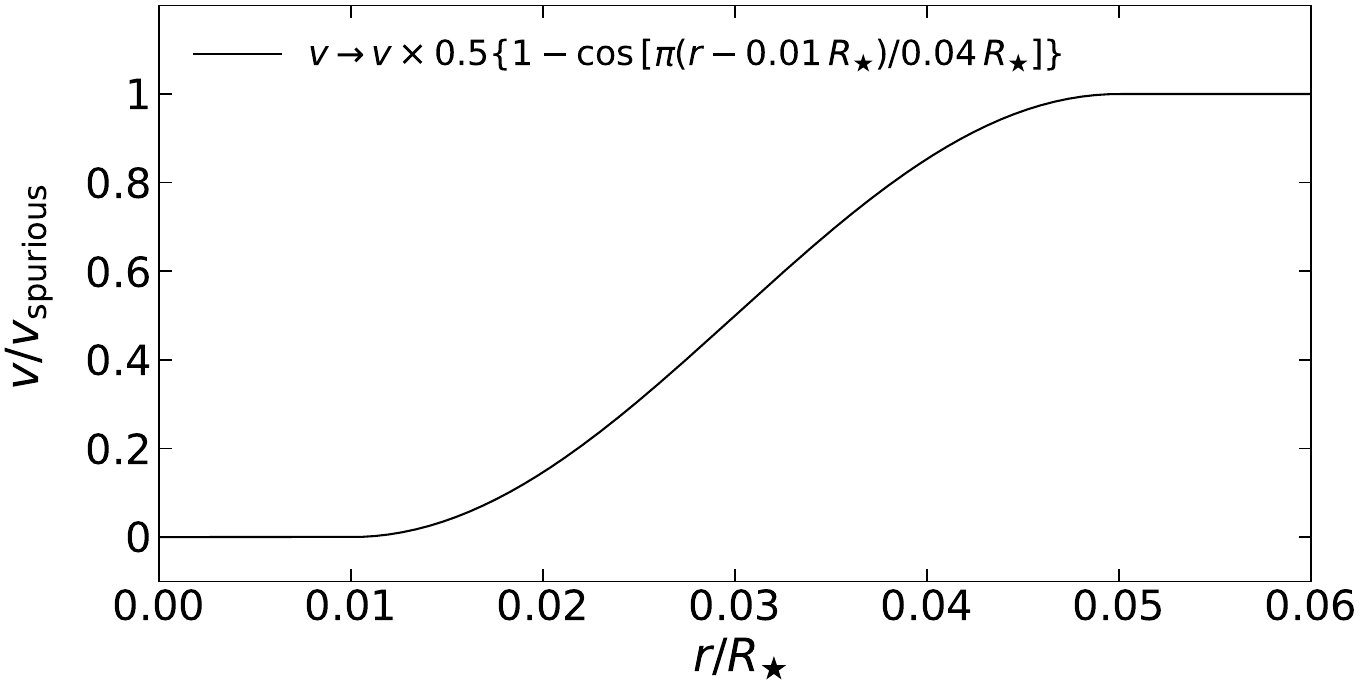}
\caption{A plot of the stabilisation function of random or spurious velocities at the very inner region of the star given by Eq.~\eqref{spongefunc}. Any such a velocity that develops randomly inside the star above the relative radius $0.05\,R_\star$ is damped only by the term $\dot{v}=-v/\tau$ (see Sect.~\ref{numthree}) while below $0.01\,R_\star$ it is zero.}
\label{fig_sponges}
\end{figure}
 In addition to what is already described in Sects.~\ref{numtwo} and \ref{numthree}, we add two explanatory figures~\ref{fig_steltemporalstability} and \ref{fig_sponges}. Figure ~\ref{fig_steltemporalstability} shows the schematic stabilization charts plotted as the 1D slice through the star at the late times of our simulation, that is, during the 9th, 11th (the last reported) and 12th passage (excluded from the presented model) through the $L_\text{j}=10^{42}\,\text{erg}\,\text{s}^{-1},\,\theta=10^{\circ}$ galactic jet. The upper panel shows the evolution of the stellar density structure up to the 10th star-jet passage. In these slices, the star appears as stable, while it shows very little increase in the matter content on the left side (the jet-incoming side) at the time shortly after the corresponding jet entry. At the approximate time of the jet exit, it shows a little drop in the mass near the stellar surface. The bottom panel shows the same 1D slice but even later, that is, at the end of the 11th star-jet passage and at the end of the 12th passage, which we did not include in the presented model due to the noticeable loss of the stellar stability. This was the main reason why we stopped the current modelling of the jet-star interaction process at this point. Identifying a more advanced and more long-term stabilisation scheme remains the biggest challenge for the future stages of this simulation. 

In Figure~\ref{fig_sponges}, we demonstrate the necessary stabilisation of the inner stellar core region. The plotted curve, see also Eq.~\eqref{spongefunc}, expresses the relative damping of spurious or random velocities induced by accumulated inaccuracies in the numerical process up to the radius $0.05\,R_\star$. We tested a large number of other possible stabilisation schemes and this one proved to be the most optimal, taking into account that this small inner stellar region does not enter the simulation process anyhow. Above this radius, the stellar structure is not stabilised artificially. We have only implemented the acceleration damping term $\dot{v}=-v/\tau$ explained in Sect.~\ref{numthree} that follows the principles described for red giants' simulations in \citet{2017A&A...599A...5O}. Following this study, we also adopted the stellar internal gravitational force profile, introduced in their Eq.~(10), in order to smoothly balance the gravitation from the massive stellar core. Such an artificial stabilisation of the inner stellar core may be problematic for future simulations, especially if the ablation process penetrates deep into the star. This then could change the studied physical processes and for this reason, we will need to significantly improve the stabilisation scheme, ideally to avoid such artificial damping functions.

\bsp	
\label{lastpage}
\end{document}